\documentclass[twocolumn,aps,superscriptaddress,pra]{revtex4}

\usepackage{amsmath,amssymb,graphicx}
\usepackage{subfigure}
\usepackage{algorithmic}
\usepackage{enumerate}
\usepackage{times}
\usepackage{color}
\usepackage{soul}
\definecolor{yblue}{rgb}{0.06, 0.3, 0.57}
\usepackage[pdftex]{hyperref}
\hypersetup{colorlinks=true,linkcolor=blue,citecolor=blue,urlcolor=blue}

\begin{document}

\title{Instabilities of vortex-ring-bright soliton in trapped binary 3D Bose-Einstein condensates}

\author{Victor P. Ruban}
\email{ruban@itp.ac.ru}
\affiliation{Landau Institute for Theoretical Physics, RAS, Chernogolovka, 
Moscow region 142432, Russia}

\author{Wenlong Wang}
\email{wenlongcmp@scu.edu.cn}
\affiliation{College of Physics, Sichuan University, Chengdu 610065, China}
\date{\today}

\author{Christopher Ticknor}
\email{cticknor@lanl.gov}
\affiliation{Theoretical Division, Los Alamos National Laboratory, Los Alamos, New Mexico 87545, USA}

\author{Panayotis G. Kevrekidis}
\email{kevrekid@umass.edu}
\affiliation{Department of Mathematics and Statistics, University of Massachusetts, Amherst MA 01003-4515, USA}

\begin{abstract}
Instabilities of vortex-ring-bright coherent structures in harmonically trapped two-component three-dimensional Bose-Einstein condensates are studied numerically
within the coupled Gross-Pitaevskii equations and interpreted analytically. 
Interestingly, the filled vortex core with a sufficiently large amount of the bright
component is observed to reduce the parametric interval of stability of the
vortex ring. We have identified the mechanisms of several linear instabilities
and one nonlinear parametric instability in this connection. Two of the linear
instabilities are qualitatively different from ones reported
earlier, to our knowledge, 
and are associated with azimuthal modes of $m=0$ 
and $m=1$, i.e., deviations of the vortex from the stationary ring shape.
Our nonlinear parametric resonance instability occurs between the $m=0$ and $m=2$ 
modes and signals the exchange of energy between them.
\end{abstract}

\maketitle

\section{Introduction}

The study of Bose-Einstein condensates (BECs) has offered  
for two and a half decades now an ideal playground
for the exploration of nonlinear phenomena~\cite{pethick,string,Panos:book}. Specifically,
the study of topological excitations has been of wide
interest to the research communities of atomic, nonlinear
and wave physics~\cite{Alexander2001,fetter2}. Indeed, relevant
reviews have focused not only on two-dimensional
vortical structures, but also on three-dimensional
vortex lines and vortex rings~\cite{Komineas}, as well as 
on more complex patterns including 
skyrmions~\cite{marzlin2000creation,mizushima2002mermin,reijnders2004rotating}, Dirac monopoles~\cite{dsh1} and quantum knots~\cite{dsh2,dsh3,Ruban:Knots,PhysRevA.99.063604}.

While the majority of the studies has naturally been directed
at the understanding of the single-component BEC setting,
the study of multi-component BECs has also attracted
considerable attention both in the two-component 
setting~\cite{KEVREKIDIS2016140}, but also in the case
of the so-called spinor condensates~\cite{kawaguchi2012spinor,stamper2013spinor}
of more than two components. Indeed, such multi-component
settings have offered an ideal framework for the exploration
of ideas of phase separation~\cite{Trippenbach_2000,Barankov2002, Lee2016, Indekeu2015}, but also for the manifestation of a
wide range of fluid-like instabilities. 
The latter include, among others, the 
 Rayleigh-Taylor instability~\cite{Sasaki,Gautam,Kadokura}, 
 the Kelvin-Helmholtz instability~\cite{Takeuchi,Suzuki,Baggaley},
 the capillary~\cite{Sasaki_K, Indekeu2018} and Richtmyer-Meshkov~\cite{Bezett} instabilities, as well as the countersuperflow~\cite{Law,Yukalov,Takeuchi_2,Hamner}, 
 but also the Rosensweig~\cite{Saito} instabilities. This wealth of findings
 clearly illustrates the fact that multi-component systems
 may possess a significant additional wealth of features, in 
 comparison with the simpler single-component ones.

It is exactly on this nexus of nonlinear topological coherent structures
and their instabilities, but focusing on the multi-component
variants thereof that our present study intends to focus.
Earlier work of different subsets of the present 
authors~\cite{Wang:hopfions,Wang:VR,Wang:DSVR,Ruban2017}
has explored the instabilities of chiefly one-component,
three-dimensional structures, such as vortex lines and vortex
rings, as well as multi-line/ring variants thereof. 
The last few years have led to a deeper and intensified consideration of 
vortical patterns bearing a second component that fills
the relevant vortex core~\cite{VB1,VB2,VB3}, as well
as their dynamics and instabilities~\cite{viktor_rec1,viktor_rec2},
and interactions with each other~\cite{richaud1,richaud2}
and with defects~\cite{wenlong_rec3}. Most of these above
studies have been centered around the somewhat less computationally
intensive, yet still quite interesting 2d realm. Our aim here
is to extend such multi-component, filled-vortex considerations
to structures arising in three-dimensional condensates. More
concretely, as our prototypical example, 
we will explore filled-core vortex rings.

\begin{figure}[htb]
\begin{center}
\includegraphics[width=\columnwidth]{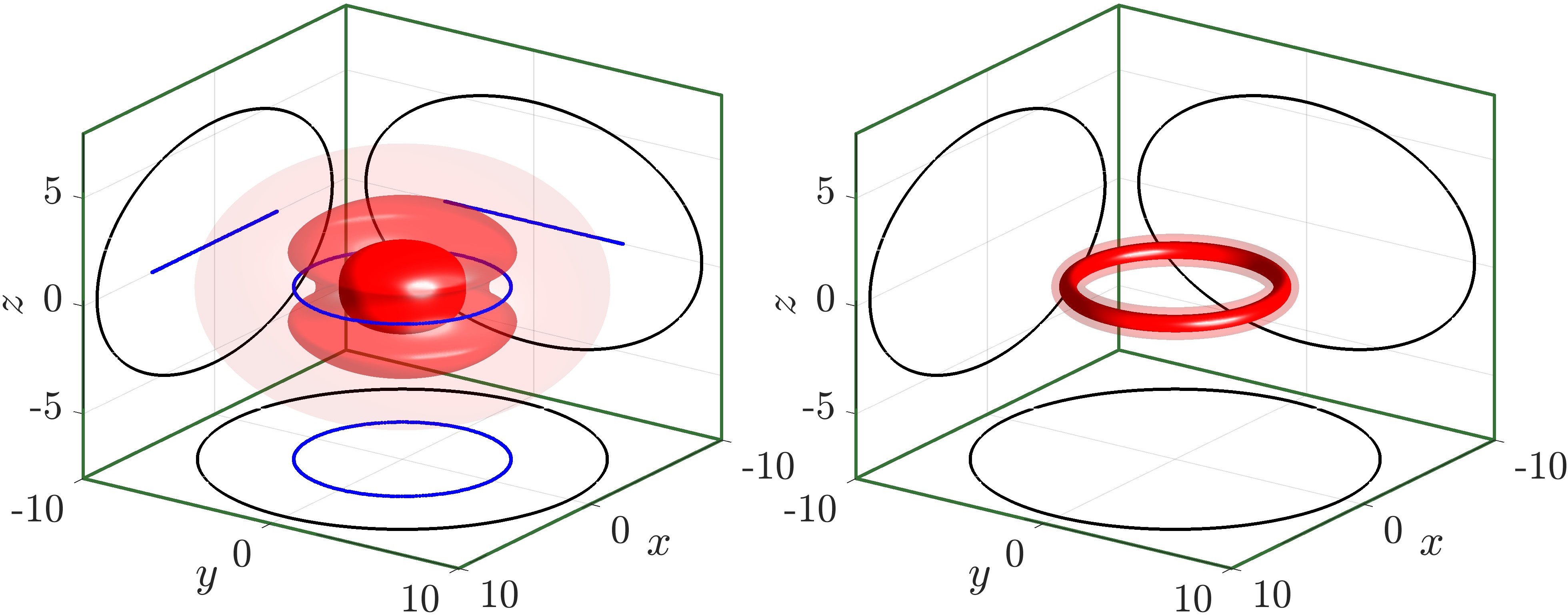}
\includegraphics[width=0.84\columnwidth]{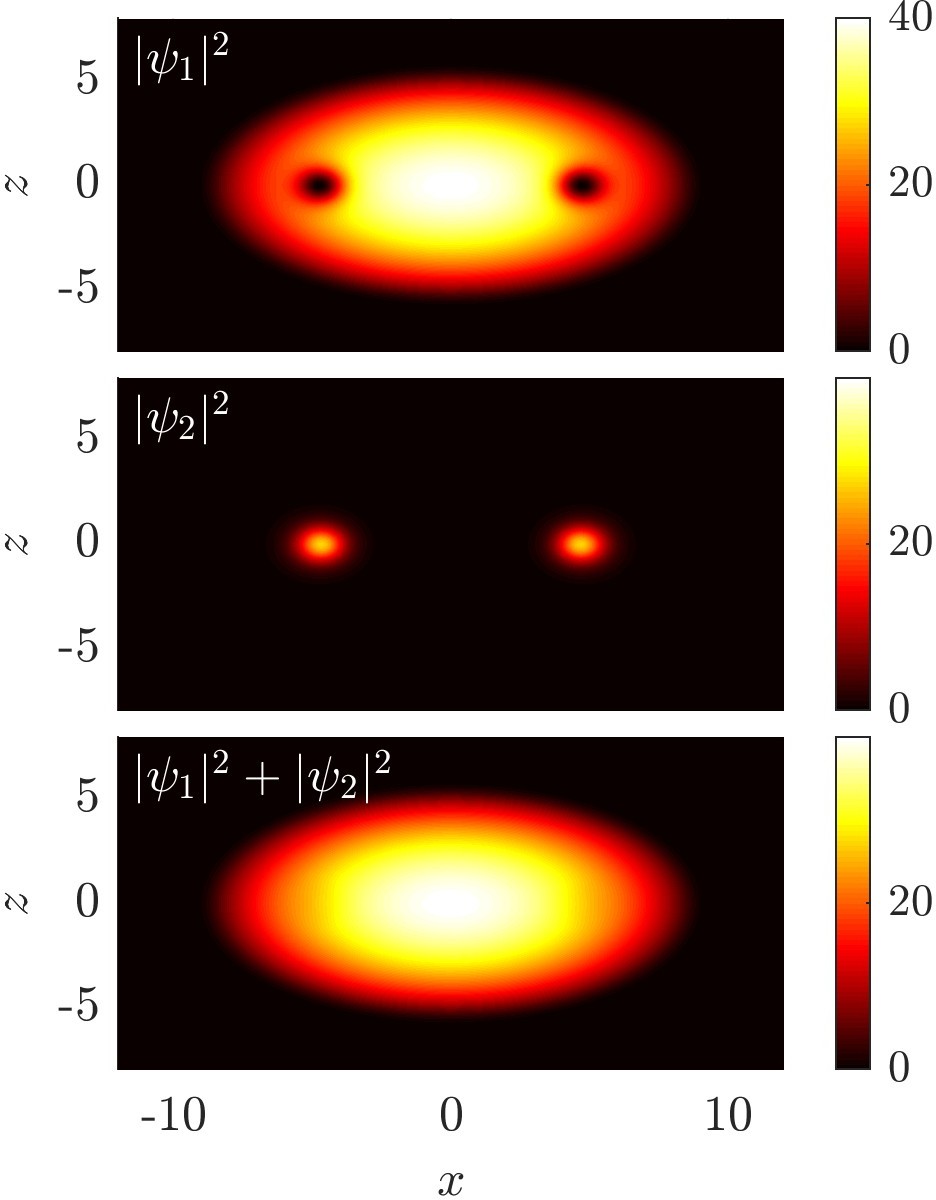}
\caption{
A typical, numerically exact fat-core VRB configuration with a relatively 
massive bright component at $\kappa=1.6, g_{12}=1.02, \mu_1=40$ and 
$\mu_2=39.6$. In the top row, the density contours are shown with the 
core of the VR highlighted in blue; the density projections
along the different planes are also shown. 
In the second-fourth row panels, the density projections 
along the $x$-$z$ plane are illustrated in detail.
Note that the configuration is rotationally symmetric
about the $z$-axis. The vortical (dipolar) pattern of the second
row and the bright (second component) pattern of the third row
add up to the total density featuring a Thomas-Fermi (TF) profile for
the density equal to $\max(\mu_1-V, 0)$ on the fourth panel.
}
\label{states3}
\end{center}
\end{figure}

\begin{figure}[htb]
\begin{center}
\includegraphics[width=\columnwidth]{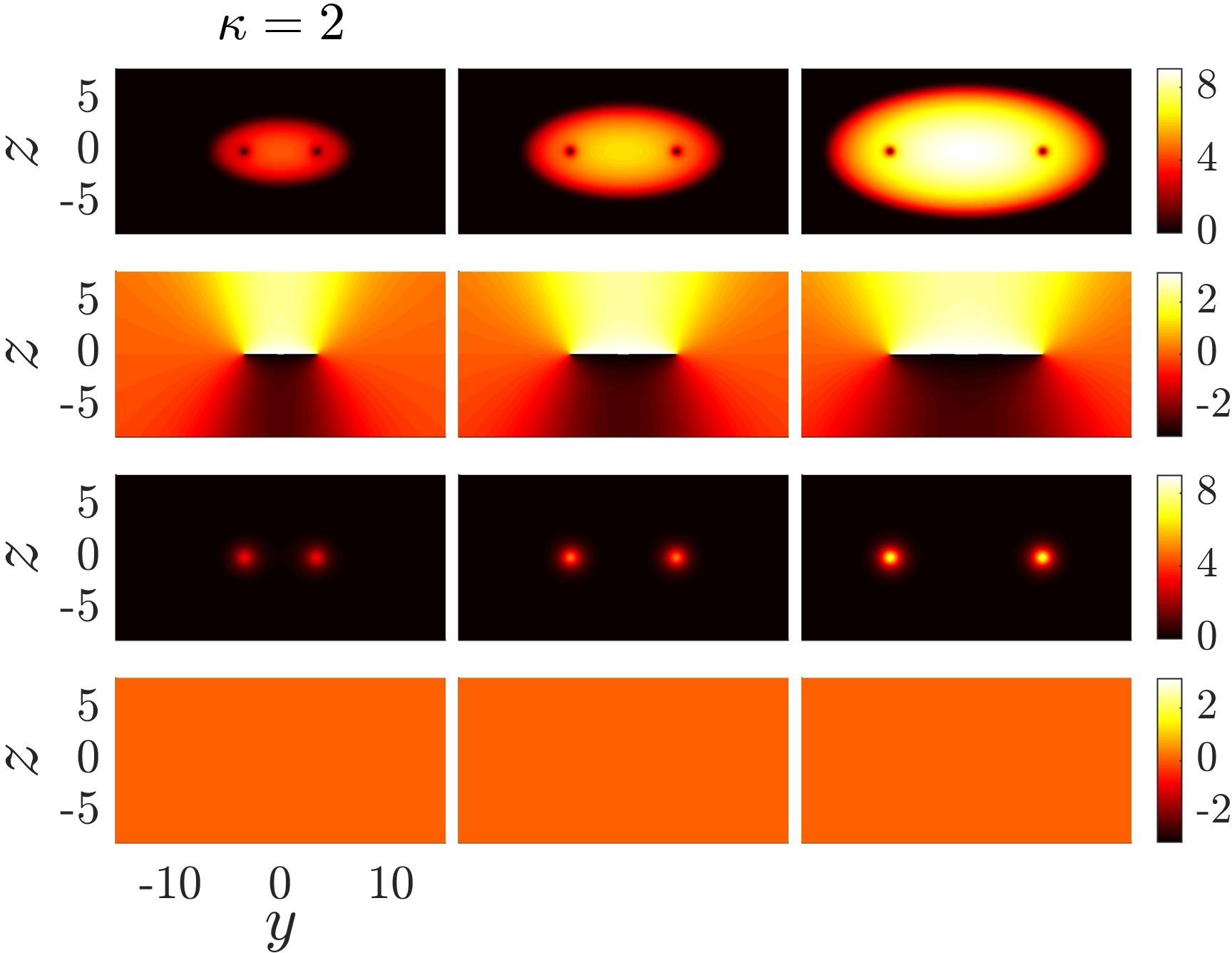}
\includegraphics[width=\columnwidth]{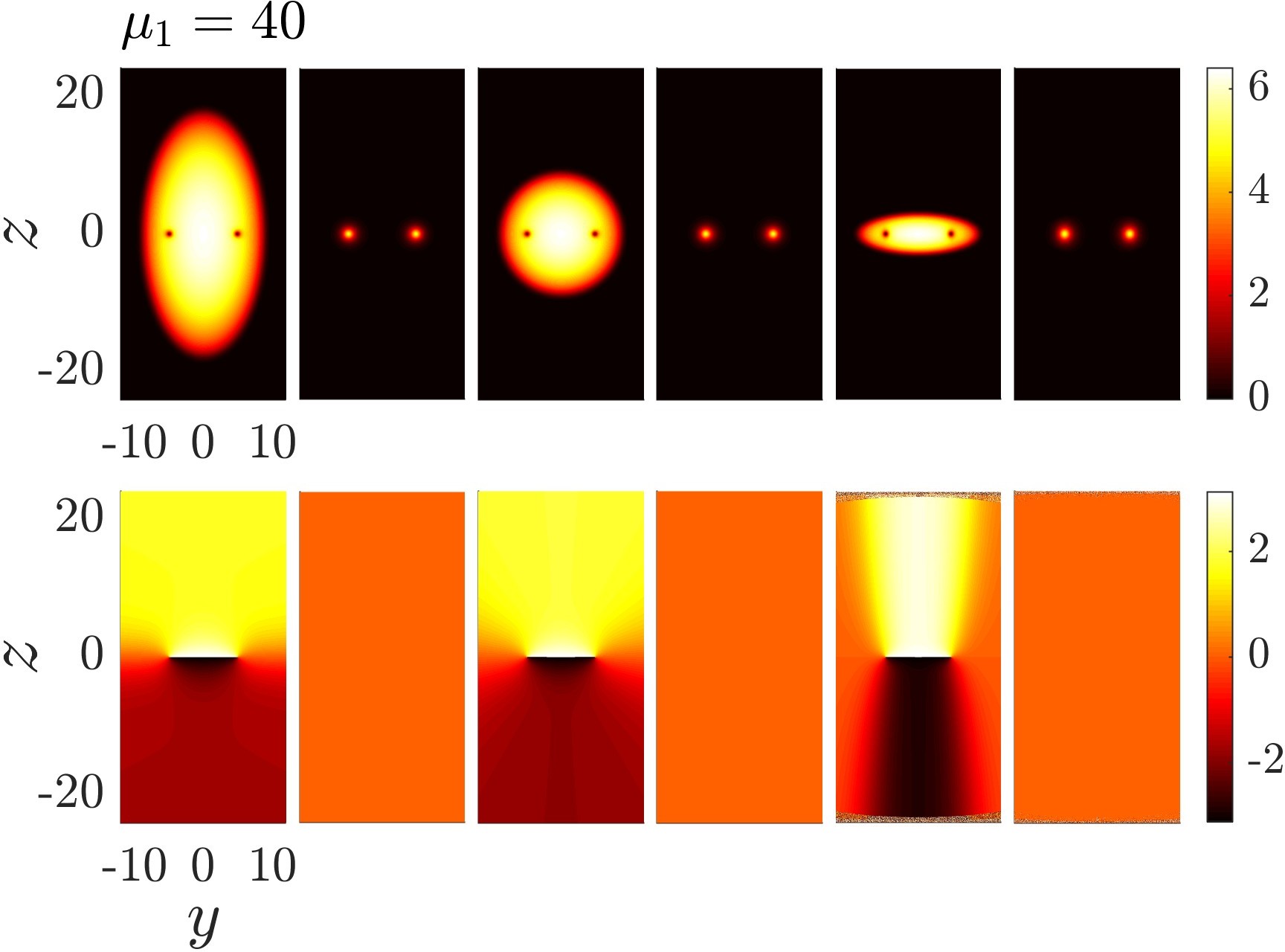}
\caption{
Typical stationary VRB configurations: the amplitude and phase profiles 
are depicted in odd and even rows, respectively. In top panels, the
three columns, in turn, are for chemical potentials $\mu_1=20, 40$ and 
$80$ at $\kappa=2$, see Eq.~(\ref{trajectory}) for the respective 
values of $\mu_2$. In the bottom panels, the state at $\mu_1=40$ above 
is continued in the aspect ratio to $\kappa=0.5, 1$ and $3$, respectively.
}
\label{states}
\end{center}
\end{figure}

The so-called vortex-ring-bright (VRB) structures 
(see some numerical examples in Figs. \ref{states3} and \ref{states}) 
are among the canonical generalizations of the widely studied
single-component vortex ring~\cite{Komineas,Wang:VR,Ruban2017}.
For the latter in our earlier work, we had explored~\cite{Wang:AI3}
numerically the theoretically predicted instabilities~\cite{Horng:VR}, 
finding that azimuthal perturbations of different modes of
azimuthal wavenumber $m$ become unstable in different regimes
of anisotropy of the condensate confinement. Indeed,
for prolate BECs, the mode with $m=1$ was found to be unstable
leading to tumbling motion of the VR structure. Weakly oblate
BECs represented the canonical regions of stability of the VR,
while sufficiently oblate ones led to the instability,
progressively of the $m=2$, $m=3$ etc. modes, splitting
the VR into two-, three- etc. vortex lines, respectively.
Here, our aim is to examine what happens to the same 
phenomenology, when the core of the ring vortex is 
filled by a second component. In line with the wealth
of phenomenology identified in earlier multi-component
studies, we find a variety of potential instabilities, including
some that are unprecedented, to the best of our knowledge. 
The  $m=2$ mode instability still occurs, but at a
wider range of anisotropies due to the presence of the
second component. For the $m=1$ mode, an oscillatory variant
of the relevant instability is newly emergent for a suitable range 
of atom numbers/chemical potentials. Finally, as concerns
linear instabilities, in a narrow parametric regime an
unprecedented instability of the $m=0$ mode is found to arise.
Lastly, we also identify a nonlinear (parametric) instability
stemming from the resonance between the $m=0$ and the $m=2$
modes, causing the exchange of energy between the two.

In what follows, our presentation of these phenomena
for the VRB structures is as follows. First, we 
qualitatively justify theoretically the origin of these instabilities,
after having presented the mathematical setup of our work. 
Then we present our computational setup, and in Sec.~\ref{results}, 
we support our theoretical analysis by means
of numerical computations of existence, stability and dynamics
of VRB structures in different parametric regimes.
Finally, in Sec.~\ref{conclusion}, we summarize our findings and 
present our conclusions. 

\section{Mathematical Setup \& Qualitative Theoretical Analysis}
\label{theory}
\subsection{The basic model}

As the basic mathematical model for our setup, we employ the widely recognized coupled 
Gross-Pitaevskii equations for two complex wave functions, $\psi_1({\bf r},t)$ 
(the vortical component), and $\psi_2({\bf r},t)$ (the bright component) \cite{pethick,string,Panos:book}. 
For simplicity, equal masses $m_1=m_2=m$ of the species atoms are considered.
Let an axisymmetric harmonic trap be characterized by a perpendicular frequency
$\omega_\perp$ and by an anisotropy $\kappa=\omega_\parallel/\omega_\perp$.
Using the trap units $\tau=1/\omega_\perp$ for the time,
$l_{\rm trap}=\sqrt{\hbar/(m \omega_\perp)}$ for the length, and  
$\varepsilon=\hbar\omega_\perp$ for the energy,
the equations of motion are written in dimensionless form 
\begin{eqnarray}
{\sf i}\dot \psi_1=-\frac{1}{2}\nabla^2 \psi_1
+\left[V({\bf r})+g_{11}|\psi_1|^2+g_{12}|\psi_2|^2\right]\psi_1,&&
\label{GP1}\\
{\sf i}\dot \psi_2=-\frac{1}{2}\nabla^2 \psi_2
+\left[V({\bf r})+g_{21}|\psi_1|^2+g_{22}|\psi_2|^2\right]\psi_2,&&
\label{GP2}
\end{eqnarray}
where  the overdot stands for the temporal partial derivative,
\begin{equation}
V({\bf r})=(x^2+y^2+\kappa^2 z^2)/2
\end{equation}
is the trap potential, while 
$g_{ij}$ is a symmetric $2\times 2$ matrix of nonlinear interactions.
Physically, the interactions are determined by the scattering lengths $a_{ij}$ 
\cite{Pu-Bigelow-1998}:
\begin{equation}
g^{\rm phys}_{ij}=2\pi \hbar^2a_{ij}(m_i^{-1}+m_j^{-1}), \qquad i,j=1,2.
\end{equation}
In this work, we consider the symmetric case $a_{11}=a_{22}=a>0$, so 
the self-repulsion coefficients are equal to each other. 
It should be noted that while the setting, e.g., of
$^{87}$Rb features slight differences between the $a_{ij}$'s~\cite{drummond}, 
it is possible to experimentally engineer the well-known, so-called Manakov case 
of equal interactions~\cite{lannig}. Without loss of generality, 
the relevant coefficients can be normalized to the unit value, 
$g_{11}=g_{22}=1$.
With this choice, the (conserved) numbers of trapped atoms are given by the relations
\begin{eqnarray}
&&N_1=\frac{l_{\rm trap}}{4\pi a}\int |\psi_1|^2 d^3{\bf r}=(l_{\rm trap}/a) n_1,\\
&&N_2=\frac{l_{\rm trap}}{4\pi a}\int |\psi_2|^2 d^3{\bf r}=(l_{\rm trap}/a) n_2.
\end{eqnarray} 
In real experiments the ratio $l_{\rm trap}/a$ ranges typically from a few hundreds
to a few thousands. In what follows, $n_1$ and $n_2$ will be important control parameters.
The cross-repulsion coefficient will be assumed as $g_{12}=1+g$, with a small positive 
parameter $g$. The condition $g>0$ is required for the phase separation regime to take 
place~\cite{tim,aochui}, as has also been experimentally
manifested in~\cite{Papp}.

Since we intend to consider soft excitations on a stationary background,
two more parameters will be used as well, namely the chemical potentials $\mu_1$ and $\mu_2$.
The numbers of particles $n_1$ and $n_2$ are dependent upon $\mu_1$ and $\mu_2$.
We are mainly interested in the Thomas-Fermi (TF) regime $\mu_1\gg 1$, when the
(single component) background 
density profile is given by a simple approximate formula
\begin{equation}
|\psi_{1}^{(0)}|^2=\rho({\bf r})\approx \mu_1 -V({\bf r}).
\end{equation}
The typical sizes of the trapped cloud are thus $R_\perp=\sqrt{2\mu_1}$ and $R_z=R_\perp/\kappa$, while 
a typical width of an empty vortex core is $\xi_*\sim 1/\sqrt{\mu_1}$.
Importantly, a filled vortex core can have a width $w$ which is essentially 
larger than $\xi_*$. Roughly $w$ can be estimated as
\begin{equation}
w\sim n_2^{1/2}\mu_1^{-3/4},
\end{equation}
since an effective volume of the bright component is  $\sim w^2\sqrt{\mu_1}$ 
(assuming a bright component that completely fills the vortex density dip;
such a regime with $\mu_1\approx\mu_2$ is typical for the critical phenomena 
under consideration; see Fig.~\ref{states3} for example),
while a typical density of the bright component is  $\sim\mu_1$.
The inequality $w > \xi_*$ can be the reason for instability of a certain kind,
as we will see further.

However, it is relevant to keep in mind that the VRB solutions may exist for a wide
range of $\mu_2$ and not just for $\mu_2$ comparable to $\mu_1$. 
A good example to draw analogies with is the dark-bright solitons of~\cite{DBS1} 
where the analytical solution makes it clear that roughly the solutions exist
for a very broad range of $\mu_2<\mu_1$.

\subsection{Variational approach and approximate Hamiltonian}

In the Thomas-Fermi regime, a typical time period for vortex motion is parametrically 
long, $\tau_{\rm vort}\sim\mu$, since the quantum of circulation is 
$2\pi\sim R_\perp^2/\tau_{\rm vort}$. On the other hand, typical periods
of potential oscillations (sound modes) are about 1 and larger. 
Thus, the latter hard degrees of freedom are well separated from the soft degrees of freedom. 
As sound modes are not excited (sitting at much higher 
frequencies in the TF limit), the dynamics of soft modes can be described 
self-consistently through an appropriate effective vortex Lagrangian \cite{Ruban2001}. 
Here we briefly discuss some basic properties of such a variational description.

Let us first recall that the commonly used inertia-free approximation for a long-scale 
dynamics of a closed vortex filament in a trapped single-component BEC corresponds 
to a Lagrangian functional of the general form \cite{Ruban2001,Ruban2017,Ruban2018}
\begin{equation}
{\cal L}=\Gamma\oint{\bf F}({\bf R})\cdot[{\bf R}_\beta \times {\bf R}_t]d\beta
-{\cal H}_v\{{\bf R}(\beta,t)\},
\label{Lagrangian}
\end{equation}
where $\Gamma=2\pi \hbar/m$ is the circulation quantum, and the unknown 
vector function ${\bf R}(\beta,t)$ describes the shape of the vortex in three dimensions, 
with $\beta$ being an arbitrary longitudinal parameter along the line. Here, the
Hamiltonian ${\cal H}_v$ is the vortex energy on the given density background. 
For self-consistency, the vector function ${\bf F}({\bf R})$ should satisfy the 
condition \cite{Ruban2017}
\begin{equation}
\mbox{div}_{\bf R}{\bf F}({\bf R})=\rho({\bf R}).
\end{equation}
The equation of motion in the vector form reads
\begin{equation}
\Gamma[{\bf R}_\beta\times{\bf R}_t]\rho({\bf R})=\delta{\cal H}_v/\delta{\bf R}.
\end{equation}
If the shape of the distorted vortex ring is given in the cylindrical coordinates 
by two real functions $R(\varphi,t)$ and $Z(\varphi,t)$ characterizing the radial 
extent of the VR and its $z$-location, then the two scalar equations of motion
are of the following non-canonical Hamiltonian form,
\begin{eqnarray}
 \Gamma\rho(R,Z)R\dot Z&=&\delta{\cal H}_v/\delta R,
 \label{eom1}
 \\
-\Gamma\rho(R,Z)R\dot R&=&\delta{\cal H}_v/\delta Z.
\label{eom2}
\end{eqnarray}

For configurations such as a moderately perturbed vortex ring, the vortex Hamiltonian
${\cal H}_v$ is often taken in the Local Induction Approximation (LIA), which 
corresponds to the deep TF regime~\cite{Ruban2001}:
\begin{eqnarray}
{\cal H}_v&\approx&{\cal H}_{\rm LIA}=
\frac{\Gamma^2}{4\pi}\Lambda \oint \rho({\bf R})|{\bf R}_\beta| d\beta\nonumber\\
&=&\frac{\Gamma^2}{4\pi}\Lambda \int \rho(R,Z)\sqrt{R^2+R'^2+Z'^2}d\varphi,
\label{H_LIA}
\end{eqnarray}
where $\Lambda=\ln(R_*/\xi_*)\approx\ln(\mu)$ is a large LIA constant, 
with $R_*$ being the equilibrium ring radius. 
It is easily derived from expression (\ref{H_LIA}) that  
in the LIA framework $R_*=R_\perp/\sqrt{3}$. 
When small deviations $\delta R(\varphi,t)$ and 
$\delta Z(\varphi,t)$ are considered, the second-order Hamiltonian is
\begin{equation}
{\cal H}^{(2)}_{\rm LIA}\propto \ln(\mu)
\sum_m \Big[(m^2-3)|\delta R_m|^2+ (m^2-\kappa^2)|\delta Z_m|^2\Big],
\end{equation}
where $m$ is the wavenumber of the azimuthal Fourier mode. 
It is this second-order contribution that leads
to the linearized motion of the VR in accordance with
Eqs.~(\ref{eom1}-\ref{eom2}).
This yields the corresponding 
eigenfrequencies of VR motion as \cite{Horng:VR,Ruban2017}
\begin{equation}
\omega^{\rm LIA}_m\propto\frac{\ln(\mu)}{\mu}\sqrt{(m^2-3)(m^2-\kappa^2)}.
\label{pred}
\end{equation}
Accordingly, the ring is stable in the parametric interval 
$1\leqslant\kappa\leqslant 2$.
The left edge of the stability interval is determined by the coefficient in front 
of $|\delta Z_1|^2$,
while the right edge is determined by the coefficient in front of $|\delta Z_2|^2$.  
This result is valid in the limit $\mu\to\infty$. 
As to finite values of $\mu$, the stability interval 
has been numerically found as $1\leqslant\kappa\leqslant \kappa_c(\mu)<2$ 
\cite{Wang:AI3}.
It is important for our present work that the critical value of the anisotropy
parameter $\kappa_c(\mu)$ increases together with $\mu$,
tending to its asymptotic limit provided by Eq.~(\ref{pred}). To interpret this fact
properly, we should recall that the LIA Hamiltonian (\ref{H_LIA}) is a limiting case 
of a more accurate non-local Hamiltonian functional \cite{Ruban2018}
\begin{equation}
{\cal H}_v=\frac{1}{2}\oint \oint 
G_{kl}({\bf R}_1,{\bf R}_2) R'_k(\beta_1)  R'_l(\beta_2) d\beta_1 d\beta_2,
\label{H_G}
\end{equation}
where $G_{kl}({\bf R}_1,{\bf R}_2)$ is the  (appropriately regularized) matrix Green
function for the auxiliary equation
\begin{equation}
\mbox{curl}\frac{1}{\rho({\bf r})}\mbox{curl}{\bf A} ={\bf\Omega}({\bf r}).
\end{equation}
Here ${\bf A}$ is a vector potential for the condensate flow around the vortex 
[the flow is nearly incompressible in the sense $(\nabla\cdot \rho{\bf v})\approx 0$,
and therefore ${\rho{\bf v}}\approx\mbox{curl}{\bf A}$], 
while ${\bf\Omega}$ is the singular vorticity distributed along 
the vortex central line.
Unfortunately, it is impossible to solve the above equation analytically with the 
density profile $\rho({\bf r})=\mu-(x^2+y^2+\kappa^2 z^2)/2$ 
(the only known 
analytical solution corresponds to the Gaussian  density profile and is expressed 
through a complicated integral \cite{Ruban2017-2}). Without an explicit  
Hamiltonian at hand, we are only able to extract just some general consequences of
this description. However, three of them are crucially important and are
briefly discussed below.

First off, it is evident that a small non-dimensional 
geometric regularization parameter is $\xi_*/R_* \sim 1/\mu$.
The corresponding second-order Hamiltonian for small deviations of the vortex ring
should take the form 
\begin{equation}
{\cal H}^{(2)}_v\propto\sum_m \Big[A_m(\kappa,\mu)|\delta R_m|^2+ 
B_m(\kappa,\mu)|\delta Z_m|^2\Big],
\end{equation}
with real coefficients $A_m=A_{-m}$ and $B_m=B_{-m}$.
At large values $\mu\gg 1$, these functions behave as 
\begin{eqnarray}
A_m(\kappa,\mu)&\approx&\tilde A_m(\kappa)+(m^2-3)\ln(\mu),\\
B_m(\kappa,\mu)&\approx&\tilde B_m(\kappa)+(m^2-\kappa^2)\ln(\mu),    
\end{eqnarray}
where $\tilde A_m$ and $\tilde B_m$ are finite regular functions corresponding to
essentially non-local parts of the interactions.
The squared eigenfrequencies are 
\begin{equation}
\omega_m^2\propto A_m B_m/\mu^2.
\end{equation}
For linear stability, this product should be positive. Due to general symmetry reasons, 
coefficient $B_1(\kappa,\mu)$ takes zero value at $\kappa=1$ for all $\mu$, 
so the left edge of the stable interval does not change and involves
the tumbling VR instability due to $m=1$ discussed for the
case of prolate condensates. But the coefficient 
$B_2(\kappa,\mu)$ becomes zero  at some critical value of the anisotropy parameter,
\begin{equation}
\kappa_c(\mu)=2-{\cal O}(1/\ln(\mu)).
\end{equation}
Hence, the latter instability associated with $m=2$
does depend on the specific value of the chemical
potential, as illustrated in~\cite{Wang:AI3}, reaching
the asymptotic limit of $\kappa_c=2$ only as $\mu \rightarrow \infty$.

The second important point is that, according to the general theory of Hamiltonian 
systems, there exist so-called normal complex variables 
\begin{equation}
{\sf a}_m\sim \frac{\sqrt{|A_m|}\delta R_m-{\sf i}\sqrt{|B_m|}\delta Z_m}
{\sqrt{2|\omega_m|}},
\end{equation}
such that the quadratic part of the Hamiltonian acquires an especially simple form,
\begin{equation}
{\cal H}^{(2)}_v=\sum_{m=-\infty}^{+\infty} \omega_m {\sf a}^*_m {\sf a}_m,
\label{H_2a}
\end{equation}
and the equations of motion are 
${\sf i}\dot{\sf a}_m=\partial{\cal H}_v/\partial{\sf a}^*_m$. 
Thus, in the linear approximation we have just a set of uncoupled harmonic oscillators. 
Their dynamics is reduced to rotation of the phases,
\begin{equation}
{\sf a}_m(t)\approx{\sf a}_m(0)\exp(-{\sf i}\omega_m t).
\end{equation}

Nonlinear interactions between the oscillators are described by cubic and 
higher-order terms in the Hamiltonian expansion in powers of ${\sf a}_m$.
In particular, the three-wave Hamiltonian is of the general form
\begin{eqnarray}
{\cal H}_v^{(3)}\!=\!\frac{1}{6\!}\sum \!\delta_{m_1+m_2+m_3}
[U_{m_1,m_2,m_3}{\sf a}^*_{m_1}{\sf a}^*_{m_2}{\sf a}^*_{m_3\!}+\! c.c.]&& \nonumber\\
+\frac{1}{2}\sum \delta_{m_1+m_2-m_3}
[V_{m_1,m_2,m_3}{\sf a}^*_{m_1}{\sf a}^*_{m_2}{\sf a}_{m_3}+c.c.],&&
\label{H_3a}
\end{eqnarray}
with some interaction coefficients $U$ and $V$.

The third important observation is that within the stability interval the coefficients
$A_0$, $B_0$, $A_1$, and $B_1$ are negative, while all the coefficients for 
$m\geqslant 2$ are positive. This fact implies that
the values $\omega_0$  and $\omega_1=\omega_{-1}$ have the negative sign \cite{Ruban2017}.
Physically this indicates the opposite direction of rotation for modes with $m=0$
and $m=\pm 1$.

\subsection{Linear $m=2$ instability}

Let us now consider a vortex ring with the core filled by the second component. 
The first instability encountered upon increasing 
anisotropy (past the spherical condensate limit) is the one
associated with $m=2$ and hence this is the one with which
we start our considerations.
We may assume that in some cases the role of the bright component is reduced mainly to
increasing an effective relative width of the vortex core. In other words, a ``fat'',
filled vortex ring can behave similarly to a vortex ring in a one-component system 
but with a smaller $\mu$, up to rescaling of the time variable. In such cases, the 
geometric regularization parameter becomes $w/R_*\sim n_2^{1/2}\mu_1^{-5/4}$ instead of 
$\xi_*/R_* \sim 1/\mu_1$, so the critical anisotropy value 
(corresponding to the $m=2$ instability) changes to 
\begin{equation}
\tilde\kappa_c(n_2,\mu_1)\approx 2-{\cal O}(1/\ln[n_2^{-1/2}\mu_1^{5/4}])< \kappa_c(\mu_1).
\end{equation}
For a given $\mu_1$ and a given anisotropy parameter within the region 
$\tilde\kappa_c(n_2,\mu_1)<\kappa <\kappa_c(\mu_1)$, 
the filled vortex ring is unstable despite the fact that 
the corresponding empty-core ring is stable.
Apparently, there should exist a critical value $n_{2,c}(\kappa,\mu_1)$, such that 
$\tilde\kappa_c(n_{2,c},\mu_1)=\kappa$. The vortex-ring-bright structure becomes
unstable when $n_2>n_{2,c}$. In our numerical computations,
that will be reported below, we will indeed observe such 
an instability near the right side of the (empty-vortex stable) 
anisotropy interval $[1:\kappa_c(\mu_1)]$. Hence, the presence of the second 
component narrows the interval of the anisotropy parameter $\kappa$,
within the oblate condensate geometry, for which the VR is dynamically stable.

\subsection{Nonlinear parametric instability}
\label{NLPI}

The negative value for $\omega_0$ and positive value for $\omega_2$ 
additionally render possible a nonlinear parametric resonance corresponding 
to explosive three-wave nonlinear interaction of
$0\longleftrightarrow 3$ type and described by terms as
$[U{\sf a}^*_0 {\sf a}^*_2 {\sf a}^*_{-2}+c.c.]$ in the Hamiltonian expansion 
on powers of the normal complex variables. In Ref.~\cite{Ruban2017}, 
this phenomenon was considered for a single-component condensate 
(with a different density profile) within the simplified LIA approach.
In the present work, it is studied for binary condensates, within the fully 
three-dimensional model of the coupled Gross-Pitaevskii equations. 
  
The condition for this nonlinear resonance is
\begin{equation}
2\omega_2+\omega_0\approx 0,
\label{param}
\end{equation}
and it is satisfied near a definite value of the anisotropy $\kappa_{\rm prm}$, 
depending upon $n_2$ and $\mu_1$. It should be noted here that { $\kappa_{\rm prm}$ 
is sensitive to the logarithm of the effective ratio $w/R_*\sim n_2^{1/2}\mu_1^{-5/4}$. 
For comparison, in the empty-core case the solution is sensitive to the logarithm of
$\xi_*/R_*\sim 1/\mu_1$. Interestingly, the 
logarithmic correction is so essential
for condition (\ref{param}) that even at quite large $\mu_1\sim 40$ we cannot safely 
use the LIA prediction $\kappa_{\rm prm}=\sqrt{16/7}$ for the resonant value~\cite{Ruban2017}.
Thus, for a filled vortex ring in a binary condensate with realistic parameters, 
the LIA prediction yields practically inaccurate results.}
In general, as $n_2$ increases, $\kappa_{\rm prm}(n_2,\mu_1)$ decreases. 
The approximate description of this instability is given by a simplified fully 
integrable Hamiltonian with just three degrees of freedom,
\begin{eqnarray}
H_{\rm prm}&=&(\delta -2\omega_2)  |{\sf a}_0|^2  
+\omega_2(|{\sf a}_2|^2+|{\sf a}_{-2}|^2)\nonumber\\
&+& U({\sf a}^*_0{\sf a}^*_2{\sf a}^*_{-2} + {\sf a}_0{\sf a}_2{\sf a}_{-2}),
\label{H_param}
\end{eqnarray}
where $\delta$ is a small detuning parameter.
The width of the resonance depends  on $\delta$ and on the
initial wave amplitudes.
The growth of the amplitudes is not exponential in time. In particular, with 
$\delta=0$ there is a simple solution of the following form,
\begin{equation}
{\sf a}_2={\sf a}_{-2}=\frac{{\sf i}\exp(-{\sf i}\omega_2 t)}{U(t_0-t)},\quad
{\sf a}_0=\frac{{\sf i}\exp(2{\sf i}\omega_2 t)}{U(t_0-t)}.
\end{equation}
This  demonstrates a power-law growth and is, in principle, 
associated with a finite-time singularity, although the dynamics
saturates prior to such an event.
To analyze the system (\ref{H_param}) in general, one has to take into account
the two additional integrals of motion,
\begin{equation}
|{\sf a}_0|^2-|{\sf a}_2|^2=D_+,\qquad |{\sf a}_0|^2-|{\sf a}_{-2}|^2=D_-.    
\end{equation}
Let us introduce a new canonical complex variable 
${\sf c}=|{\sf a}_0|\exp[{\sf i}(\mbox{Arg}({\sf a}_0)+\mbox{Arg}({\sf a}_2)
+\mbox{Arg}({\sf a}_{-2}))]$. Accordingly, the dynamical system (\ref{H_param})
is reduced to just one degree of freedom, with an effective Hamiltonian
\begin{equation}
H_{\rm eff}=\delta |{\sf c}|^2
+U\sqrt{(|{\sf c}|^2-D_+)(|{\sf c}|^2-D_-)}({\sf c}^*+{\sf c}).
\end{equation}
The phase trajectories are level contours for the above expression in coordinates
$\xi=\mbox{Re}({\sf c})$ and $\eta=\mbox{Im}({\sf c})$. In particular, with $D_+=D_-=D$ 
we have the family of cubic curves
\begin{equation}
(\xi^2+\eta^2-D)(\delta/U +2\xi)=\mbox{const}.
\end{equation}
The above-mentioned analytic solution corresponds to $\delta=0$ and $\xi=0$.

The fully nonlinear three-dimensional system of coupled Gross-Pitaevskii equations 
behaves, as may be expected, in a more complicated manner. For instance, when 
shifted from the exact resonance condition sufficiently far by $\delta$, it may
demonstrate a recurrent behavior. However, an accurate theoretical description 
of the recurrence is impossible without taking into account the terms from the 
four-wave Hamiltonian ${\cal H}^{(4)}_v$ and higher orders. The above simplified
three-wave model is only valid at an initial low-amplitude stage (as discussed also
above), while the recurrence actually occurs at considerably larger wave amplitudes,
when the higher order nonlinear terms dominate the dynamics.

\subsection{Linear $m=1$ instability}

Besides the $m=2$ instability, which is qualitatively the same as in the 
single-component case, we have detected numerically a qualitatively different 
linear instability caused by the presence of the bright component. This instability
involves 3D modes with azimuthal number $m=1$, and it is 
particularly relevant in 
the middle of the parametric interval $1<\kappa<\kappa_c(\mu_1)$. 
It is important to highlight that this is a distinct
instability scenario than the $m=1$ case occurring for $\kappa<1$. 
Basically, the unstable mode is a mix of deviation of the vortex central line 
from the perfect axially symmetric circular (annular) 
shape and a nonuniform 
cross-section of the core along the vortex. For this new instability, we have not yet 
developed an accurate quantitative theoretical description. However, we believe 
that a proper qualitative explanation can be provided as follows.

The point is that the above described mechanism of instability for the $m=2$ mode was 
based on the assumption that the filling component does not present its own dynamics.
Such a regime is only possible if the corresponding degrees of freedom remain hard.
However, the assumption is apparently incorrect if nonuniform longitudinal oscillations 
of the bright component along the vortex filament become essentially excited due to 
softening of their effective potential energy. Qualitatively, the longitudinal flows 
are similar to a one-dimensional gas dynamics with an effective ``equation of state''.
The softening corresponds to an effective decrease in ``speed of sound'' at sufficiently
large ``gas density'', and it is presumably similar to the mechanism of the ``sausage''
instability for a classical hollow columnar vortex due to the presence of 
surface tension \cite{Ponstein}. In the case of binary BECs, the effective surface
tension between the two components is roughly proportional to $\sqrt{g_{12}-1}=\sqrt{g}$,
while the width of a domain wall between the components is proportional to
$1/\sqrt{g}$ \cite{tension}. {Of course, the applicability of the analogy with 
the classical picture assumes that the wall is narrow in comparison with an effective
vortex core radius. In our case, since we deal with small values of $g$, this assumption
is not valid, so the analogy with a classical vortex is quite distant.} 
Nevertheless, the softening of the longitudinal flows does take place as the amount 
of filling component is increased \cite{viktor_rec1}. In the trapped system, they couple
with the $m=1$ mode of ring shape oscillations and produce an oscillatory instability.
Mathematically, this coupling can be expressed by a quadratic Hamiltonian of the 
following form,
\begin{equation}
H_{\rm cpl}=-\frac{1}{2}p^2-\frac{\omega_1^2}{2}q^2 
+\frac{1}{2}P^2+\frac{\Omega_1^2}{2}Q^2 +\zeta q Q,
\label{coupling}
\end{equation}
where canonical variables $q$ and $p$ represent the ring deviations 
(they are proportional to $\delta R_1$ and $\delta Z_1$, respectively),
while $Q$ and $P$ represent the first Fourier harmonic of the longitudinal
oscillations of the filling component. The corresponding function $\tilde Q(\varphi,t)$
[``gas density''] is proportional to the density integral of the second component over
the cross-section $\arctan(y/x)=\varphi$ of the filled vortex. More precisely,
\begin{equation}
\tilde Q(\varphi,t)\propto\int |\psi_2(r,z,\varphi,t)|^2 r dr dz.
\end{equation}
A canonically conjugate function $\tilde P(\varphi)$ is basically proportional to 
the potential of longitudinal velocity of the bright component. 

The parameter $\omega^2_1$ here is the squared frequency of $m=1$ mode of transverse
oscillations as determined by coefficients $A_1$ and $B_1$ with a given ratio $w/R_*$. 
The frequency of the longitudinal mode is denoted as $\Omega_1$. It basically 
coincides with the longitudinal frequency of a straight filled vortex
at wavelength  $2\pi R_*$, for the same mean filling per unit length,
and for periodic boundary conditions.
There is also a coupling coefficient $\zeta$ between these two degrees of freedom
(the only admissible by symmetry reasons, but actually unknown due to the absence 
of explicit expression for the Hamiltonian).

It is important that the two negative signs in expression (\ref{coupling}) 
are in accordance with the opposite direction of rotation for the first mode 
of the ring shape deviations, while the longitudinal-flow degree of freedom,
when taken separately,
behaves as an ordinary oscillator with positively defined self-energy.

The eigenvalues of Hamiltonian (\ref{coupling}) are
\begin{equation}
\lambda_{1,2}^2=\frac{1}{2}\Big[-\omega_1^2-\Omega_1^2
\pm\sqrt{\big(\Omega_1^2-\omega_1^2\big)^2-4\zeta^2}\phantom{.}\Big].
\end{equation}
It is clear from the above expression that with $\Omega_1$ sufficiently close to 
$|\omega_1|$, the eigenvalues become complex, and an oscillatory instability takes 
place. It is interesting to note that with sufficiently small $\zeta$ the above 
formula predicts a finite unstable interval in $\Omega_1$, so a re-stabilization 
may occur when the softening of the longitudinal mode is too deep (small $\Omega_1$). 
This will be explored in our numerical computations below.

\subsection{Massive-core transverse instability}

Another important dynamical effect not covered by the Lagrangian (\ref{Lagrangian})
is the transverse inertia of the vortex core. For two-dimensional flows, this effect 
has been considered in recent works \cite{richaud1,richaud2}.
The three-dimensional case studied here is more complicated. In general, the local 
cross-section of a filled vortex core is not circular  but rather elliptic or 
even more distorted, and therefore it is less straightforward to perform an exhaustive
theoretical analysis for a ``fat'' distorted massive core. However, roughly we may take
into account only the most important parameter, i.e., the transverse mass (see below).
This is different from the longitudinal mass coinciding with the bright component 
mass per unit length of the vortex (the corresponding longitudinal flows have been
briefly discussed in the previous subsection). Here we consider the effect of 
transverse mass upon stability of a filled vortex ring in a trap. The transverse 
mass $M_\perp$ (a local characteristic per unit length)  includes (as a part)
the mass of the bright component, but also the added mass of the vortex component 
caused by the density depletion. This added mass is qualitatively similar to the well-known added mass in classical hydrodynamics, since every transverse motion of 
the core is accompanied by an additional potential flow in the vortex component. 
That flow is effectively localized on the scale of the core width.
Accordingly, an additional kinetic energy is associated with such flow. 
This kinetic energy is a quadratic functional in time derivatives of the vortex 
configuration. Therefore, it has to be added to the Lagrangian (\ref{Lagrangian}).  
Together with the transverse kinetic energy of the bright component, we have the term
$$
{\cal K}_\perp=
\frac{1}{2}\oint M_\perp(\beta,t) |{\bf R}_{t,\perp}|^2 |{\bf R}_\beta| d\beta. 
$$  
The above expression essentially contains the definition of the transverse mass.
Unfortunately, it is very difficult to calculate the added mass analytically, but 
it is of the same order as the longitudinal mass of a non-weakly filled core.
Therefore, as a simple estimate we may use the following formula,
$$
M_\perp \sim M_\parallel \sim n_2/\sqrt{\mu_1}.
$$

Linearized  equations of motion for small deviations of the ring now are
\begin{eqnarray}
&&-M_\perp R_*\ddot R +\Gamma \rho_* R_* \dot Z=\delta{\cal H}_v^{(2)}/\delta R,\\
&&-M_\perp R_*\ddot Z -\Gamma \rho_* R_* \dot R=\delta{\cal H}_v^{(2)}/\delta Z.
\end{eqnarray}
When written in Fourier representation, these equations take the simple form
\begin{eqnarray}
&&-\frac{M_\perp}{\Gamma\rho_*}\ddot R_m + \dot Z_m=\frac{\Gamma}{4\pi R_*^2}A_m R_m,\\
&&-\frac{M_\perp}{\Gamma\rho_*}\ddot Z_m - \dot R_m=\frac{\Gamma}{4\pi R_*^2}B_m Z_m.
\end{eqnarray} 
With a fixed mass, the mathematical structure of these equations is the same as for 
a particle in a constant magnetic field in the presence of an external quadratic potential.
The eigenfrequencies for the above system are determined by a bi-quadratic equation,
\begin{equation}
(\tau \omega_m^2-\hat A_m)(\tau \omega_m^2-\hat B_m)=\omega_m^2,
\label{PPspectra}
\end{equation}
where $\tau=M_\perp/(\Gamma\rho_*)$, while $\hat A_m=\Gamma/(4\pi R_*^2)A_m$ and
$\hat B_m=\Gamma/(4\pi R_*^2)B_m$. It is easy to see that if $\hat A_m$ and $\hat B_m$
are both negative, as is the case for $m=0$ and $m=1$, then at sufficiently
large values of $\tau$ the discriminant of the above equation becomes negative.
This signals the appearance of an oscillatory instability. 

For a ``completely filled'' VRB in the deep TF limit, we can roughly put 
$R_*=\sqrt{2\mu_1/3}$, $\rho_*=2\mu_1/3$, 
$M_\perp=4\pi n_2/(2\pi R_*)$, and
\begin{eqnarray}
A_m&=&(m^2-3)\ln(C_{\rm fit}\mu_1^{5/4}n_2^{-1/2}),\nonumber\\
B_m&=&(m^2-\kappa^2)\ln(C_{\rm fit}\mu_1^{5/4}n_2^{-1/2}),\nonumber
\end{eqnarray}
where $C_{\rm fit}$ is a fitting constant.
Then, since $\Gamma=2\pi$, we will have in Eq.~(\ref{PPspectra}):
$\tau=n_2/[\pi (2\mu_1/3)^{3/2}]$,
$\hat A_m=3A_m/(4\mu_1)$, $\hat B_m=3B_m/(4\mu_1)$.
With fixed parameters $\mu_1$ and $n_2$, we can solve Eq.~(\ref{PPspectra})
for $\omega_m(\kappa)$ and compare them to the numerical results,
similarly to Fig.~3 in the work \cite{Wang:AI3} on VRs. 
To explore this potential instability for $m=0$, as well as
more generally the above analytical predictions, we
will now turn to numerical computations.

\begin{figure*}[htb]
\begin{center}
\subfigure[]{\includegraphics[width=0.45\textwidth]{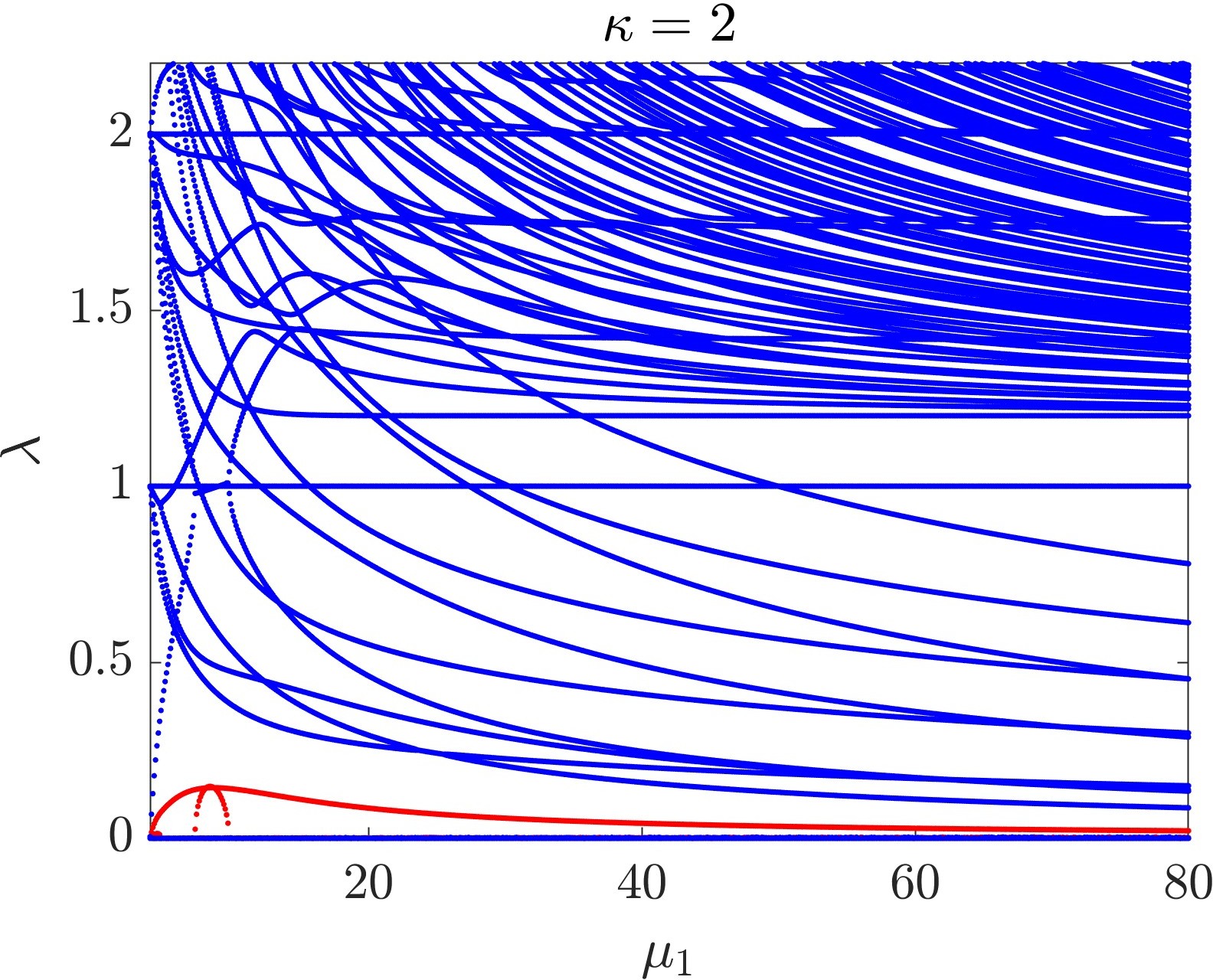}\label{VRBrun1}}
\hspace{10mm}
\subfigure[]{\includegraphics[width=0.45\textwidth]{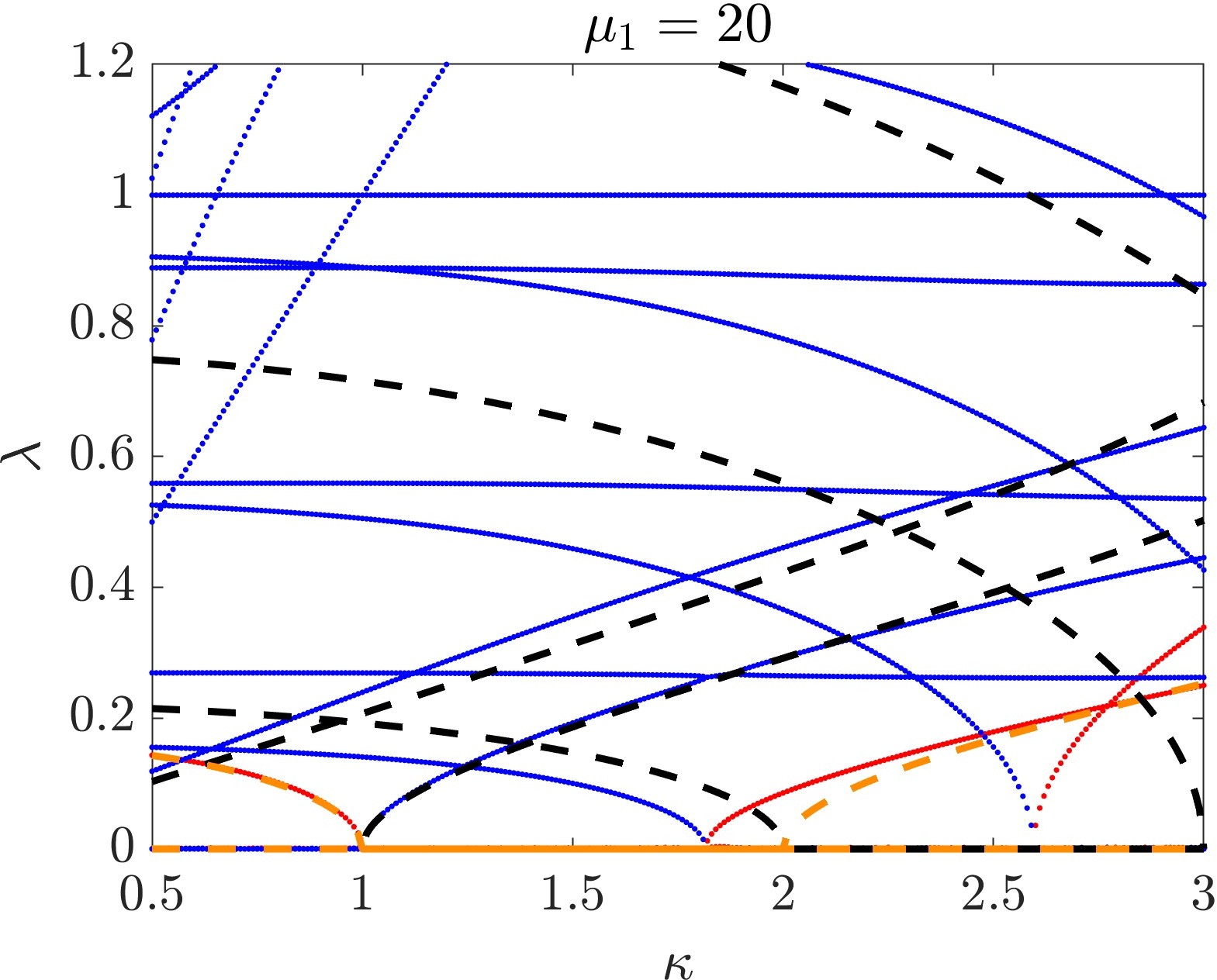}\label{VRBrun2}}
\subfigure[]{\includegraphics[width=0.45\textwidth]{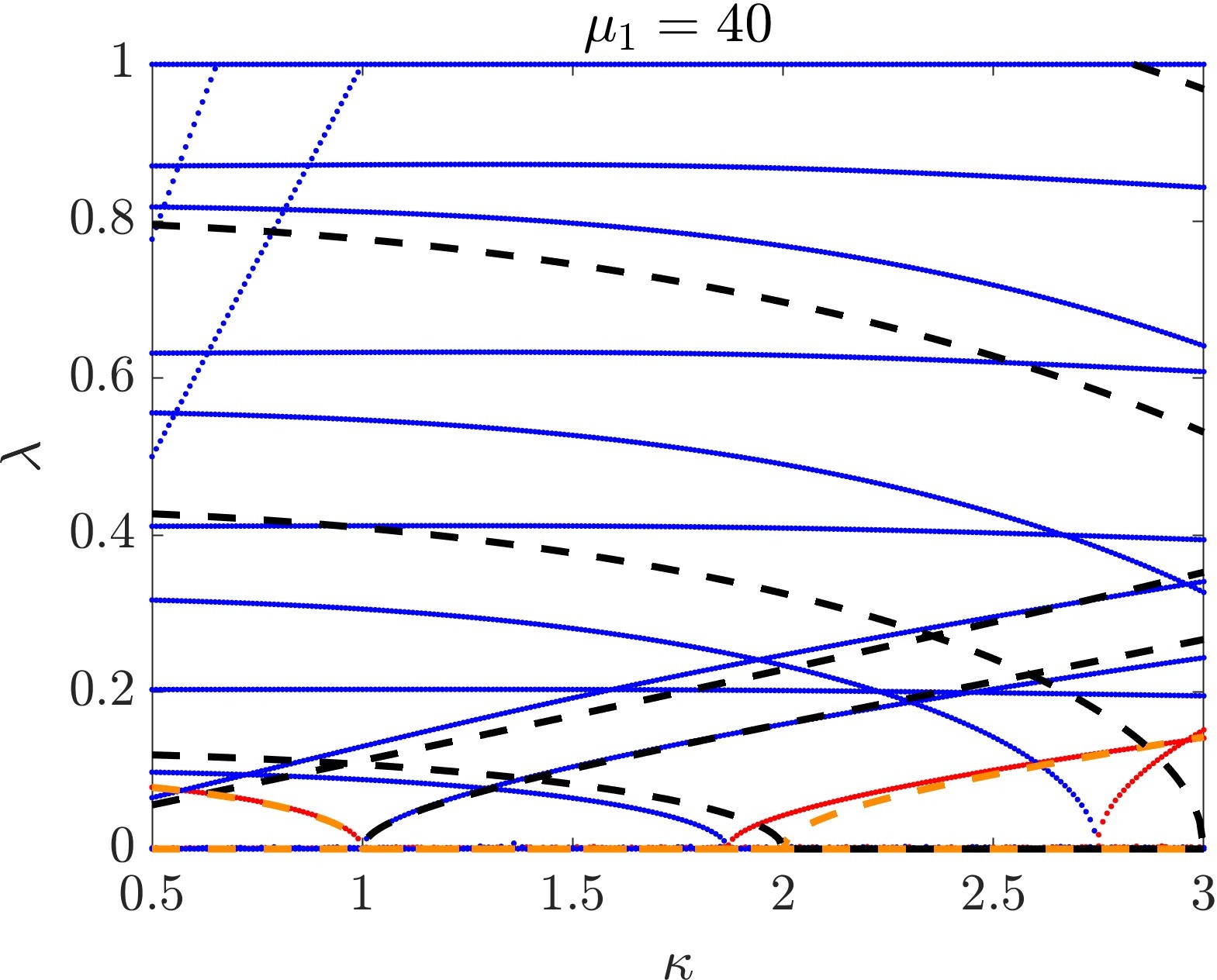}\label{VRBrun3}}
\hspace{10mm}
\subfigure[]{\includegraphics[width=0.45\textwidth]{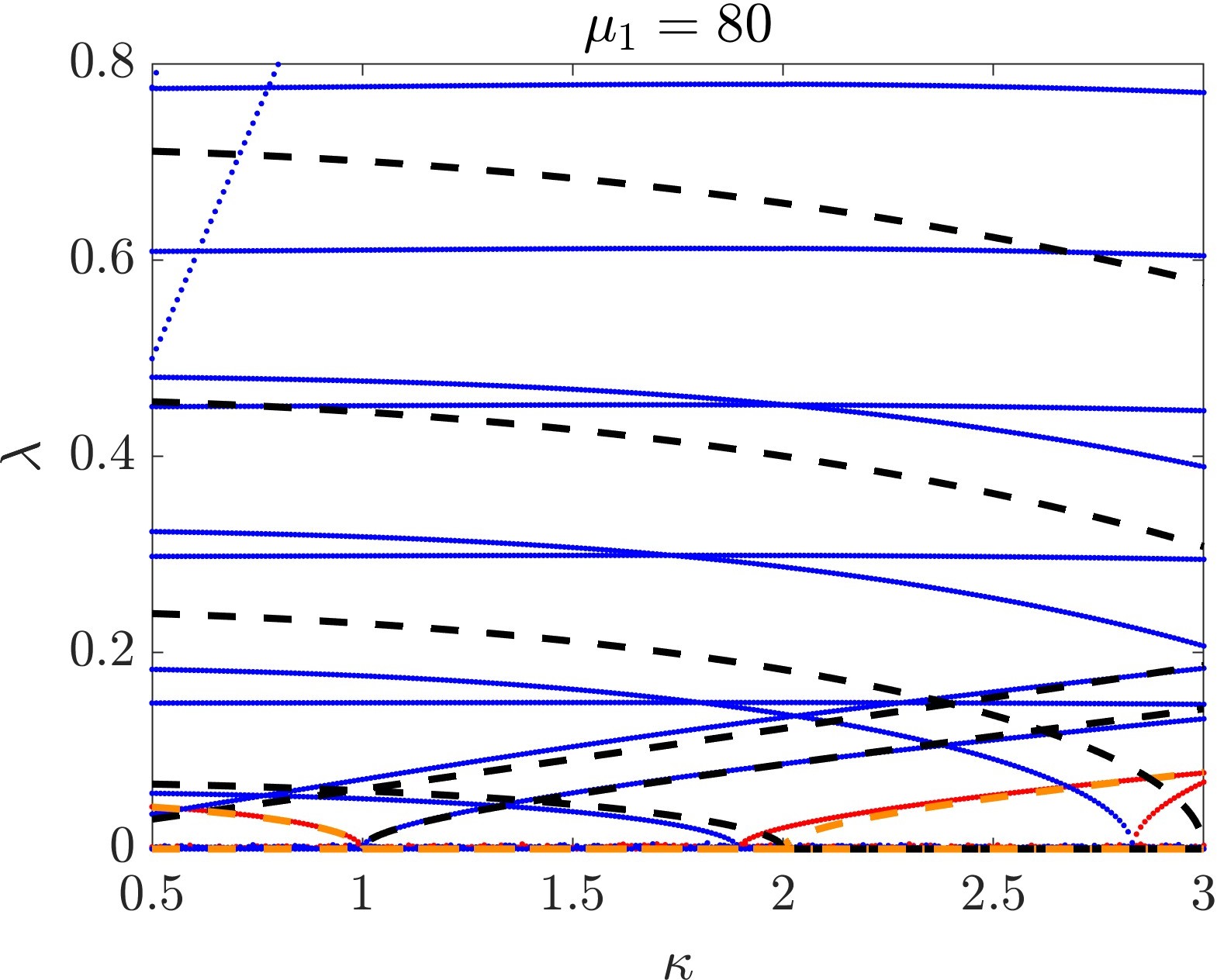}\label{VRBrun4}}
\caption{
The BdG spectra of the VRB along the continuation path (of Eq.~\eqref{trajectory}) by varying the chemical potentials at fixed $\kappa=2$ (a) the real and imaginary parts of $\lambda$ are shown in red and blue, respectively. In (b-d), typical states in the TF regime are further continued in $\kappa$ from the above path at fixed chemical potentials $\mu_1=20, 40$ and $80$, respectively. The gross feature is similar to that of the single-component VR, and the numerical spectra compare favourably with the analytically predicted spectra (of Eq.~\eqref{PPspectra}) shown as dashed gold (real part) and dashed black (imaginary part) lines. Notice the improving nature of the 
analytical predictions as $\mu_1$ increases. See the text for more details and an analogous comparison with the VR spectra in~\cite{Wang:AI3}.
}
\label{VRBspectrum}
\end{center}
\end{figure*}

\section{Computational Setup}
\label{setup}

Our numerical computation includes identifying numerically exact VRB stationary states, characterizing their stability properties through the Bogolyubov-de Gennes (BdG) spectral analysis~\cite{becbook1,Panos:book}, as well as exploring the VRB unstable dynamics. To find a stationary state, we apply the finite element method for a spatial discretization of the fields and subsequently utilize the Newton's method for convergence given a suitable initial guess; see the parametric continuation below for the numerical continuation of the VRB states from its underlying analytic linear limit. Since the 
 states are rotationally symmetric about the $z$-axis, we have 
 identified them in the reduced $(\rho, z)$ cross section, where $\rho=\sqrt{x^2+y^2}$. Upon convergence, the BdG spectrum of the state is collected using the partial-wave method \cite{Wang:DSS,Wang:DBS,10.1093/amrx/abr007}, again in the $(\rho, z)$ plane. In this work, we have collected the spectra of the pertinent low-lying angular Fourier modes $m=0, 1, 2, ..., 5$~\cite{Wang:DBS}. It is worth mentioning that this $2d$ reduced computation is far more efficient than a direct $3d$ computation, enabling us to investigate the VRB deep in the TF regime. Indeed, we have explored up to chemical potentials
 as large as $\mu_1=80$ for this three-dimensional structure in this work. Finally, when a genuinely $3d$ state is needed for dynamics, it is initialized from the $2d$ state using the cubic spline interpolation. Our dynamics is integrated using either the regular fourth-order Runge-Kutta method or a third-order operator splitting Fourier spectral method.

The VRB stationary states are parametrically continued from the underlying analytic linear limit at $(\mu_1=4, \mu_2=2)$ and $\kappa=2$. In this setting, the two small-amplitude fields decouple. The VR state is a complex mixing of the degenerate states of the ring dark soliton $(|200\rangle+|020\rangle)/\sqrt{2}$ and the planar dark soliton $|001\rangle$ with a relative phase of $\pi/2$. The bright component is in the $|000\rangle$ ground state. Here, $|mnp\rangle$ represents a $3d$ quantum harmonic oscillator state in the Cartesian coordinates. These linear states are used as the initial guess for slightly larger chemical potentials $(\mu_1, \mu_2)$ at fixed $\kappa=2$. When a solution is found near the linear limit, it is then used as the initial guess for larger chemical potentials at $\kappa=2$, and so on in this parametric 
continuation. By trial and error, we find that the following continuation path is both typical (the bright mass is neither too large nor small) and robust:
\begin{eqnarray}
\mu_2 = \mu_1 - 1.2 - 0.8\exp(-0.5(\mu_1-4)).
\label{trajectory}
\end{eqnarray}
In addition, when the chemical potentials are sufficiently large 
(i.e., $\mu_1 \gtrsim 10$), the existence of the VRB around this particular continuation path becomes robust and we can readily continue the states therein further in all other parameters, i.e., $\mu_2$, $\kappa$, and $g_{12}$. For example, for a typical $\mu_1$, $\mu_2$ can be tuned in a wide parametric range. The lower bound increases with increasing $\mu_1$ due to the tighter confinement of the vortex core and the upper bound is typically slightly below $\mu_1$. Here, the bright mass in the stationary state is tuned by adjusting $\mu_2$.

\section{Numerical results}
\label{results}

\subsection{Stationary states and the BdG spectra}

Several typical numerically exact VRB configurations are depicted in Fig.~\ref{states}. In this work, we have continued the VRB state following the continuation path of Eq.~\eqref{trajectory} from the linear limit all the way to $\mu_1=80$ which is deep in the TF regime at $\kappa=2$. Because the trap geometry is expected to have a strong influence on the VRB stability, we have explored the effect of $\kappa \in [0.5, 3]$ at three typical chemical potentials at $\mu_1=20, 40, 80$.

The BdG spectra of the VRB along these four continuation paths are summarized in Fig.~\ref{VRBspectrum}. Overall, the gross structure is qualitatively very similar to that of the single-component VR \cite{Wang:AI3}. The $m=1$ mode becomes unstable exactly below $\kappa=1$,
i.e., for prolate condensates. In the opposite direction, the $m=2$ mode becomes unstable at $\kappa >2$ (theoretically) and there is a finite chemical potential effect narrowing down the stability interval. The critical $\kappa$ moves closer to the theoretical limit $\kappa=2$ as $\mu_1$ is increased. Then, there is a similar trend for the $m=3$ mode, where the critical $\kappa$ moves towards the theoretical
limit of $\kappa=3$ with an even stronger finite chemical potential effect. Between the two regimes, the VRB has a stable regime in the interval $1<\kappa\lesssim 2$. Therefore, the VRB is most stable in a slightly oblate condensate.

The numerical spectra also compare favourably with the corresponding
theoretical prediction of Eq.~\eqref{PPspectra}, shown in dashed gold and dashed black lines for the real and imaginary parts, respectively. In Eq.~(\ref{PPspectra}), there is a fitting parameter $C_{\rm fit}$, which number is chosen to match the real part of the spectra, as motivated by the single-component VR work~\cite{Wang:AI3}. The $C_{\rm fit}$ is reasonably 
robust: our best fit at $\mu_1=20$ yields $C_{\rm fit}=2.5$, the ones at $\mu_1=40$ and $80$ are only slightly larger, yielding $C_{\rm fit}=2.6$ in both cases. The increasing stable mode with increasing $\kappa$ is the $m=0$ mode, the modes of $m=1, 2, 3$ can be readily identified due to their stability-instability transitions. Then, the yet higher-lying ones in increasing order correspond to $m=4$ and $5$, respectively. As $\mu_1$ increases, it is noted that the numerical spectra move closer to the theoretical predictions. This parallels the corresponding findings in the single-component VR comparison~\cite{Wang:AI3}, yet here we systematically generalize the results to the two-component VRB structure. 
Indeed, the latter is richer due to the modes associated 
with the presence of the second component.
For example at $\mu_1=80$, the almost horizontal lines around 
Im$(\lambda)=0.15, 0.3, 0.45, 0.6, 0.75$ are due to $m=1, 2, ..., 5$, respectively. 
Such almost horizontal and nearly equidistant lines in Figs.~\ref{VRBrun2}-\ref{VRBrun4} correspond 
to the longitudinal modes. Mathematically, they are similar to sound modes in a 
1D gas dynamics. In particular, the $m=1$ longitudinal mode, together with the $m=1$ 
transverse mode, produce the $m=1$ instability at larger $n_2$,
as we see in Fig.~\ref{VRBspectrum5}.

\subsection{Effect of the bright mass}
To understand the effect of the filled bright mass, it is helpful to compare the spectra of the VRB and VR in detail. To this end, we compare their stability intervals and typical growth rates qualitatively using several representative observables as summarized in Table~\ref{para}. From these details, we identify the following features:
\begin{enumerate}
\item The critical $\kappa_c=1$ for the mode $m=1$ appears to be exact for both the VRB and VR. For higher Kelvin modes (i.e., modes of
higher $m$), the critical $\kappa_c(m) \lesssim m$ and approaches $m$ as $\mu_1$ increases. In addition, large chemical potentials tend to 
decrease the unstable growth rates for both states.
\item The stability interval systematically shrinks when the VR is filled with a bright component, i.e., the bright component tends to narrow down the stability regime. 
\item Interestingly, when both the VRB and VR are unstable, the bright component tends to decrease the corresponding
growth rates. I.e., while the presence of the bright component expands the
region of instability, it concurrently weakens the instability growth
rates in the cases where the instability was already present.
However, this is not strictly satisfied, e.g., the bright component enhances $\lambda_r(m=3, \kappa=3)$ at small chemical potentials, and then lowers it at larger ones.
\end{enumerate}

\begin{table}[htb]
\caption{
Some observables comparing the spectra of the VRB in Fig.~\ref{VRBspectrum} and the corresponding ones of VR \cite{Wang:AI3}. Here, $\mu_1$ for the VR is understood as its chemical potential, and $\lambda_r=\mathrm{Re}(\lambda)$.
\label{para}
}
\begin{tabular*}{\columnwidth}{@{\extracolsep{\fill}} l c c c r}
\hline
\hline
States &observables &$\mu_1=20$  &$\mu_1=40$  &$\mu_1=80$ \\ \hline
VR  &$\kappa_c(m=1)$  &1 &1 &1 \\
\hline
VR  &$\kappa_c(m=2)$  &1.86 &1.9 &1.92 \\
\hline
VR  &$\kappa_c(m=3)$  &2.755 &2.845 &2.89 \\
\hline
VR  &$\lambda_r(m=1, \kappa=0.5)$  &0.1572 &0.09331 &0.05435 \\
\hline
VR  &$\lambda_r(m=2, \kappa=2)$  &0.08632 &0.04619 &0.02445 \\
\hline
VR  &$\lambda_r(m=3, \kappa=3)$  &0.2931 &0.1426 &0.07254 \\
\hline
VRB  &$\kappa_c(m=1)$  &1 &1 &1 \\
\hline
VRB  &$\kappa_c(m=2)$  &1.815 &1.87 &1.895 \\
\hline
VRB  &$\kappa_c(m=3)$  &2.595 &2.75 &2.83 \\
\hline
VRB  &$\lambda_r(m=1, \kappa=0.5)$  &0.1434 &0.07857 &0.04306 \\
\hline
VRB  &$\lambda_r(m=2, \kappa=2)$  &0.08574 &0.04279 &0.02149 \\
\hline
VRB  &$\lambda_r(m=3, \kappa=3)$  &0.3391 &0.1518 &0.06821 \\
\hline
\hline
\end{tabular*}
\end{table}

The narrowing of the stability interval by the bright mass suggests a stability-instability transition when $\mu_2$ is tuned, at least for the $m \geq 2$ modes. This is indeed observed as we shall discuss below. However, it is interesting that we find the $m=0$ and $m=1$ modes can also have such transitions in this scenario. In addition, these instabilities are 
unprecedented in the VR context, to our knowledge, as the eigenvalues are complex ones. By contrast, the eigenvalue of the unstable $m=1$ mode below $\kappa=1$ is purely real. Our dynamic simulations confirm that the instabilities are indeed distinct ones, i.e., they appear to be genuinely caused by the interplay between the VR and the bright core and therefore they are not present in the single-component VR.

\begin{figure*}[htb]
\begin{center}
\includegraphics[width=0.24\textwidth]{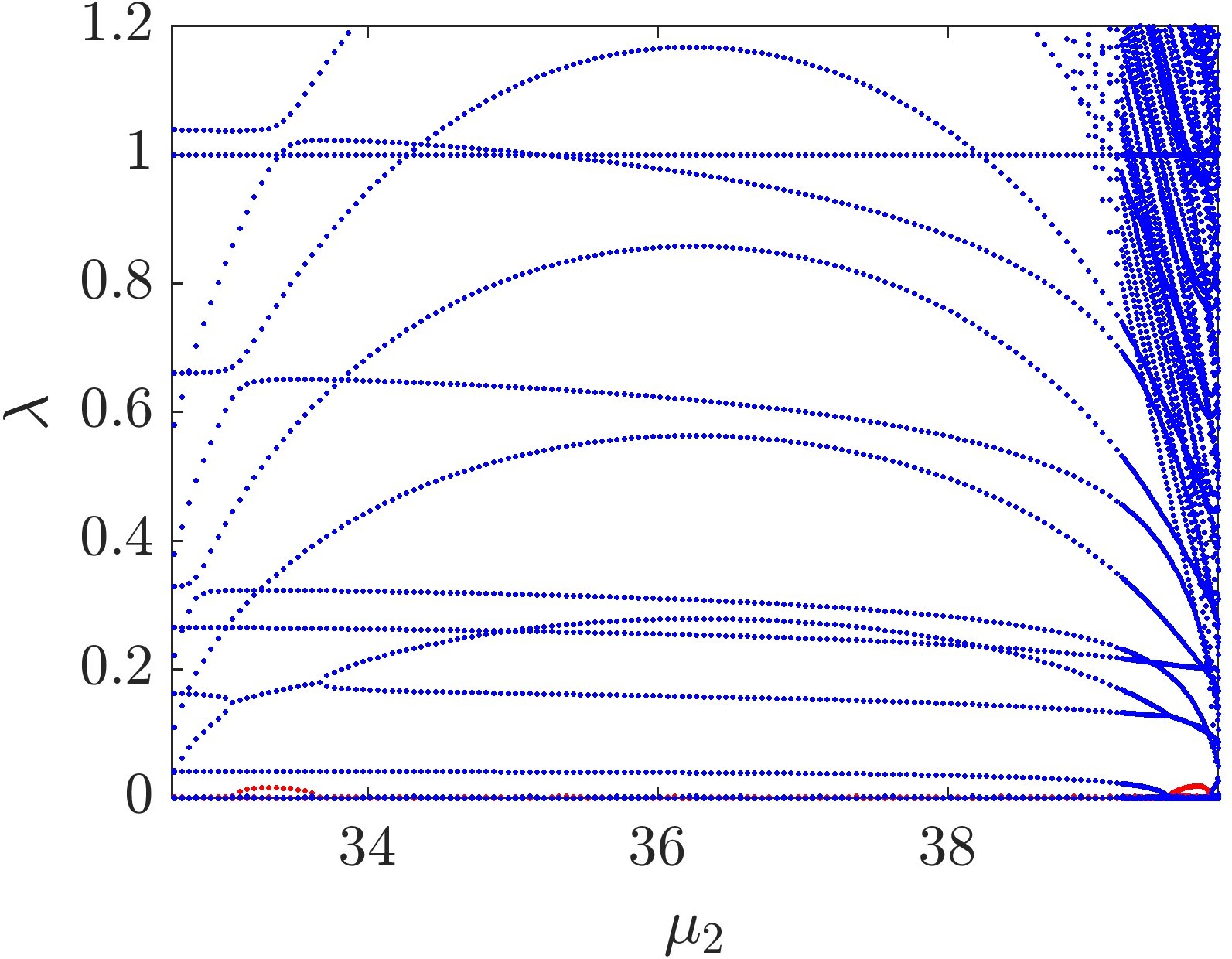}
\includegraphics[width=0.24\textwidth]{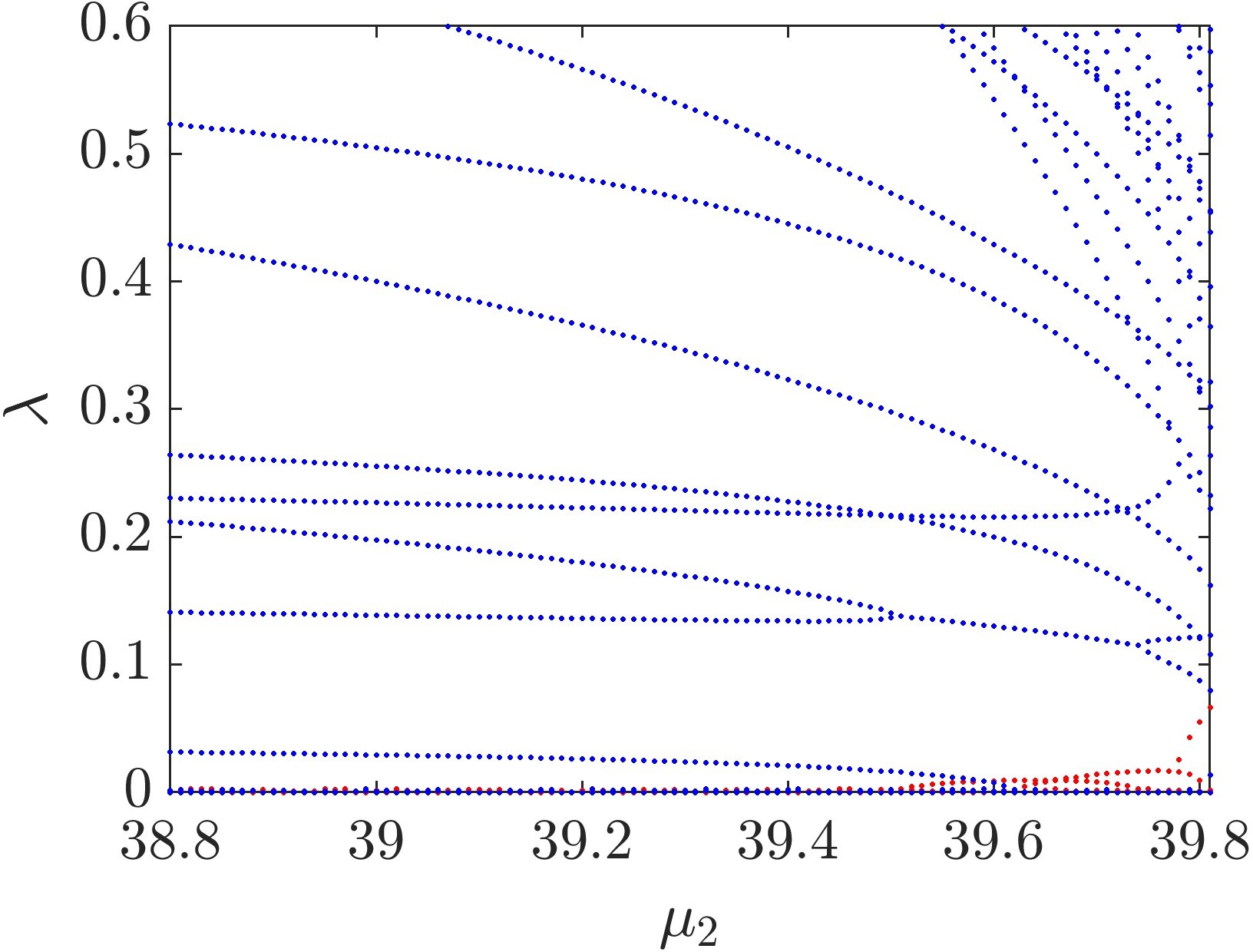}
\includegraphics[width=0.24\textwidth]{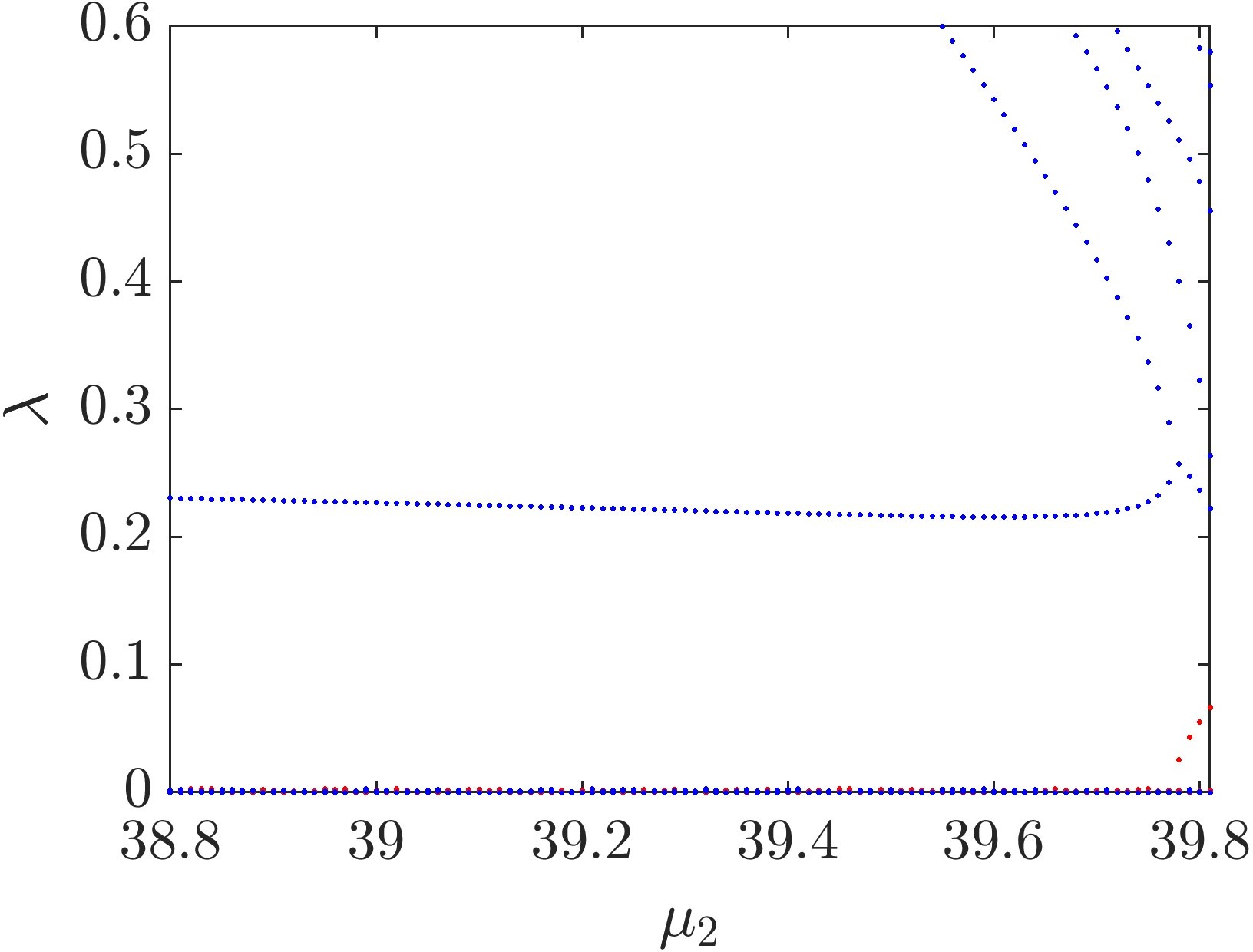}
\includegraphics[width=0.24\textwidth]{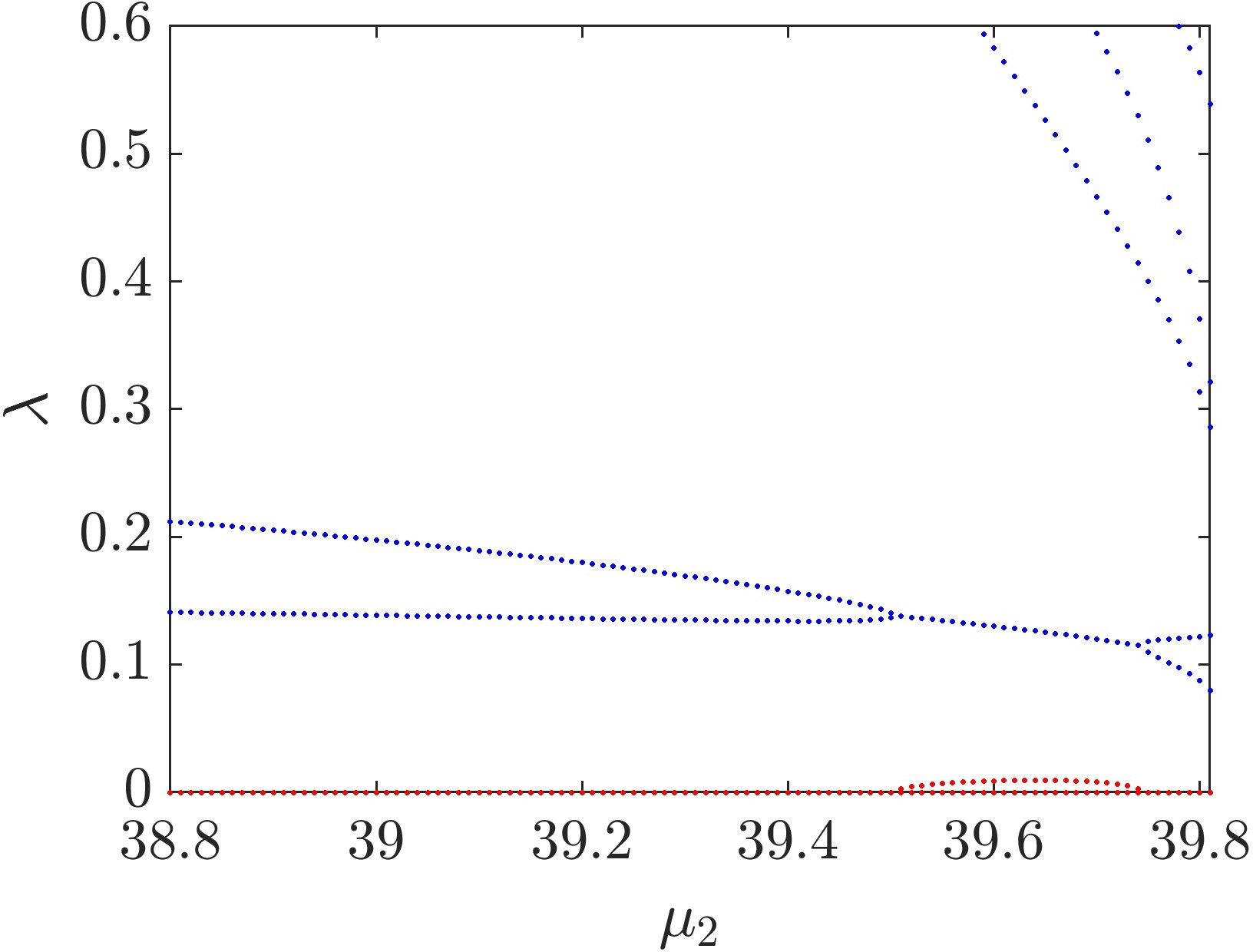}
\includegraphics[width=0.24\textwidth]{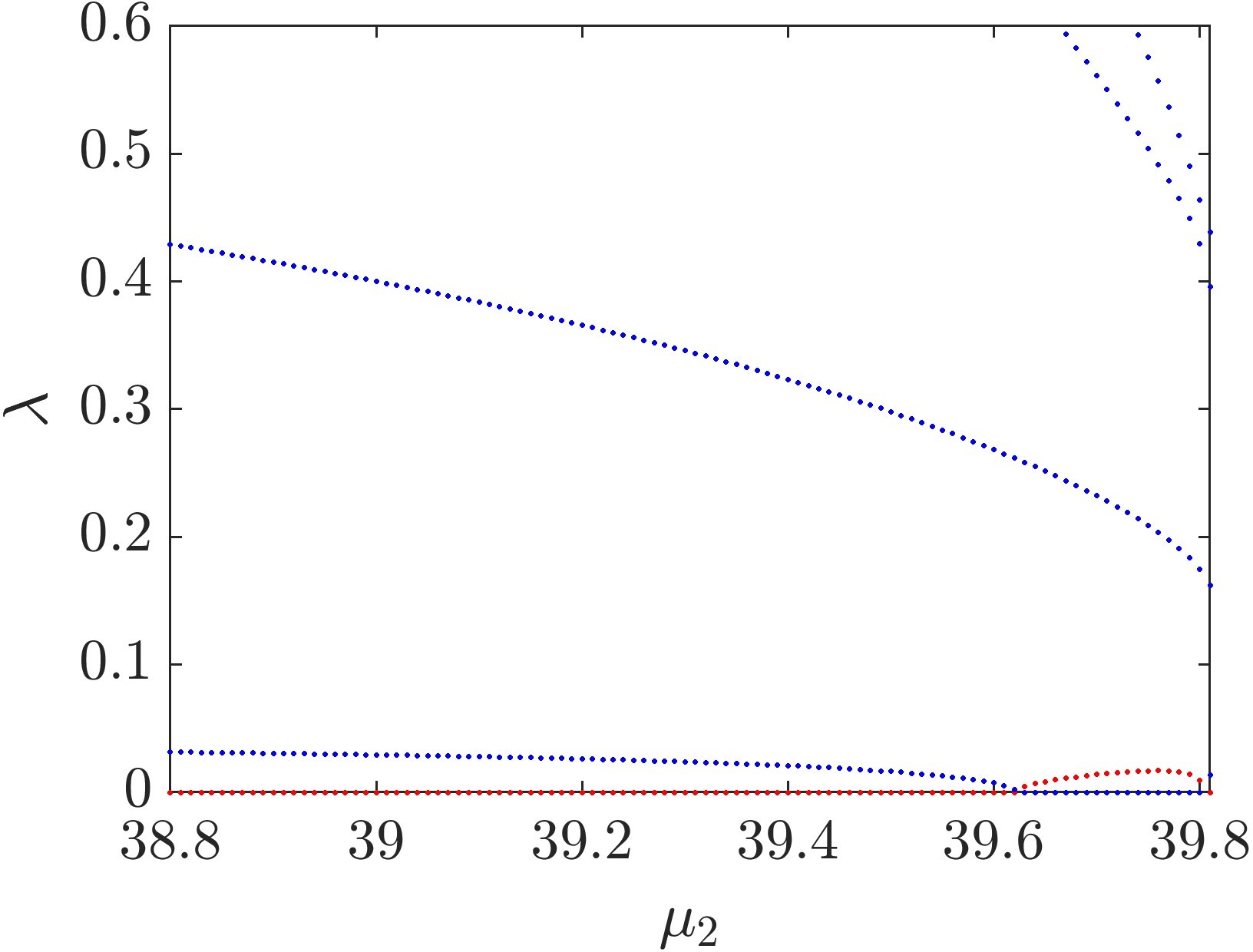}
\includegraphics[width=0.24\textwidth]{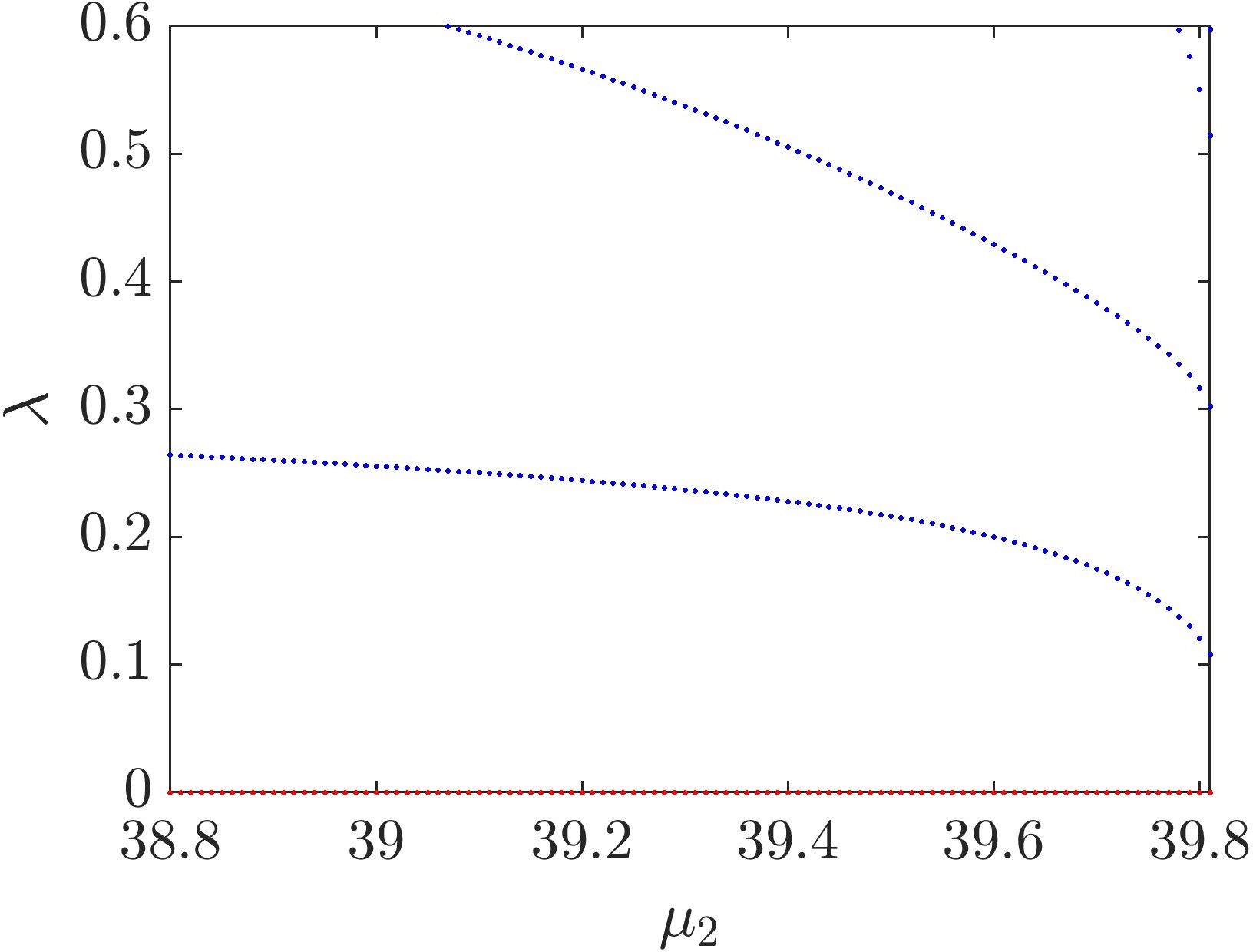}
\includegraphics[width=0.24\textwidth]{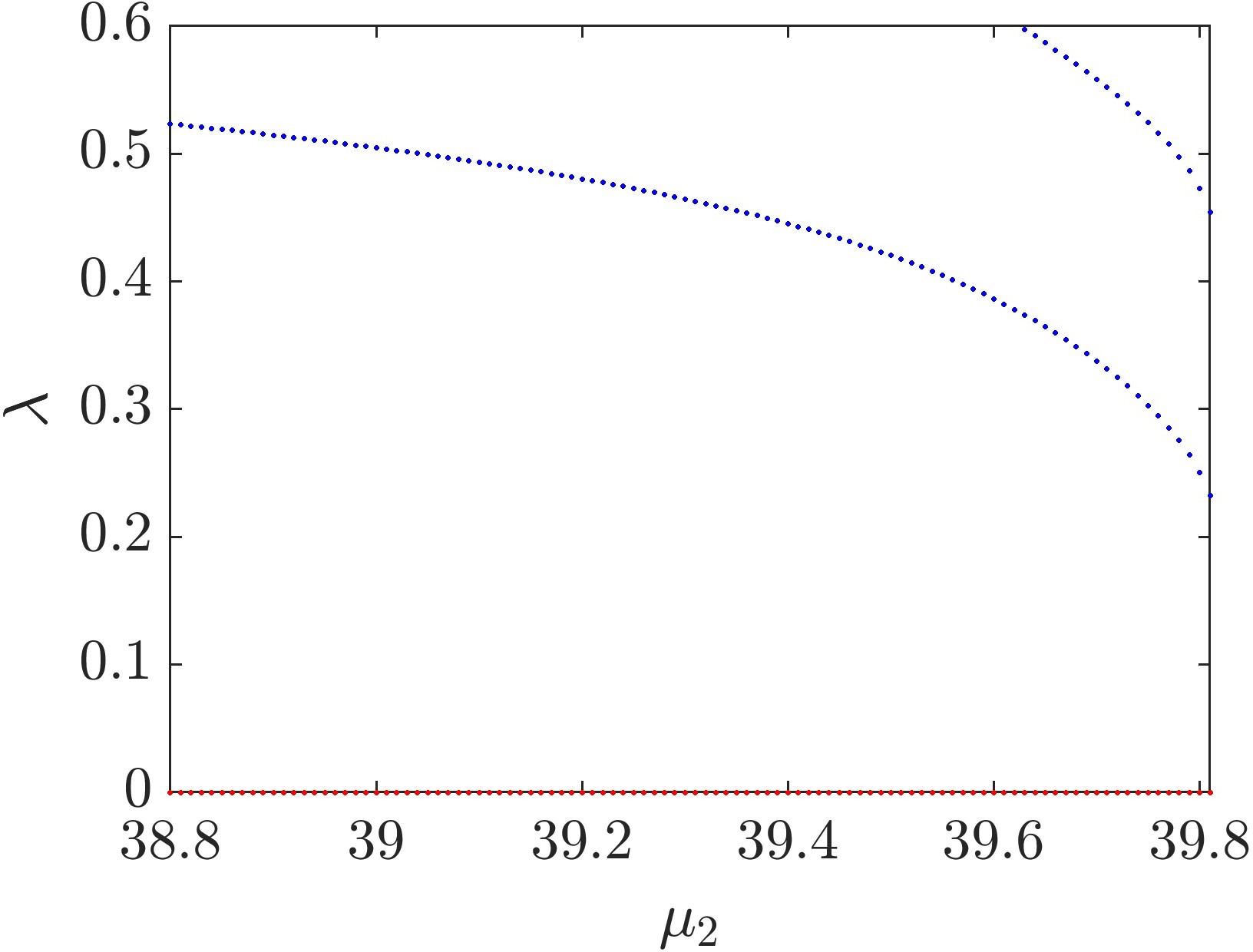}
\includegraphics[width=0.24\textwidth]{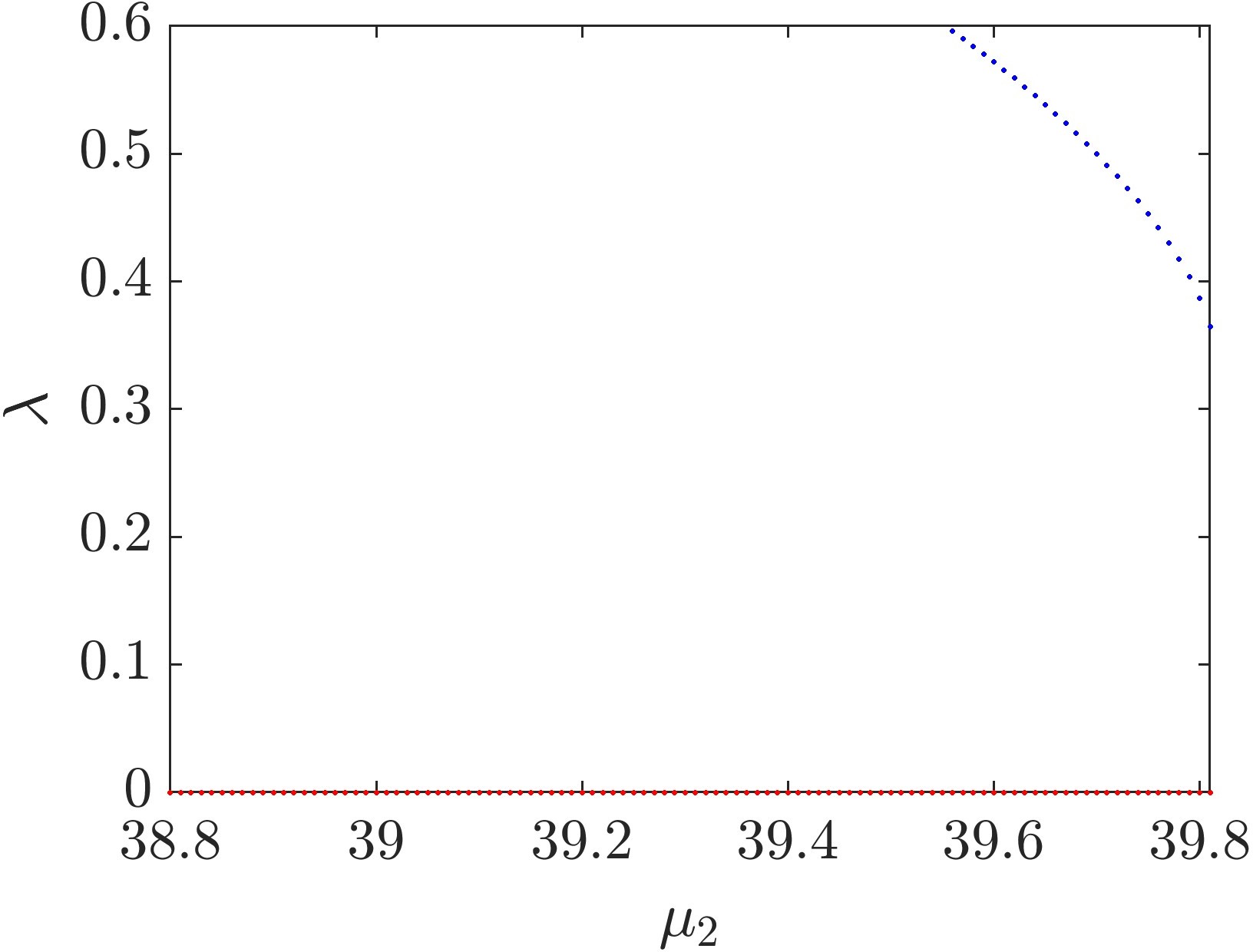}
\caption{
The first panel showcasing the BdG spectrum as a function of
$\mu_2$ for a VRB illustrates the feature that $\mu_2$ can be tuned in a wide range in the TF regime, spanning from a vanishing number of atoms to a system-size bright component. When the bright mass is varied, the VRB may suffer intervals of instability. Here, $\mu_1=40$, $32.65<\mu_2<39.87$, $\kappa=1.8$, $g_{12}=1$. The second panel focuses on the relatively large $\mu_2$ regime for another typical case $\mu_1=40$, $\kappa=1.8$, but $g_{12}=1.01$. Detailed inspection shows that the $m=0, 1, 2$ modes can potentially become unstable as illustrated in the following panels depicting the spectra of the Fourier modes $m=0, 1, ..., 5$, respectively. It should be noted that the $m=1$, $2$ instabilities are very common but the $m=0$ instability is only restricted to large fillings if it exists and frequently occurs in a rather narrow parametric regime.
}
\label{VRBspectrum5}
\end{center}
\end{figure*}

\begin{table}[htb]
\caption{
Some critical $\alpha$ and $\mu_2$ values when the bright mass is tuned at different $g_{12}$. Here, $\kappa=1.8$ and $\mu_1=40$.
\label{para2}
}
\begin{tabular*}{\columnwidth}{@{\extracolsep{\fill}} l c c c r}
\hline
\hline
$g_{12}$ &$\alpha_c(m=1)$ &$\alpha_c(m=2)$ &$\mu_{2c}(m=1)$ &$\mu_{2c}(m=2)$ \\ \hline
1  &0.1440  &0.1440 &39.52 &39.52 \\
\hline
1.01  &0.0958  &0.1345 &39.5 &39.62 \\
\hline
1.02  &0.0774  &0.1238 &39.53 &39.71 \\
\hline
1.03  &0.0669  &0.1158 &39.58 &39.8 \\
\hline
1.04  &0.0588  &0.1061 &39.63 &39.88 \\
\hline
\hline
\end{tabular*}
\end{table}

We have conducted a systematic study of the VRB spectra by tuning $\mu_2$ at various values of $\kappa$ and $g_{12}$.  The first panel of Fig.~\ref{VRBspectrum5} illustrates a Manakov case, and $\mu_2$ is studied in its full range at $\mu_1=40$ and $\kappa=1.8$. This is achieved by starting from the state at $(\mu_1,\mu_2)=(40,38.8)$ and $\kappa=1.8$ in Fig.~\ref{VRBspectrum}, and then tuning $\mu_2$ by either increasing or decreasing its value until the state no longer exists. As such, we find that the VRB exists for this particular case in a wide interval $32.65<\mu_2<39.87$, ranging from a faint bright soliton to a very fat core and also showing the robustness of the existence of the VRB structure at large $\mu_1$. The spectrum reveals an interesting feature that the bright component can introduce unstable intervals, and both the $m=1$ and $m=2$ modes can become unstable.
To gain more intuition about the impact of the bright component
mass, we introduce the mass ratio observable :
\begin{eqnarray}
\alpha=N_2/N_1
\end{eqnarray}
complementing the more abstract $\mu_2$. Detailed inspection shows that the left instability (the one below $\mu_2=34$) is due to the mode $m=1$, but the mass ratio there is extremely small $\alpha \in [0.0004247, 0.001128]$. On the other hand, both $m=1$ and $m=2$ modes become unstable when the filling is reasonably large $\alpha \gtrsim 0.1440$. The $m=1$ instability in the two intervals is oscillatory in nature, therefore differs from the $m=1$ instability in Fig.~\ref{VRBspectrum}. By contrast, the $m=2$ instability is real and appears to be a regular one. 

The instabilities in the relatively large $\alpha$ regime appear to be common for a wide range of $g_{12}$ and $\kappa$, and a typical spectrum at $g_{12}=1.01$ and $\kappa=1.8$ is also shown in Fig.~\ref{VRBspectrum5}. While the $m=1$ and $m=2$ modes become unstable simultaneously in the Manakov case, this appears to be an interesting coincidence and in general they do not bifurcate together at other $g_{12}$ and $\kappa$ values. Furthermore, an unstable $m=0$ mode is found despite the fact
that it only occurs in the highly filled regime. However, it should be noted that the unstable interval thereof can be much narrower in other parametric regimes.

The critical $\alpha$ and $\mu_2$ of the modes with
$m=1$ and $m=2$ for a few $g_{12}$ values are summarized in Table~\ref{para2}, and we shall explore the effect of $\kappa$ in the next subsection.
As $g_{12}$ increases, the critical $\alpha$ decreases for both modes, and the $m=1$ mode bifurcates earlier than the $m=2$ mode at a smaller mass ratio. The decreasing of $\alpha$ here is reasonable, as the inter-component repulsion becomes stronger, it presumably does not take much bright mass to produce the same effective strength of interaction for the induced instabilities.

\begin{figure*}[htb]
\begin{center}
\includegraphics[width=0.24\textwidth]{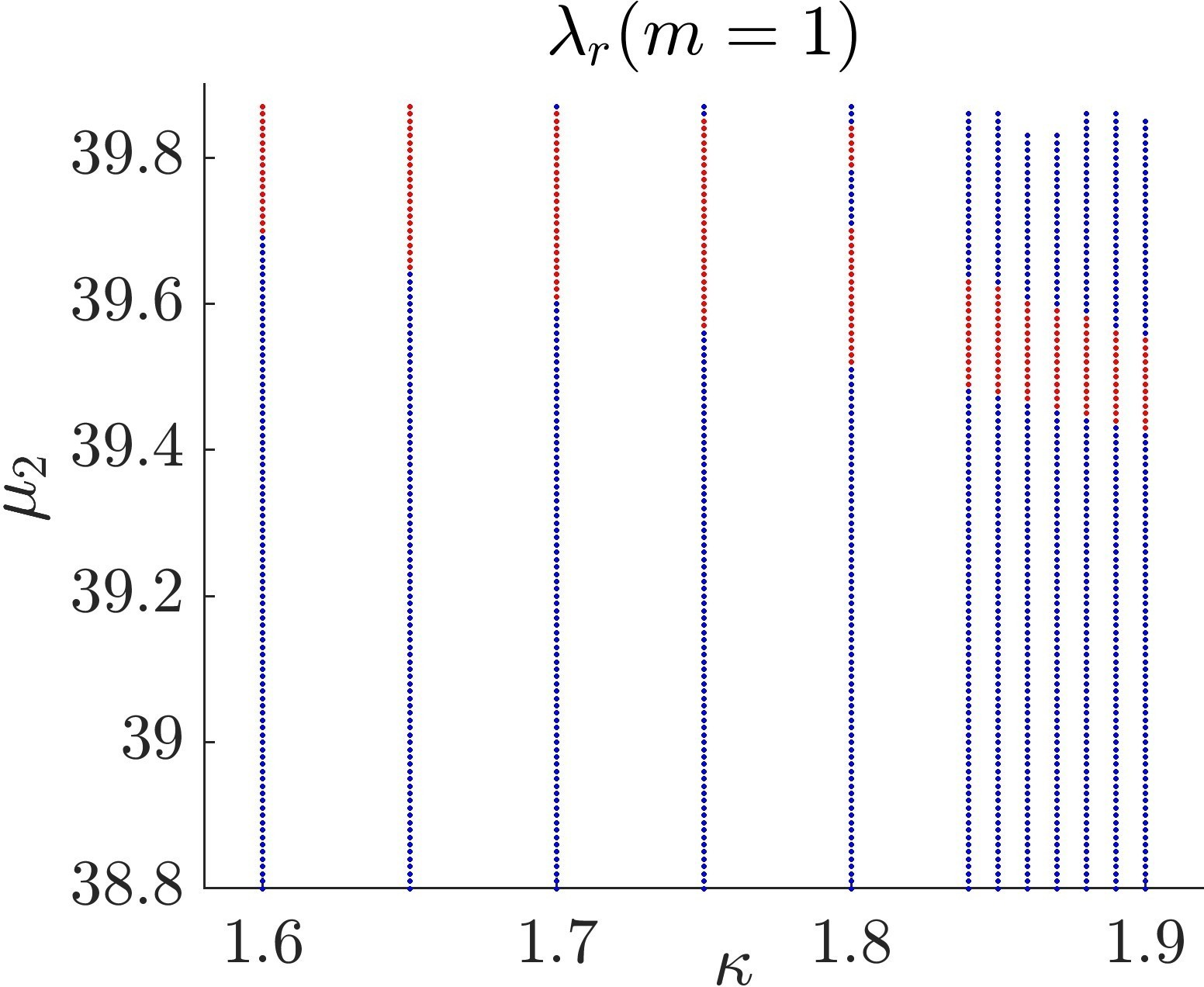}
\includegraphics[width=0.24\textwidth]{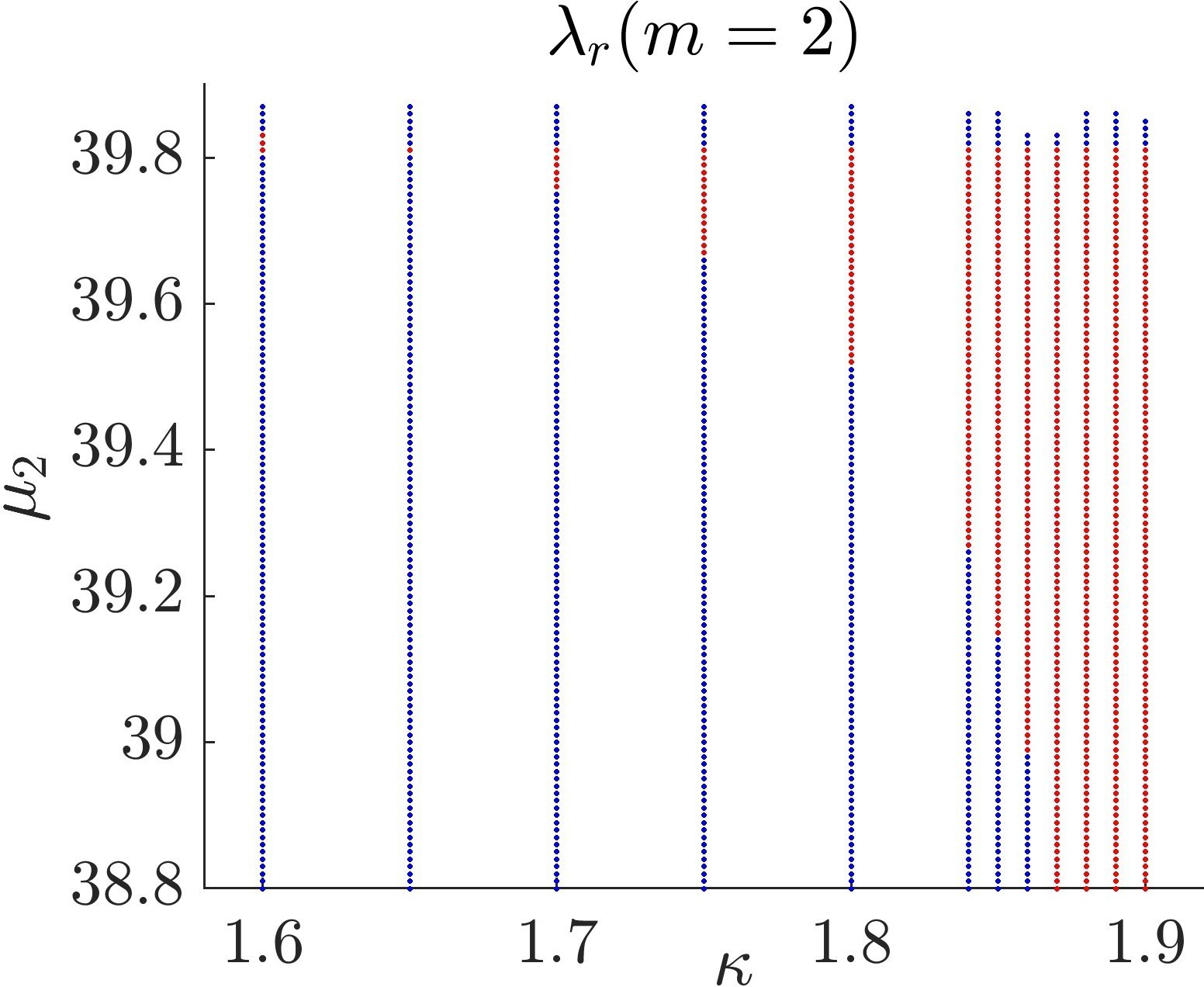}
\includegraphics[width=0.24\textwidth]{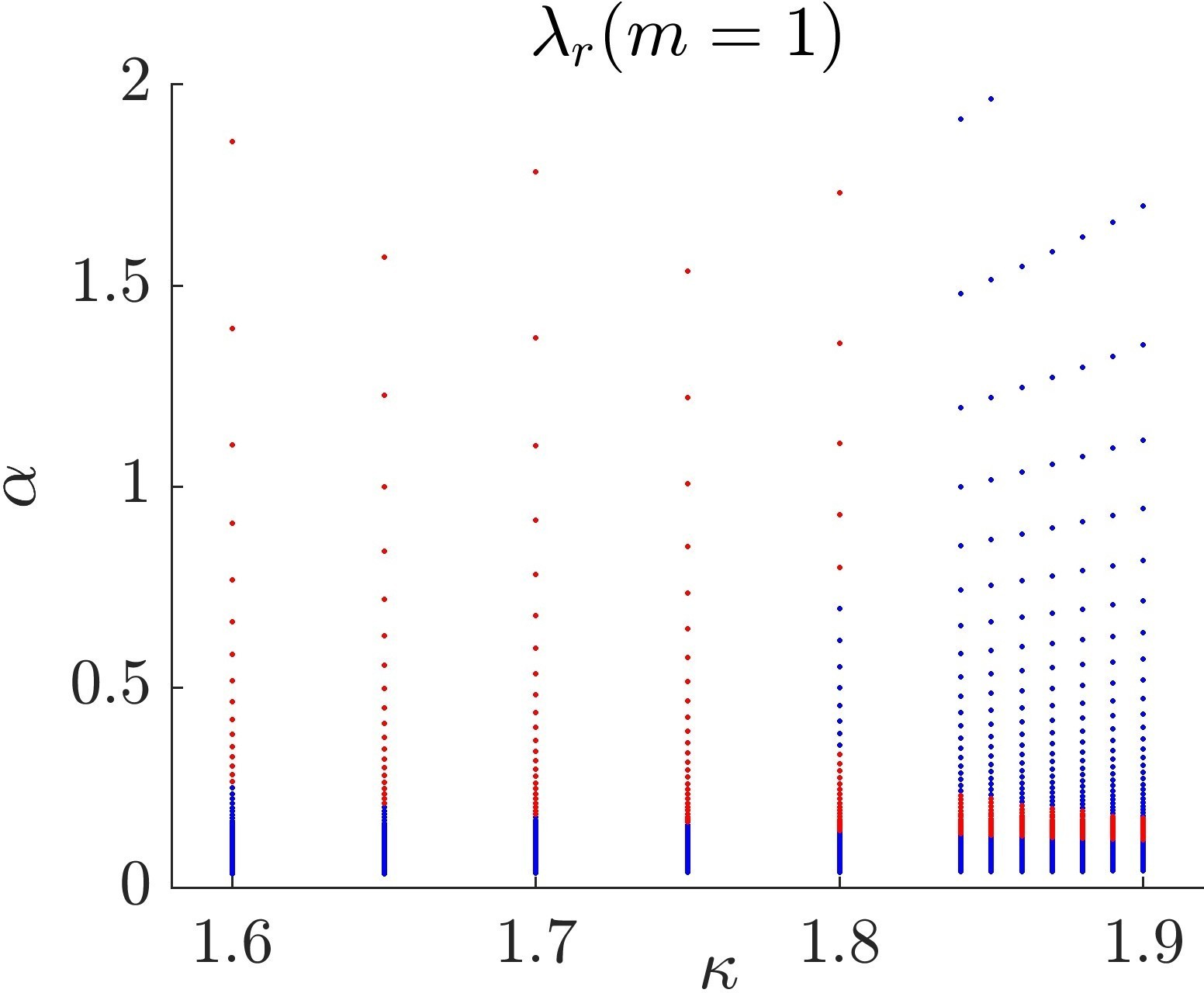}
\includegraphics[width=0.24\textwidth]{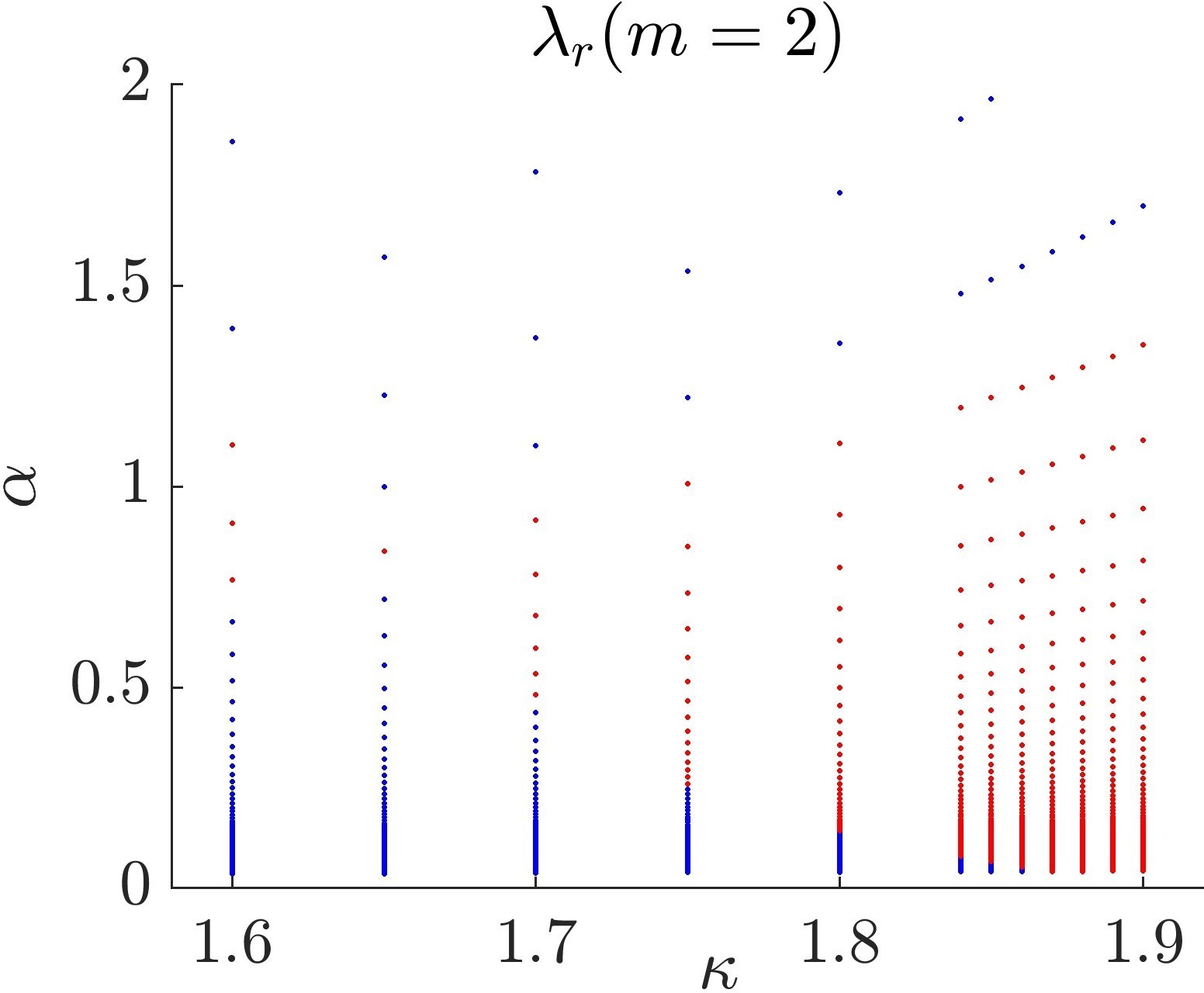}
\includegraphics[width=0.24\textwidth]{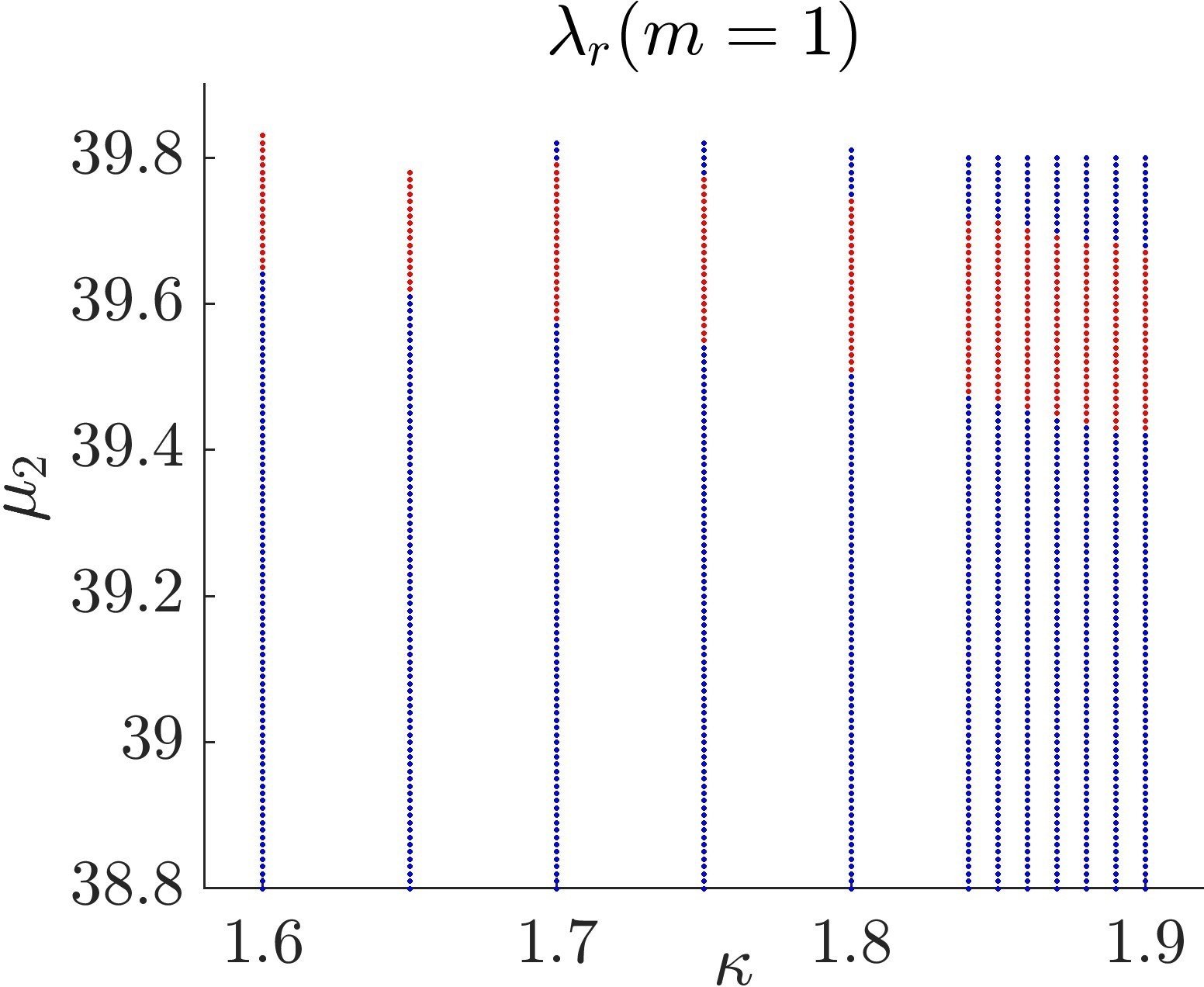}
\includegraphics[width=0.24\textwidth]{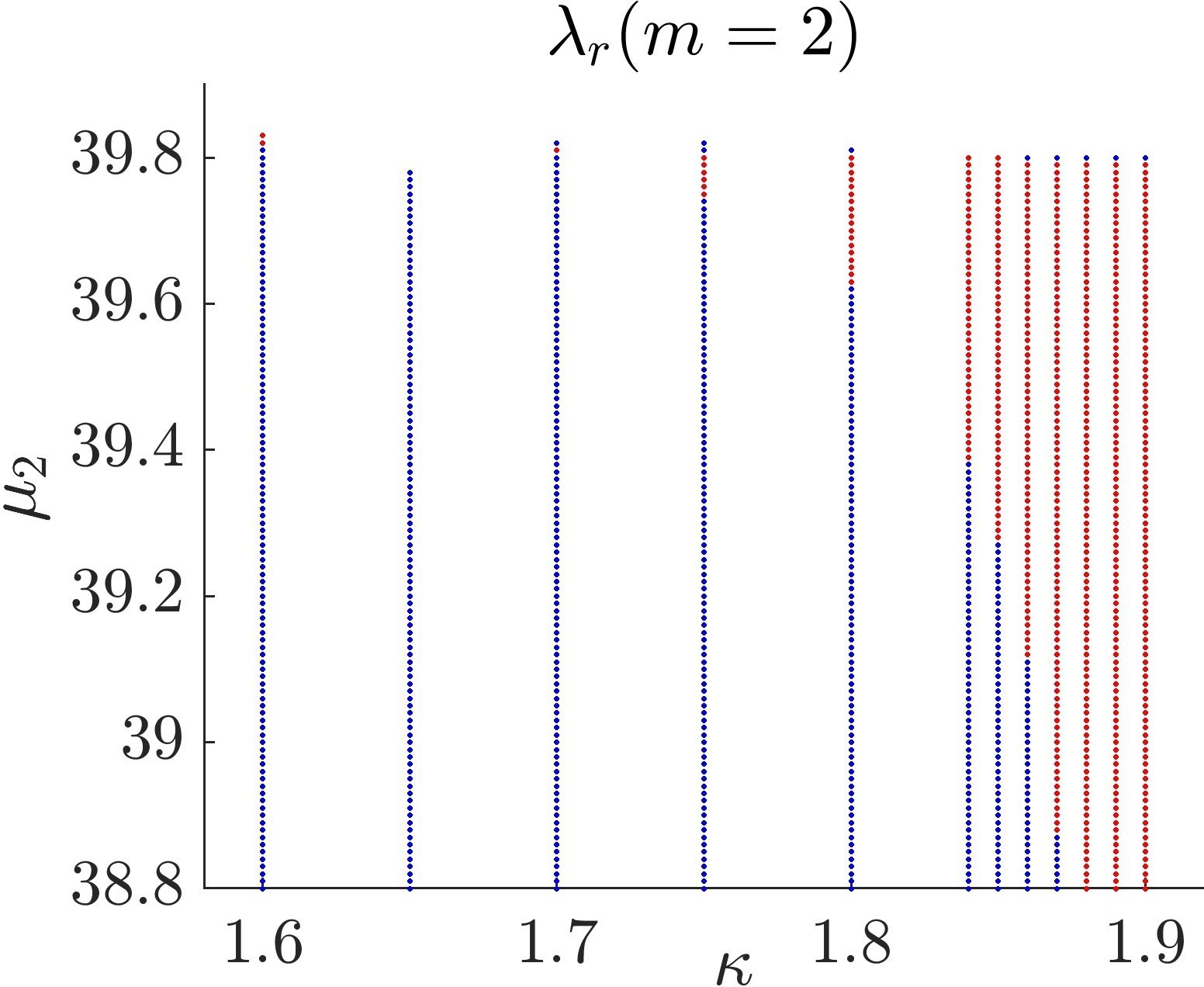}
\includegraphics[width=0.24\textwidth]{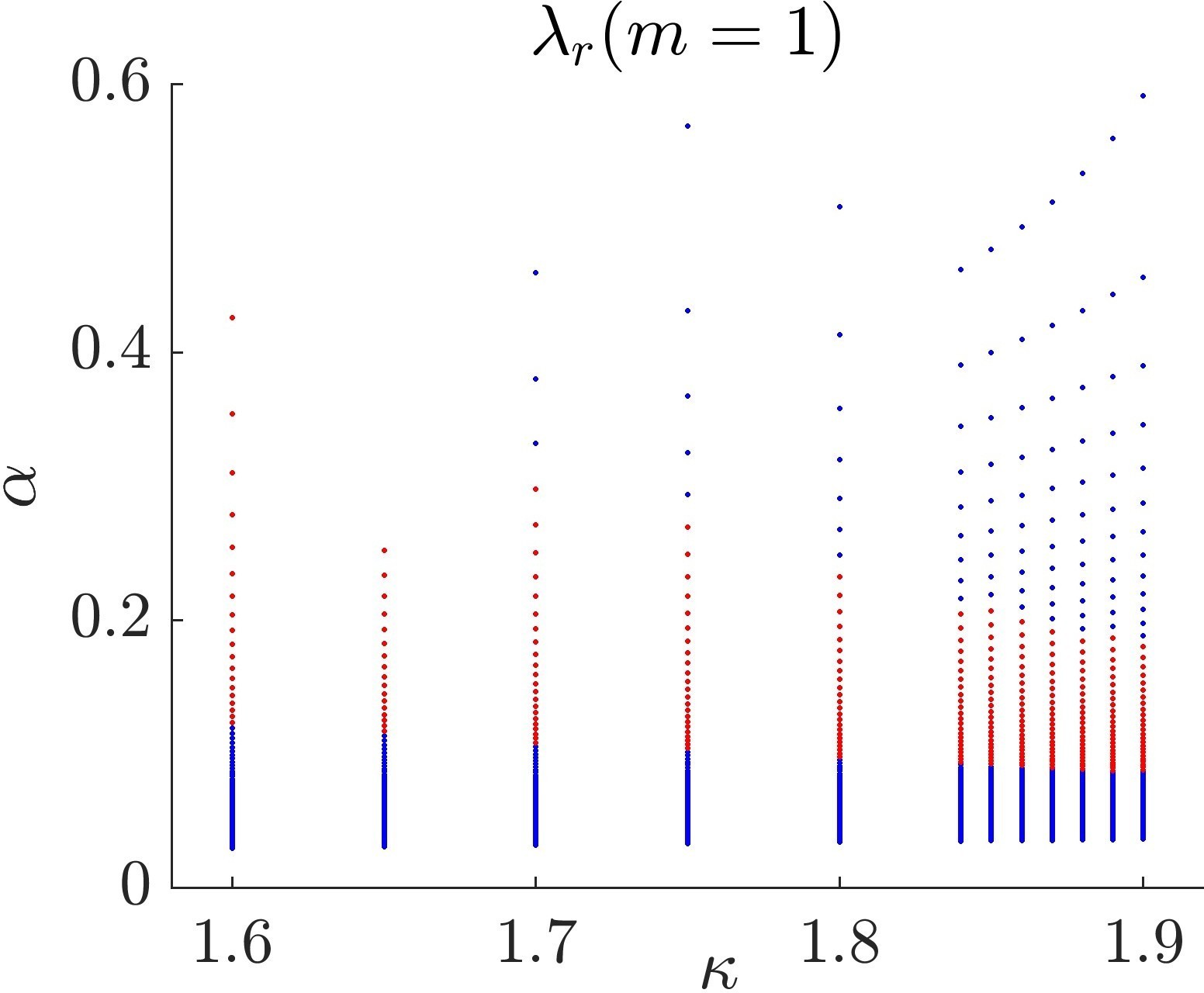}
\includegraphics[width=0.24\textwidth]{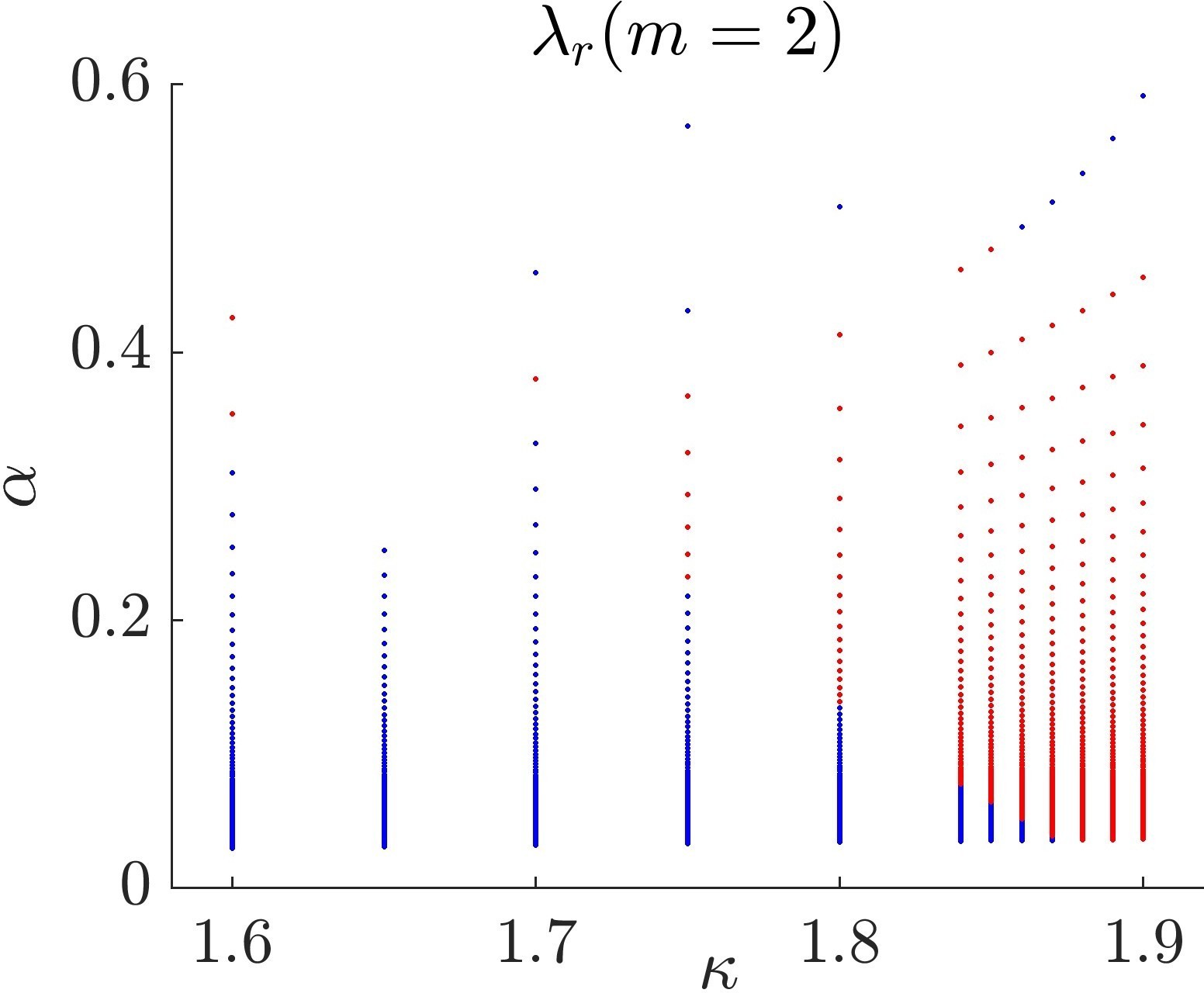}
\includegraphics[width=0.24\textwidth]{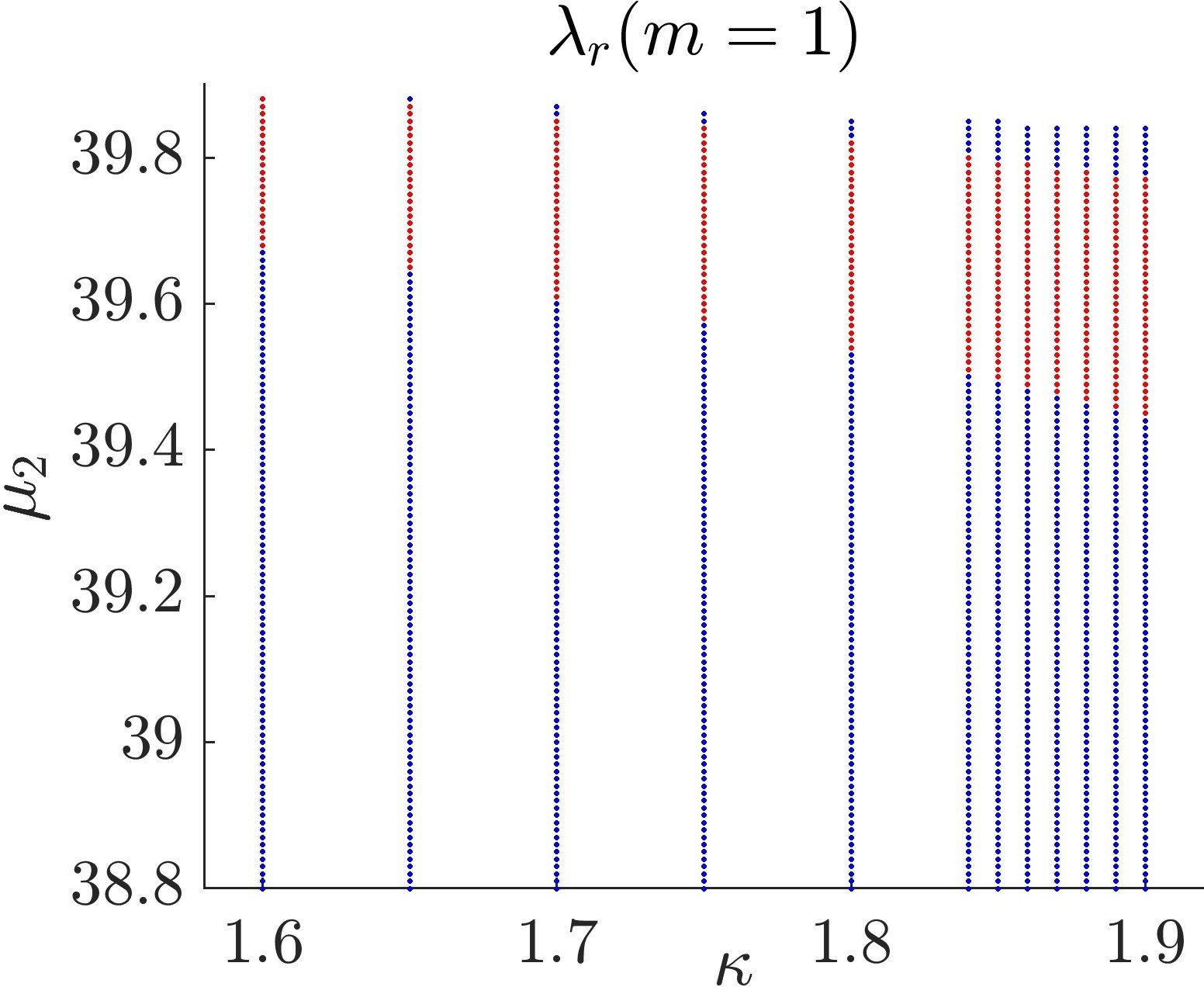}
\includegraphics[width=0.24\textwidth]{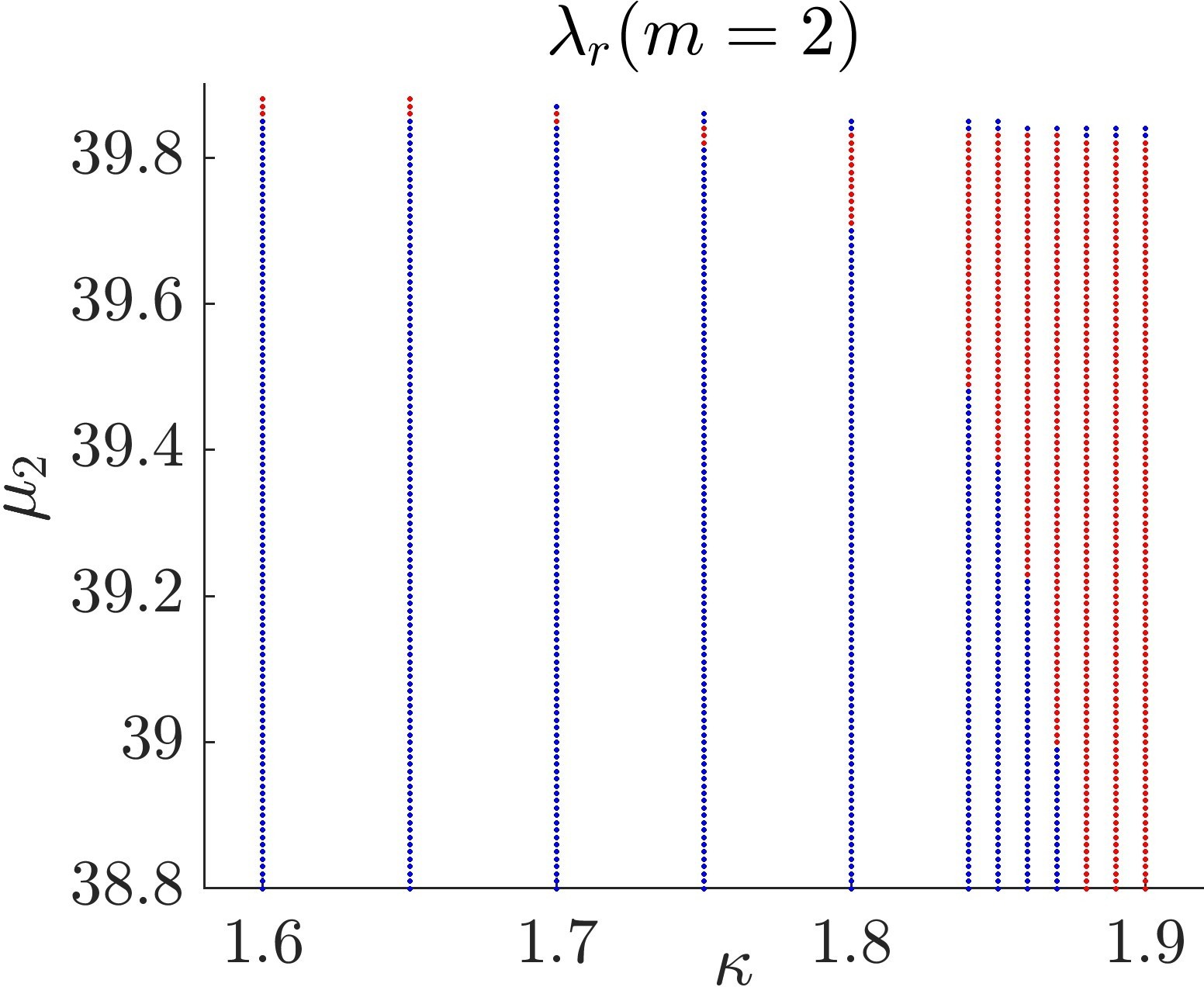}
\includegraphics[width=0.24\textwidth]{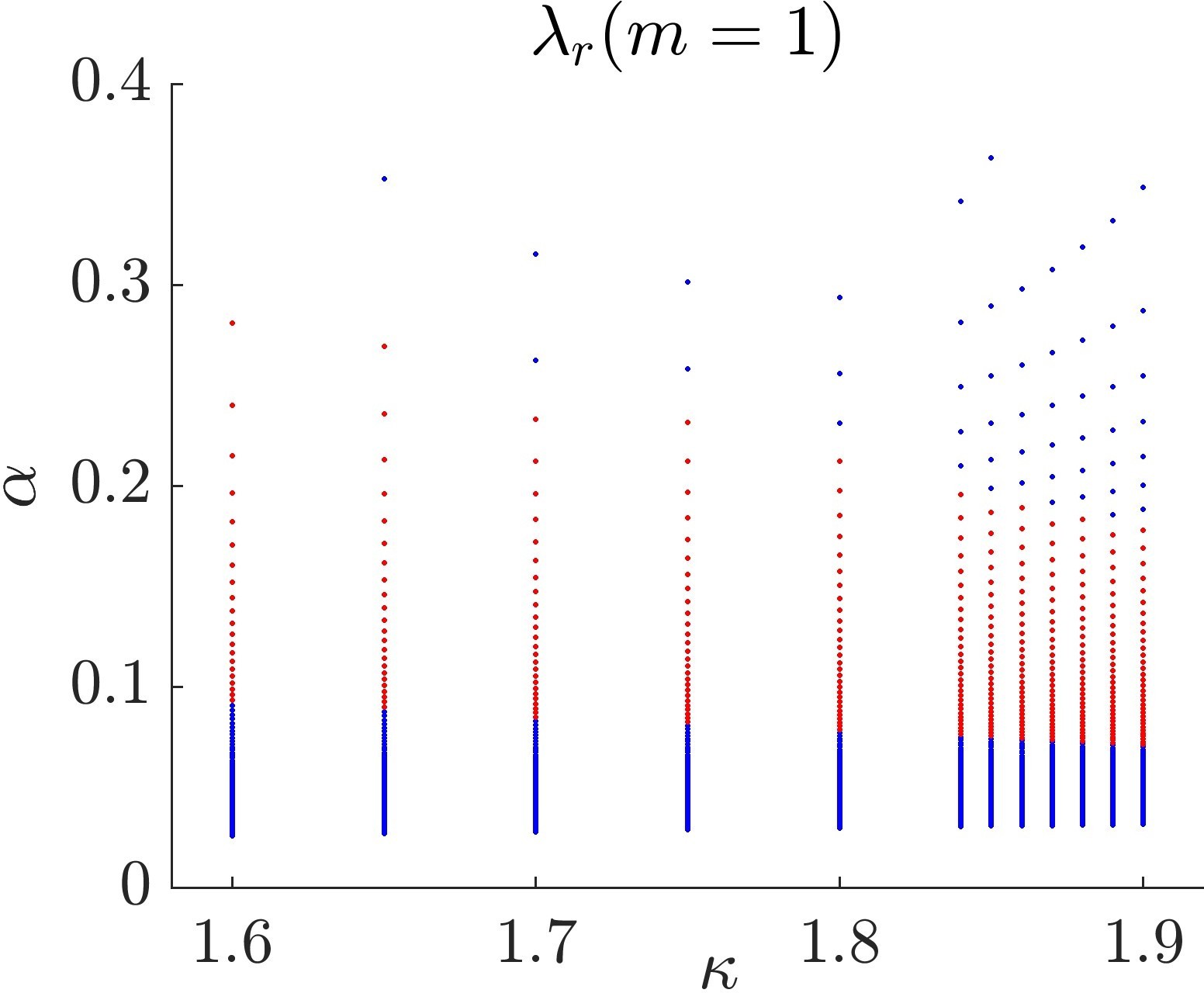}
\includegraphics[width=0.24\textwidth]{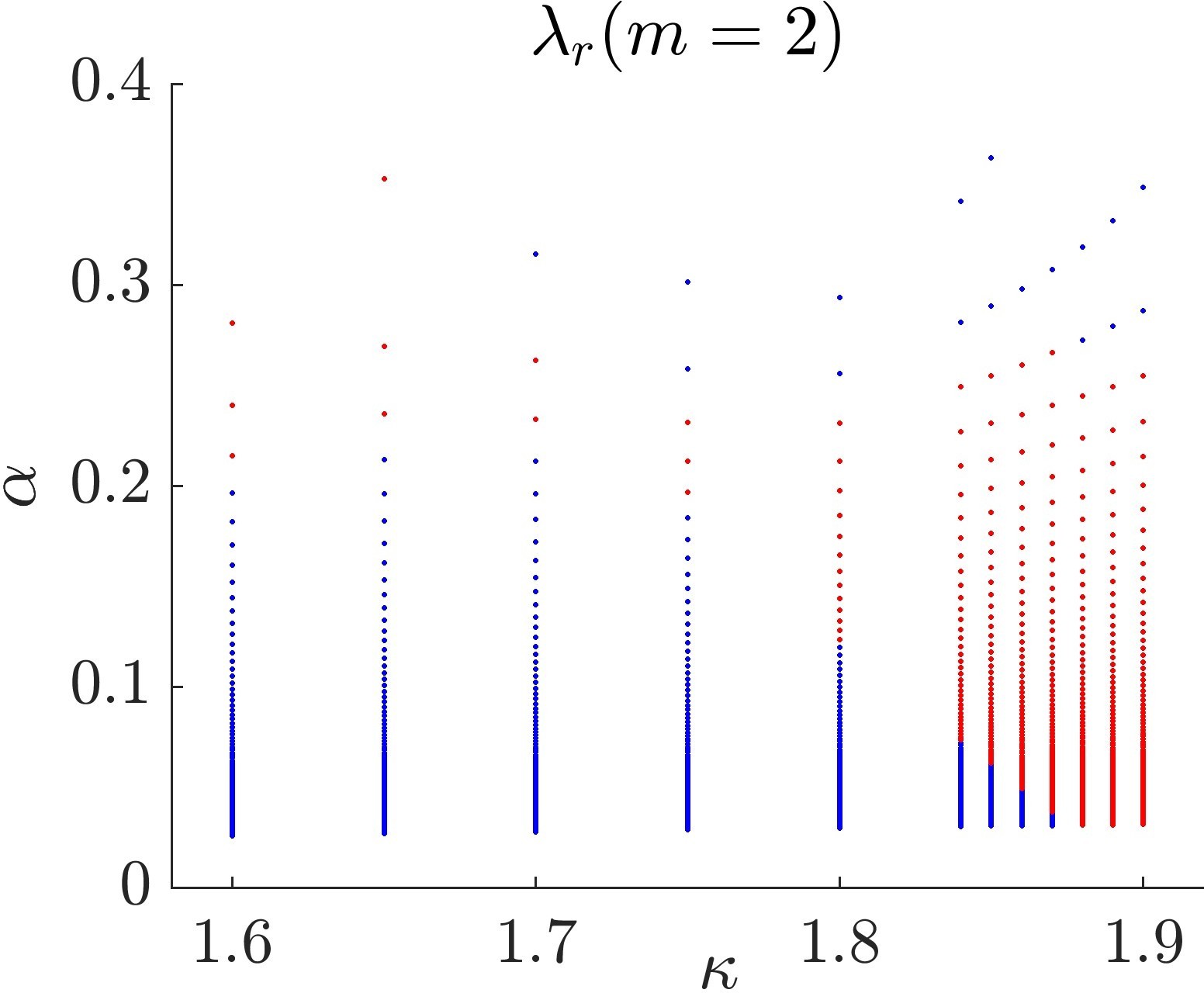}
\caption{
The effect of the bright mass on the VRB stability for various $\kappa$ and $g_{12}$ values. The rows in turn are for $g_{12}=1$, $1.01$, and $1.02$. Here, the blue points are for stable regimes and the red points are for unstable regimes. The maximum growth rate in an unstable interval is typically on the order of $O(0.01)$. 
}
\label{VRB_mu2}
\end{center}
\end{figure*}

\subsection{Effect of $\kappa$ on the critical transitions}
Here, we investigate the $\kappa$-dependence of the bright-induced instability in the large $\alpha$ regime. The examples shown in the previous subsection suggest that the critical transitions clearly depend on $\kappa$. To this end, we compute the spectra with respect to $\mu_2$ for a series of $\kappa$ values by further increasing $\mu_2$ to its upper limit from the appropriate states in Fig.~\ref{VRBspectrum}. The results are illustrated in Fig.~\ref{VRB_mu2} and some typical numbers for the Manakov case are summarized in Table~\ref{para3}.

Since Fig.~\ref{VRBspectrum} shows that the $m=2$ instability is already present at $\kappa=1.87$ before $\mu_2$ is increased, this suggests that the critical chemical potential thereof should decrease when $\kappa$ is increased. This is found to be essentially the case numerically in Fig.~\ref{VRBspectrum5} for both the $m=1$ and $m=2$ modes. However, the effect of $\kappa$ on the $m=2$ mode is stronger: note that the critical mass has a much larger slope for the $m=2$ mode. This produces an interesting exchange of bifurcation order as $\alpha$ increases at, e.g., $\kappa=1.8$ for the Manakov case. Above this value the $m=2$ instability bifurcates first, and below this value the $m=1$ instability bifurcates earlier instead.

\begin{table}[htb]
\caption{
Some critical $\alpha$ and $\mu_2$ values with respect to $\kappa$ when the bright mass is tuned for the typical Manakov case in Fig.~\ref{VRB_mu2}. The ones marked with a star (*) have rather narrow instability intervals. The ``peculiar'' trend of $\alpha_c(m=2)$ at $\kappa \lesssim 1.65$ is because in this regime, $\mu_{2c}$ essentially saturates, the drop of $\alpha$ at $\kappa=1.6$ is therefore dominated by the larger background size.
\label{para3}
}
\begin{tabular*}{\columnwidth}{@{\extracolsep{\fill}} l c c c r}
\hline
\hline
$\kappa$ &$\alpha_c(m=1)$ &$\alpha_c(m=2)$ &$\mu_{2c}(m=1)$ &$\mu_{2c}(m=2)$ \\ 
\hline
1.60  &0.2651  &0.7675$^*$  &39.70  &39.81 \\
\hline
1.65  &0.2112  &0.8379$^*$  &39.65  &39.81 \\
\hline
1.70  &0.1842  &0.4370  &39.61  &39.75 \\
\hline
1.75  &0.1583  &0.2606  &39.56  &39.67 \\
\hline
1.80  &0.1440  &0.1440  &39.52  &39.52 \\
\hline
1.84  &0.1315  &0.0789 &39.48 &39.26 \\
\hline
1.85  &0.1288  &0.0658 &39.47 &39.15 \\
\hline
1.86  &0.1262  &0.0525 &39.46 &38.99 \\
\hline
1.87  &0.1237  &$<$0.042 &39.45 &$<$38.8 \\
\hline
1.88  &0.1213  &$<$0.042 &39.44 &$<$38.8 \\
\hline
1.89  &0.1191  &$<$0.043 &39.43 &$<$38.8 \\
\hline
1.90  &0.1169  &$<$0.043 &39.42 &$<$38.8 \\
\hline
\hline
\end{tabular*}
\end{table}

It is noted that restabilization can occur when $\alpha$ is increased further, in line with the theoretical expectation, and more than one unstable intervals can also exist. Increasing $g_{12}$ has an effect of lowering the critical $\alpha$, but other than that, the general features remain largely similar to those of the Manakov case. It is worth mentioning that these instabilities are, however, quite weak and a typical maximum growth rate in an unstable interval is only approximately $O(0.01)$. 
In line with our theoretical expectation, no unstable modes for $m \geq 3$ are found in these wide parametric regimes.

\begin{figure*}[htb]
\begin{center}
\includegraphics[width=\columnwidth]{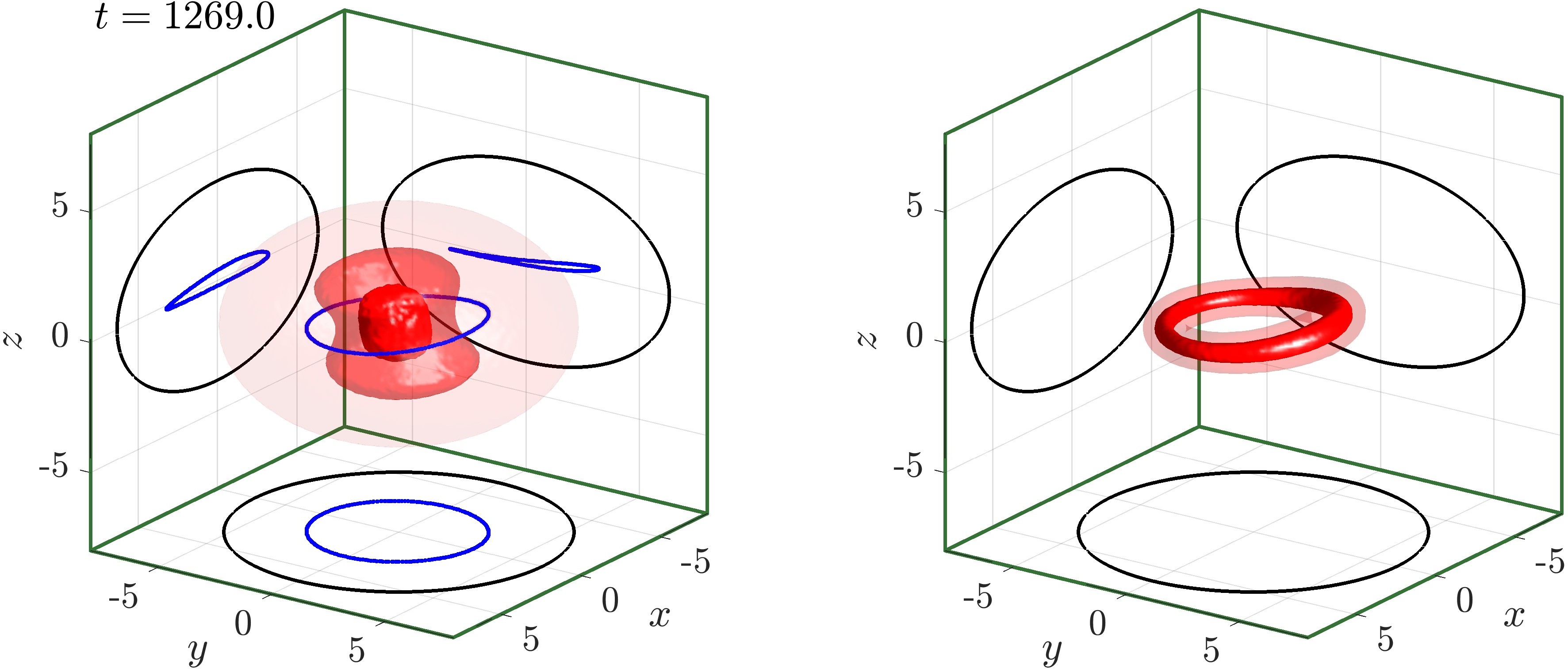}
\includegraphics[width=\columnwidth]{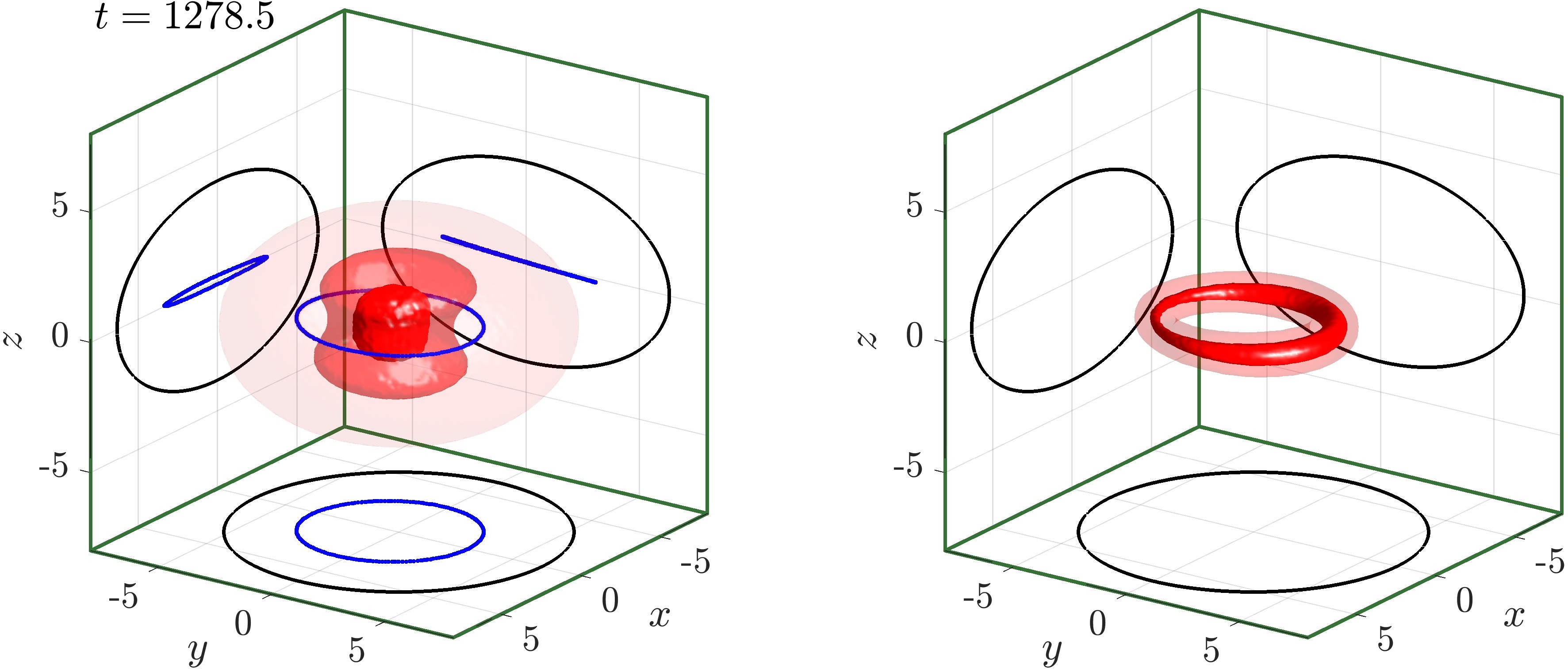}
\includegraphics[width=\columnwidth]{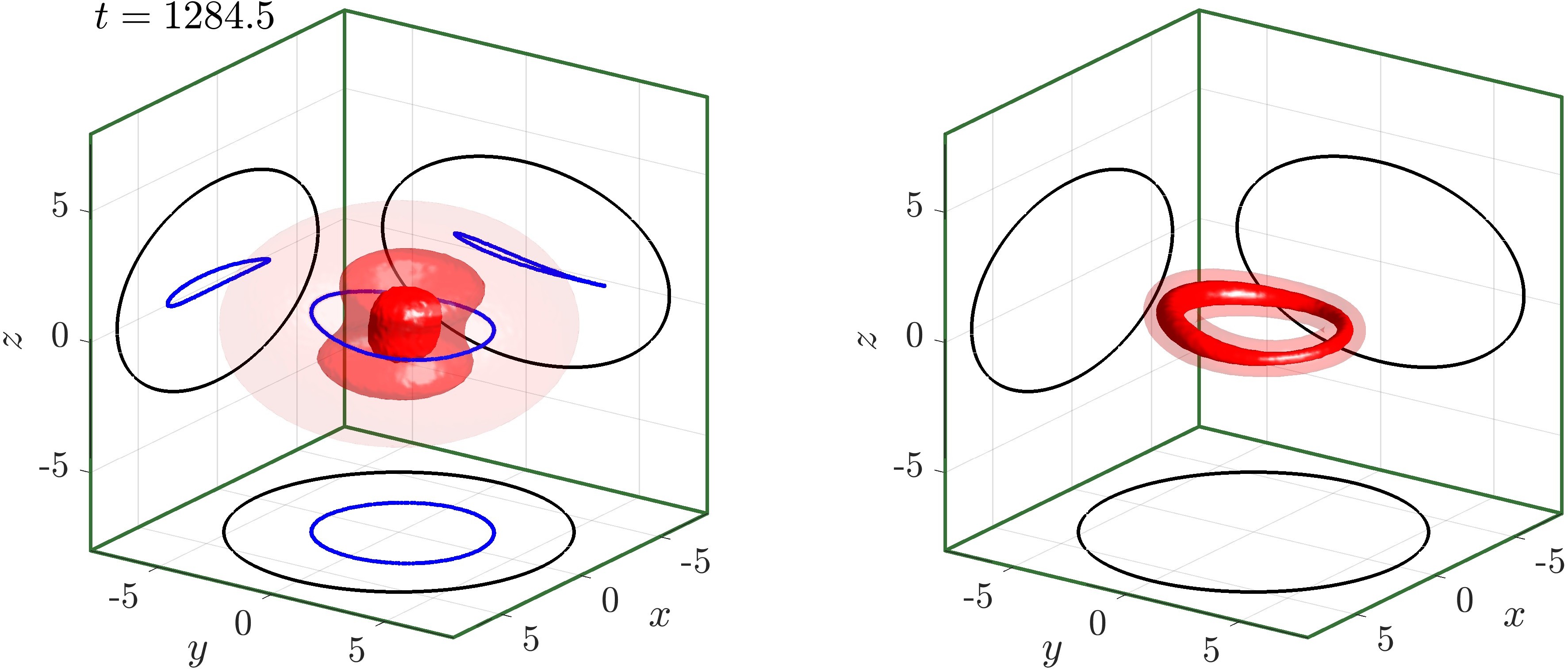}
\includegraphics[width=\columnwidth]{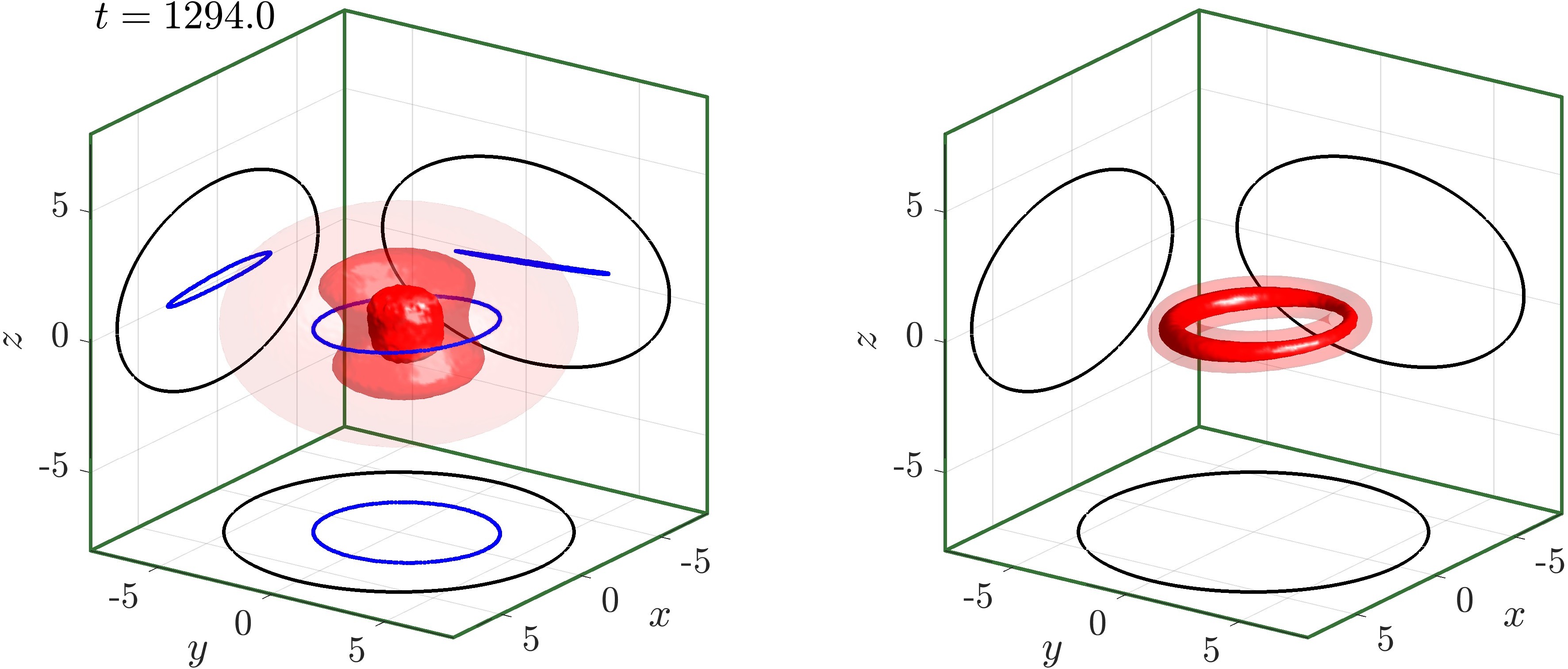}
\caption{
In the oscillatory $m=1$ unstable mode, the VRB first tilts and then propagates off the center, here in the approximately $-y$ direction at $t=1269$. The VRB then reaches the edge of the cloud and reverses its tilting direction. The VRB finishes the reversing at $t=1278.5$ and propagates in the opposite direction, passing through the center at approximately $t=1284.5$, and reaching the opposite end of the
cloud. It similarly finishes reversing its tilting direction at $t=1294$ and then runs back to the trap center. The VRB reaches the trap center at $t=1300$, in a state similar to the one we started with at $t=1269$, and then continues the cycle. The oscillation amplitude, however, gradually increases and the VRB eventually breaks into VLB filaments; see the full movie for details \cite{VRBmv1}.
}
\label{Dym1}
\end{center}
\end{figure*}

\subsection{Dynamical Simulations}

We have conducted several typical dynamical simulations following the spectra. The unstable modes in Fig.~\ref{VRBspectrum} are similar to those of the single-component VR. For example, we have run three VRB dynamics with random perturbations at $\mu_1=20$ and $\kappa=0.9, 2, 3$, where the dominant unstable modes are $m=1, 2, 3$, respectively. First, it is worth mentioning that the bright mass essentially follows the VR core. At $\kappa=0.9$, the VRB tends to flip and meanwhile it also extends one of its ends towards the edge of the condensate. The VRB breaks into two pieces before almost making a full flip.
At $\kappa=2$, the VRB breaks into two parallel vortex-line-bright (VLB) filaments extending outside the BEC. Then, the two VLBs can contract and reconnect into a full VRB. The VRB can repeat this process multiple times, with the two VLBs oriented towards different directions, before finally getting highly excited and disordered into vortical filaments. 
The instability at $\kappa=3$ is triggered much faster, and the VRB is broken into three VLBs following the $m=3$ mode as expected. Because these dynamics are similar to the VR counterparts, we shall not discuss them further here.

As mentioned earlier, we have also identified two unstable modes that are not available to the single-component VR, the oscillatorily unstable $m=1$ and $m=0$ modes. The former is very common but the latter is restricted to rather narrow chemical potential intervals of a large bright mass. We now discuss these two modes in detail.

The $m=1$ oscillatorily unstable mode is illustrated in Fig.~\ref{Dym1}.
It might be appropriate to call it the VRB sloshing mode, as the VRB sloshes back and forth in the condensate. The VRB first tilts in the condensate and then propagates off the center, here in the approximately $-y$ direction at $t=1269$. The VRB then collides with the 
edge of the cloud and reverses its tilting direction. The VRB finishes the reversal at $t=1278.5$ and propagates in the opposite direction, passing through the center at approximately $t=1284.5$, colliding with the trap again on the other side and similarly finishes reversing its tilting direction at $t=1294$ and then runs back to the trap center. The VRB reaches the trap center at $t=1300$, in a state similar to the one we started with, and then continues this cycle. The oscillation amplitude, however, increases and the VRB eventually breaks into VLB filaments; see the full movie for details \cite{VRBmv1}. The simulation parameters are $\mu_1=20, \mu_2=19.4, \kappa=1.65, \alpha=0.2596, \lambda_r=0.006895$, and the $m=1$ mode is the only unstable mode. 

The $m=0$ oscillatorily unstable mode is simpler to describe: this is  the regular precessional mode of $m=0$ around its equilibrium state but with a growing amplitude. As the amplitude grows, Kelvin modes are naturally excited, leading to instabilities. Since the mode is relatively familiar, we shall not discuss it in further detail here, but a full movie thereof
is available \cite{VRBmv2}. In this case, the $m=3$ mode is excited, and it is noticed that the VRB can break and reform for many cycles before the VRB is finally ejected out of the condensate. It is not very straightforward to find a parameter regime where $m=0$ is the only unstable mode, but it is possible to find a suitable regime where it dominates. For the state in the movie, $\mu_1=20, \mu_2=19.61, \kappa=1.7, \alpha=0.8306, \lambda_r(m=0)=0.02388, \lambda_r(m=1)=0.008185, \lambda_r(m=2)=0.006596$. Note that the bright mass is very large, with a remarkable filling fraction as large as $83$ percent.

\begin{figure}[htb]
\begin{center}
\includegraphics[width=42mm]{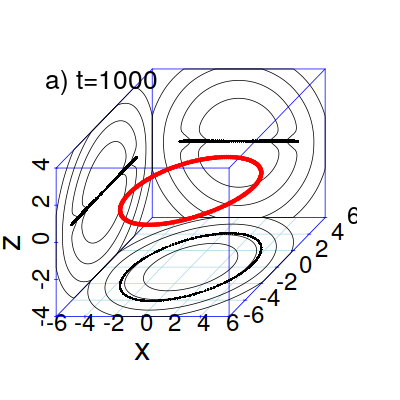}
\includegraphics[width=42mm]{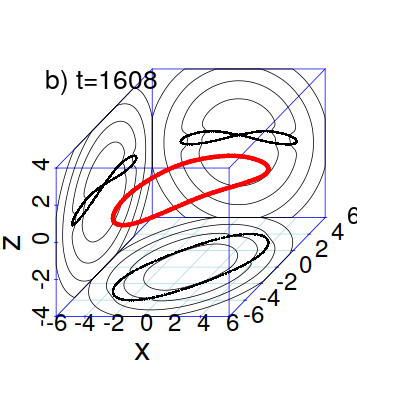}
\includegraphics[width=42mm]{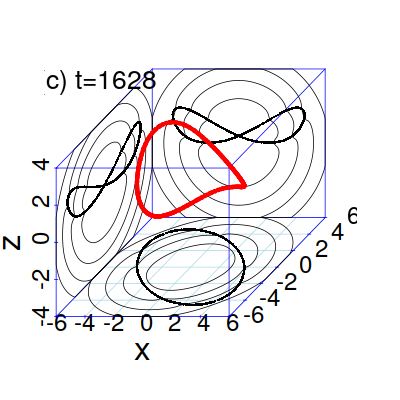}
\includegraphics[width=42mm]{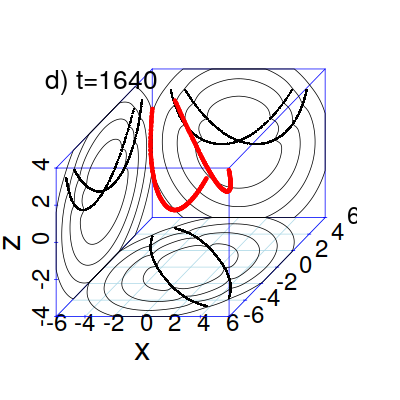}
\caption{
Snapshots of a vortex ring are shown for a parametric instability.  
The positions of the extracted vortex cores are shown as red points and their 
projections are shown on each plane along with the density contours of the BEC. 
In this example, we have picked $\mu=30$ with $\kappa=1.25$.
A small displacement from a circular structure was 
initially induced with an $m=2$ mode in the radial direction.
}
\label{param_gpe}
\end{center}
\end{figure}

\begin{figure}[htb]
\begin{center}
\includegraphics[width=0.37\textwidth]{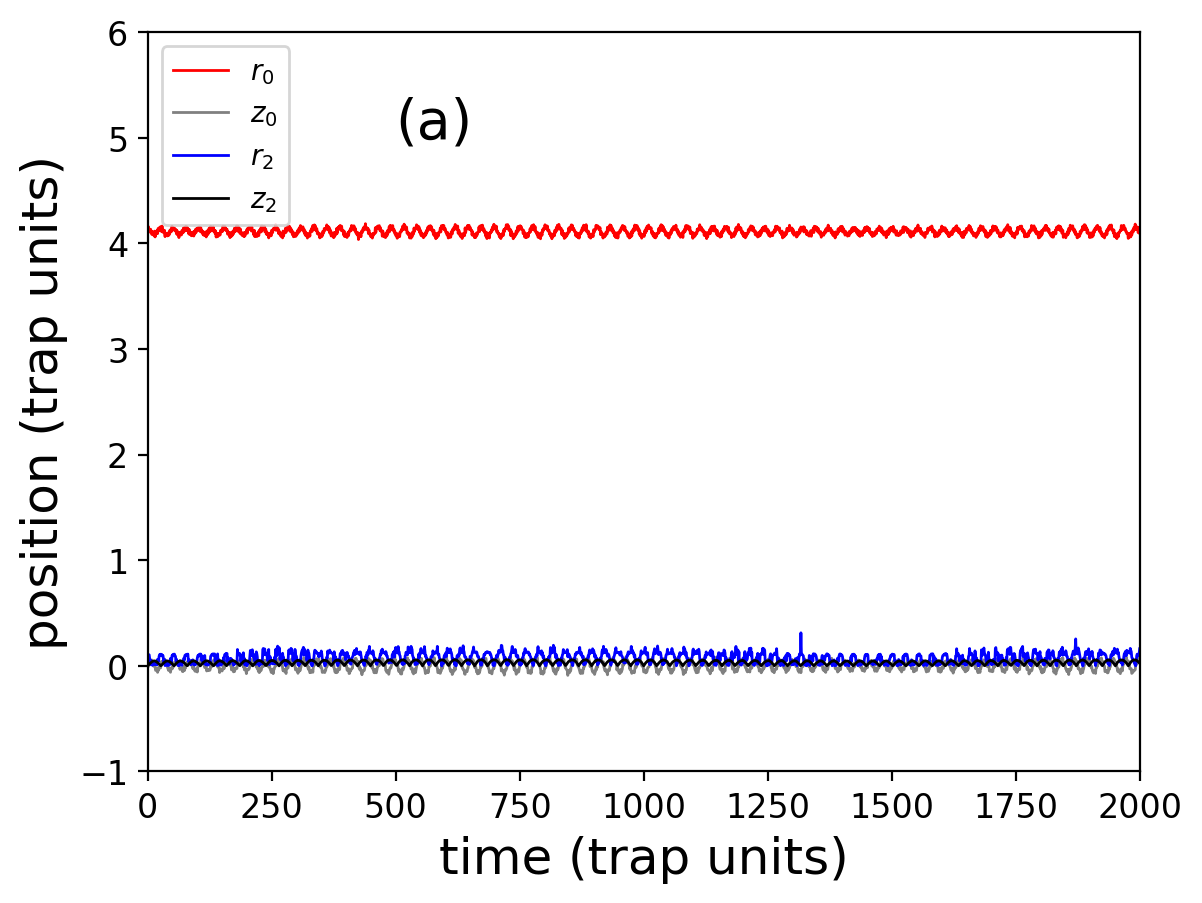}
\includegraphics[width=0.37\textwidth]{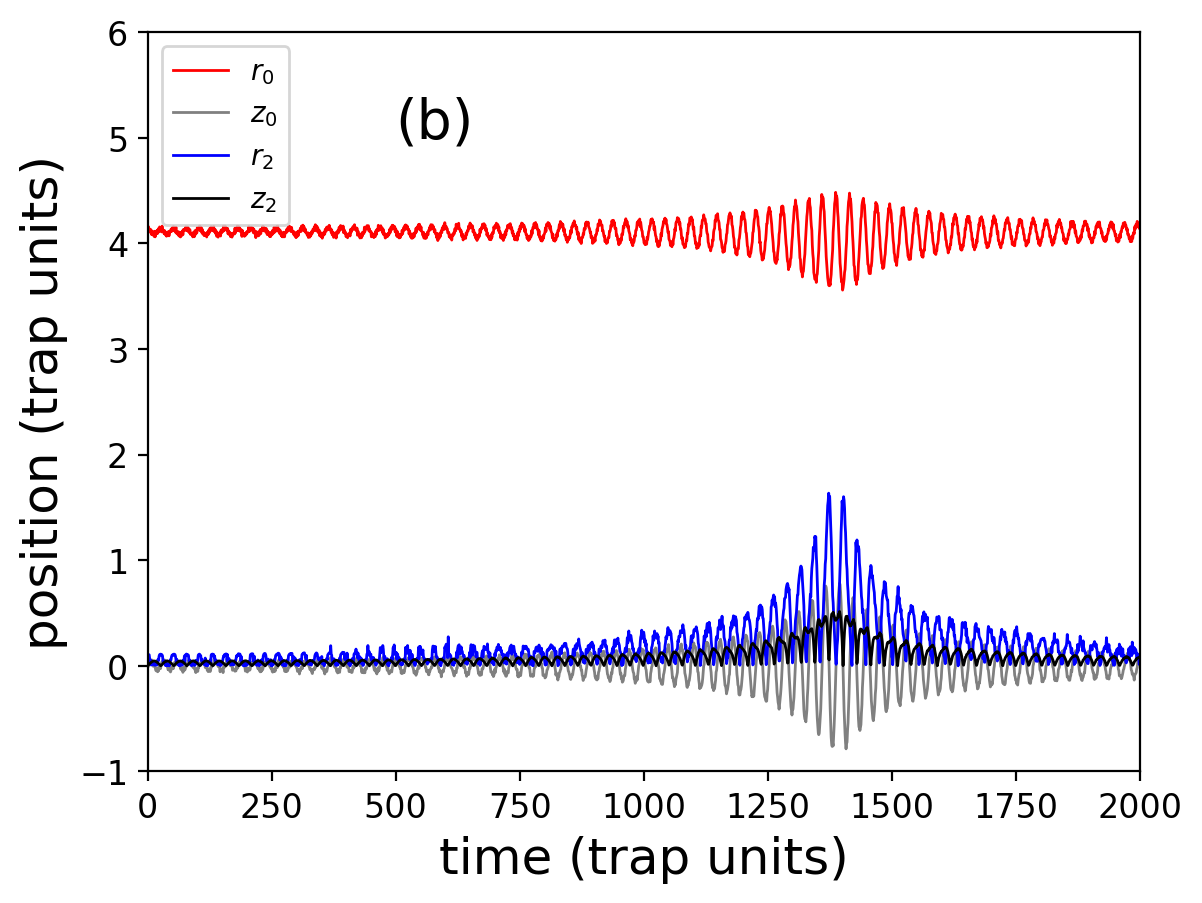}
\includegraphics[width=0.37\textwidth]{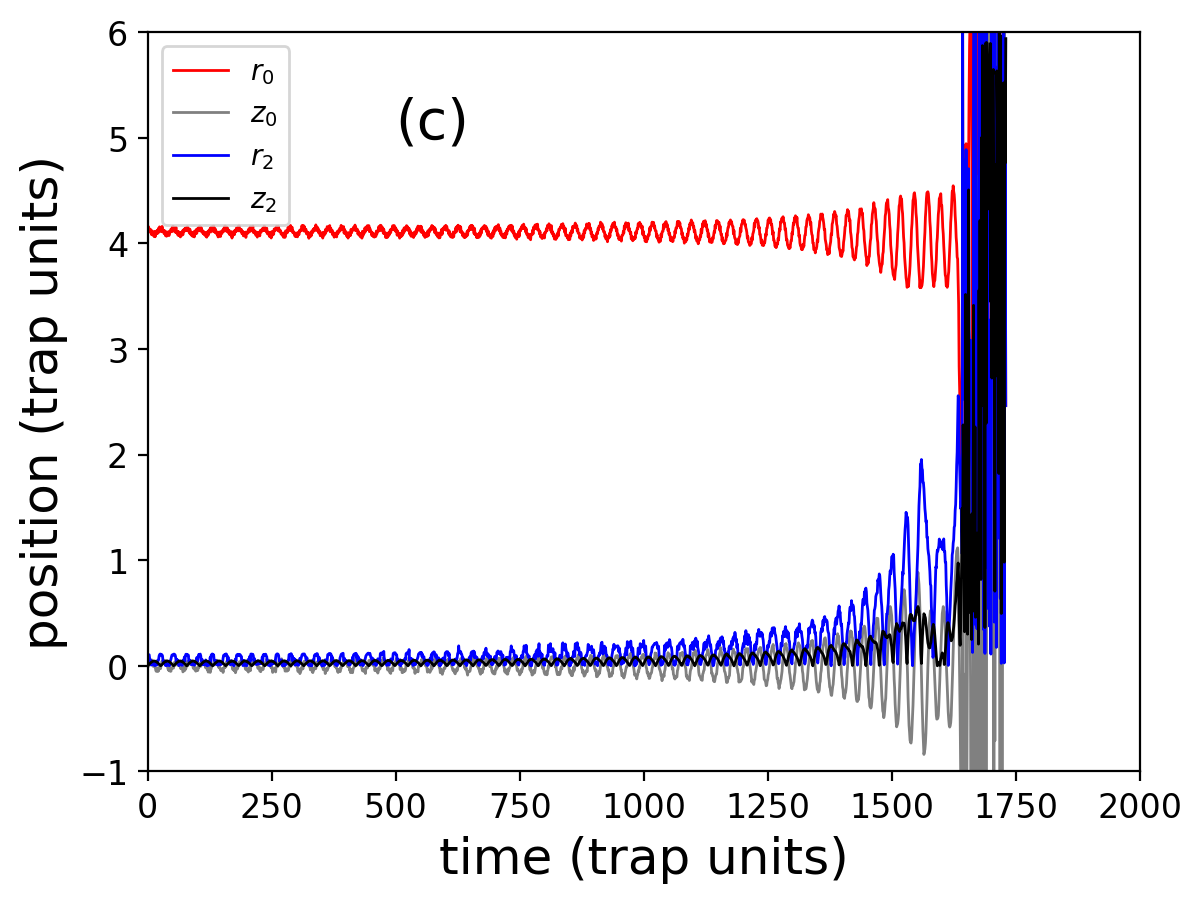}
\caption{
Several examples of the extracted modes associated with
the vortex core positions (see Eq.~(\ref{distorted_knot}))
are shown for different geometries:
(a) $\kappa=1.27$, (b) $\kappa=1.255$, and (c) $\kappa=1.25$.
The extracted vortex ring mode amplitudes shown are $r_0$ (red), 
$r_2$ (blue), $z_0$ (grey), and $z_2$ (black).
In (a) a typical stable configuration with a seeded $m=2$ 
perturbation is shown.
In (b) a configuration near the maximal response 
associated with the parametric instability has a growth and decay of a mode 
of the vortex ring. Another revival was observed (for longer times,
not shown here) without the ring being broken.
In (c) at the point of resonance, 
the ring breaks after being unstable at a time of around 1640 trap units.
All simulations are for a single-component BEC with $\mu=30$. 
}
\label{param_rz}
\end{center}
\end{figure}

\subsection{Nonlinear Parametric Instability}

So far we have explored the linear instabilities of the VRB structure.
Now, we turn to the nonlinear parametric instability analyzed qualitatively in Sec.~\ref{NLPI}. Some numerical examples showcasing this instability 
are shown in Figs.~\ref{param_gpe} and~\ref{param_rz}.
In Fig.~\ref{param_gpe} we show the vortex core locations by means
of red points.
Their projected positions are shown on each plane. 
In addition to the projected cores, we show 
the BEC density as thin contours at 0.2, 0.4, 0.6 and 0.8 of the maximum density.
The BEC is started with the ring at 0.92 $R_*$
with a small amount of $m=2$ perturbation. 
After a long time, in this case about 1640 trap units of time, the $m=2$ mode grows and becomes unstable. 
The vortex ring stretches until the ring is broken by part of the ring leaving the BEC.
This can be seen in Fig. \ref{param_gpe}(c) and (d).
Notice that while the results here are shown for the case of a 
single-component VR, similar features arise in the case of the
VRB.

Mode analysis, associated with this instability, is 
performed by extracting the core positions. 
These are sorted to be in azimuthal order. 
Then, analysis on the displacement and shape of the vortex ring was done
assuming the following form for the vortex ring:
\begin{eqnarray}
r(\varphi)=r_0+\sum_m r_m e^{i m \varphi},  \quad
z(\varphi)=z_0+\sum_m z_m e^{i m\varphi}. 
\label{distorted_knot}
\end{eqnarray}
Here $r$, $\varphi$, and $z$ are the cylindrical projections of $\bf R$.
We have taken the real part of $z_m$ and the real and positive part of $r_m$.  These have been plotted in Fig. \ref{param_rz}. Here, the nonlinear parametric instability can be identified as  occurring around $\kappa=1.25$. For$\kappa=1.27$, i.e., some distance parametrically away from this instability, the ring is stable. As $\kappa$ approaches the 
relevant parametric resonance point (from above) at $\kappa=1.255$, it is noted that the ring tends to become unstable but the instability can decay and thereby a stable ring is restored. Another revival was observed until the instability amplifies and eventually the ring breaks. At the resonance of $\kappa=1.25$, the ring becomes unstable quickly.

In other simulations, for example with $\kappa=1.14$, $n_2=78.5$, and $n_1=1570$,
we observed similar behavior.

The parametric instability might be (partially) responsible for why the VR decays along some modes that are predicted to be linearly stable from the BdG spectrum \cite{Wang:AI3}. It is therefore highly interesting to explore this type of resonance more systematically in the future, as it concerns
a system that is {\it linearly stable but nonlinearly unstable}. 


\section{Conclusions and Future Challenges}
\label{conclusion}

The properties of the VRB show significant similarities with the 
single VR, but also nontrivial differences. The bright component 
has an effect towards slowing down the unstable dynamics, i.e., 
decreasing the instability growth rate when the structure is 
(already at the single-component VR level) unstable.
Nevertheless, at the same time, the presence of the second 
(bright, filling) component
also typically narrows the stability regime in the parameter space,
especially as concerns the higher wavenumber instabilities, 
such as $m=2$, $m=3$ and so on. Perhaps more importantly,
the latter component can endow the structure with
additional instabilities in a regime where the VR would be
structurally stable.
These instabilities may be weak, yet they are of interest
given their diverse possible origin, both in terms of
the mode responsible and the linear vs. nonlinear
nature of the instability. For instance, we
discussed herein (and qualitatively justified) the oscillatory
instability arising from the $m=1$ mode, as well as the rather
narrow instability of the $m=0$ mode at the linear (spectral)
level. There are also nonlinear, parametric instabilities that
are present, arising from the resonance of modes such as 
$m=0$ and $m=2$, as we have also elucidated.

Naturally, there are numerous future directions that
are emerging as a result of the present work. 
One important aspect of consideration of such topologically
charged states is the consideration of not only density-induced
(spin-independent) effects as is the case herein, 
but also of spin-dependent
ones that arise in $F=1$ and $F=2$ spinor condensates~\cite{kawaguchi2012spinor,stamper2013spinor}. Hence,
extensions of considerations to similar (but also Skyrmion)
structures in 3-component BECs and beyond constitute a 
natural extension of the present work. Additionally, the
recent exploration of advanced numerical methods, such
as deflation~\cite{deflation} has enabled the characterization
of far more elaborate topological structures in single-component,
three-dimensional
condensates. It would be of particular interest to examine
generalizations of such states in multi-component settings
and to examine their associated mechanisms of instability.
Lastly, and similarly to the extensions of a single VR
to multiple ones in single-component BECs~\cite{Wang:VR}, 
the interaction
of a VRB with a VR or with another VRB would be interesting to 
quantify from an analytical and numerical perspective.
Such studies will be reported in future publications.

\acknowledgments 
W.W.~acknowledges supports from the National Science Foundation 
of China under Grant No. 12004268, the Fundamental Research Funds 
for the Central Universities, China, and the Science Speciality 
Program of Sichuan University under Grant No. 2020SCUNL210.
The work of V.P.R. was supported by the Russian Federation 
Program No. 0033-2019-0003. This material is based upon work
supported by the US National Science Foundation under 
Grant No. PHY-2110030 (P.G.K.).
C.T. was supported by the US Department of Energy through the Los Alamos National Laboratory. Los Alamos National Laboratory is operated by Triad National Security, LLC, for the National Nuclear Security Administration of U.S. Department of Energy (Contract No. 89233218NCA000001). 
We thank the Emei cluster at Sichuan university for providing HPC resources.

\bibliography{Refs}

\begin{thebibliography}{68}
\expandafter\ifx\csname natexlab\endcsname\relax\def\natexlab#1{#1}\fi
\expandafter\ifx\csname bibnamefont\endcsname\relax
  \def\bibnamefont#1{#1}\fi
\expandafter\ifx\csname bibfnamefont\endcsname\relax
  \def\bibfnamefont#1{#1}\fi
\expandafter\ifx\csname citenamefont\endcsname\relax
  \def\citenamefont#1{#1}\fi
\expandafter\ifx\csname url\endcsname\relax
  \def\url#1{\texttt{#1}}\fi
\expandafter\ifx\csname urlprefix\endcsname\relax\def\urlprefix{URL }\fi
\providecommand{\bibinfo}[2]{#2}
\providecommand{\eprint}[2][]{\url{#2}}

\bibitem[{\citenamefont{Pethick and Smith}(2002)}]{pethick}
\bibinfo{author}{\bibfnamefont{C.~J.} \bibnamefont{Pethick}} \bibnamefont{and}
  \bibinfo{author}{\bibfnamefont{H.}~\bibnamefont{Smith}},
  \emph{\bibinfo{title}{Bose-Einstein Condensation in Dilute Gases}}
  (\bibinfo{publisher}{Cambridge University Press},
  \bibinfo{address}{Cambridge, United Kingdom}, \bibinfo{year}{2002}).

\bibitem[{\citenamefont{Stringari and Pitaevskii}(2003)}]{string}
\bibinfo{author}{\bibfnamefont{S.}~\bibnamefont{Stringari}} \bibnamefont{and}
  \bibinfo{author}{\bibfnamefont{L.}~\bibnamefont{Pitaevskii}},
  \emph{\bibinfo{title}{Bose-Einstein Condensation}}
  (\bibinfo{publisher}{Oxford University Press}, \bibinfo{address}{Oxford,
  United Kingdom}, \bibinfo{year}{2003}).

\bibitem[{\citenamefont{Kevrekidis et~al.}(2015)\citenamefont{Kevrekidis,
  Frantzeskakis, and Carretero-Gonz\'{a}lez}}]{Panos:book}
\bibinfo{author}{\bibfnamefont{P.~G.} \bibnamefont{Kevrekidis}},
  \bibinfo{author}{\bibfnamefont{D.~J.} \bibnamefont{Frantzeskakis}},
  \bibnamefont{and}
  \bibinfo{author}{\bibfnamefont{R.}~\bibnamefont{Carretero-Gonz\'{a}lez}},
  \emph{\bibinfo{title}{The Defocusing Nonlinear Schr\"{o}dinger Equation: From
  Dark Solitons to Vortices and Vortex Rings}} (\bibinfo{publisher}{SIAM,
  Philadelphia}, \bibinfo{year}{2015}).

\bibitem[{\citenamefont{Fetter and Svidzinsky}(2001)}]{Alexander2001}
\bibinfo{author}{\bibfnamefont{A.~L.} \bibnamefont{Fetter}} \bibnamefont{and}
  \bibinfo{author}{\bibfnamefont{A.~A.} \bibnamefont{Svidzinsky}},
  \emph{\bibinfo{title}{{Vortices in a trapped dilute Bose-Einstein
  condensate}}}, \bibinfo{journal}{Journal of Physics: Condensed Matter}
  \textbf{\bibinfo{volume}{13}}, \bibinfo{pages}{R135} (\bibinfo{year}{2001}).

\bibitem[{\citenamefont{Fetter}(2009)}]{fetter2}
\bibinfo{author}{\bibfnamefont{A.~L.} \bibnamefont{Fetter}},
  \emph{\bibinfo{title}{{Rotating trapped Bose-Einstein condensates}}},
  \bibinfo{journal}{Rev. Mod. Phys.} \textbf{\bibinfo{volume}{81}},
  \bibinfo{pages}{647} (\bibinfo{year}{2009}).

\bibitem[{\citenamefont{Komineas}(2007)}]{Komineas}
\bibinfo{author}{\bibfnamefont{S.}~\bibnamefont{Komineas}},
  \emph{\bibinfo{title}{{Vortex rings and solitary waves in trapped
  {B}ose--{E}instein condensates}}}, \bibinfo{journal}{The European Physical
  Journal Special Topics} \textbf{\bibinfo{volume}{147}}, \bibinfo{pages}{133}
  (\bibinfo{year}{2007}).

\bibitem[{\citenamefont{Marzlin et~al.}(2000)\citenamefont{Marzlin, Zhang, and
  Sanders}}]{marzlin2000creation}
\bibinfo{author}{\bibfnamefont{K.-P.} \bibnamefont{Marzlin}},
  \bibinfo{author}{\bibfnamefont{W.}~\bibnamefont{Zhang}}, \bibnamefont{and}
  \bibinfo{author}{\bibfnamefont{B.~C.} \bibnamefont{Sanders}},
  \emph{\bibinfo{title}{{Creation of skyrmions in a spinor Bose-Einstein
  condensate}}}, \bibinfo{journal}{Phys. Rev. A} \textbf{\bibinfo{volume}{62}},
  \bibinfo{pages}{013602} (\bibinfo{year}{2000}).

\bibitem[{\citenamefont{Mizushima et~al.}(2002)\citenamefont{Mizushima,
  Machida, and Kita}}]{mizushima2002mermin}
\bibinfo{author}{\bibfnamefont{T.}~\bibnamefont{Mizushima}},
  \bibinfo{author}{\bibfnamefont{K.}~\bibnamefont{Machida}}, \bibnamefont{and}
  \bibinfo{author}{\bibfnamefont{T.}~\bibnamefont{Kita}},
  \emph{\bibinfo{title}{{Mermin-Ho vortex in ferromagnetic spinor Bose-Einstein
  condensates}}}, \bibinfo{journal}{Phys. Rev. Lett.}
  \textbf{\bibinfo{volume}{89}}, \bibinfo{pages}{030401}
  (\bibinfo{year}{2002}).

\bibitem[{\citenamefont{Reijnders et~al.}(2004)\citenamefont{Reijnders,
  Van~Lankvelt, Schoutens, and Read}}]{reijnders2004rotating}
\bibinfo{author}{\bibfnamefont{J.~W.} \bibnamefont{Reijnders}},
  \bibinfo{author}{\bibfnamefont{F.~J.~M.} \bibnamefont{Van~Lankvelt}},
  \bibinfo{author}{\bibfnamefont{K.}~\bibnamefont{Schoutens}},
  \bibnamefont{and} \bibinfo{author}{\bibfnamefont{N.}~\bibnamefont{Read}},
  \emph{\bibinfo{title}{{Rotating spin-1 bosons in the lowest Landau level}}},
  \bibinfo{journal}{Phys. Rev. A} \textbf{\bibinfo{volume}{69}},
  \bibinfo{pages}{023612} (\bibinfo{year}{2004}).

\bibitem[{\citenamefont{Ollikainen et~al.}(2017)\citenamefont{Ollikainen,
  Tiurev, Blinova, Lee, Hall, and M\"ott\"onen}}]{dsh1}
\bibinfo{author}{\bibfnamefont{T.}~\bibnamefont{Ollikainen}},
  \bibinfo{author}{\bibfnamefont{K.}~\bibnamefont{Tiurev}},
  \bibinfo{author}{\bibfnamefont{A.}~\bibnamefont{Blinova}},
  \bibinfo{author}{\bibfnamefont{W.}~\bibnamefont{Lee}},
  \bibinfo{author}{\bibfnamefont{D.~S.} \bibnamefont{Hall}}, \bibnamefont{and}
  \bibinfo{author}{\bibfnamefont{M.}~\bibnamefont{M\"ott\"onen}},
  \emph{\bibinfo{title}{{Experimental Realization of a Dirac Monopole through
  the Decay of an Isolated Monopole}}}, \bibinfo{journal}{Phys. Rev. X}
  \textbf{\bibinfo{volume}{7}}, \bibinfo{pages}{021023} (\bibinfo{year}{2017}).

\bibitem[{\citenamefont{Hall et~al.}(2016)\citenamefont{Hall, Ray, Tiurev,
  Ruokokoski, Gheorghe, and M{\"o}tt{\"o}nen}}]{dsh2}
\bibinfo{author}{\bibfnamefont{D.~S.} \bibnamefont{Hall}},
  \bibinfo{author}{\bibfnamefont{M.~W.} \bibnamefont{Ray}},
  \bibinfo{author}{\bibfnamefont{K.}~\bibnamefont{Tiurev}},
  \bibinfo{author}{\bibfnamefont{E.}~\bibnamefont{Ruokokoski}},
  \bibinfo{author}{\bibfnamefont{A.~H.} \bibnamefont{Gheorghe}},
  \bibnamefont{and}
  \bibinfo{author}{\bibfnamefont{M.}~\bibnamefont{M{\"o}tt{\"o}nen}},
  \emph{\bibinfo{title}{Tying quantum knots}}, \bibinfo{journal}{Nat. Phys.}
  \textbf{\bibinfo{volume}{12}}, \bibinfo{pages}{478} (\bibinfo{year}{2016}).

\bibitem[{\citenamefont{Lee et~al.}(2018)\citenamefont{Lee, Gheorghe, Tiurev,
  Ollikainen, M{\"o}tt{\"o}nen, and Hall}}]{dsh3}
\bibinfo{author}{\bibfnamefont{W.}~\bibnamefont{Lee}},
  \bibinfo{author}{\bibfnamefont{A.~H.} \bibnamefont{Gheorghe}},
  \bibinfo{author}{\bibfnamefont{K.}~\bibnamefont{Tiurev}},
  \bibinfo{author}{\bibfnamefont{T.}~\bibnamefont{Ollikainen}},
  \bibinfo{author}{\bibfnamefont{M.}~\bibnamefont{M{\"o}tt{\"o}nen}},
  \bibnamefont{and} \bibinfo{author}{\bibfnamefont{D.~S.} \bibnamefont{Hall}},
  \emph{\bibinfo{title}{Synthetic electromagnetic knot in a three-dimensional
  skyrmion}}, \bibinfo{journal}{Science Advances} \textbf{\bibinfo{volume}{4}}
  (\bibinfo{year}{2018}).

\bibitem[{\citenamefont{Ruban}(2018{\natexlab{a}})}]{Ruban:Knots}
\bibinfo{author}{\bibfnamefont{V.~P.} \bibnamefont{Ruban}},
  \emph{\bibinfo{title}{{Three-Dimensional Numerical Simulation of Long-Lived
  Quantum Vortex Knots and Links in a Trapped Bose Condensate}}},
  \bibinfo{journal}{JETP Letters} \textbf{\bibinfo{volume}{108}},
  \bibinfo{pages}{605} (\bibinfo{year}{2018}{\natexlab{a}}).

\bibitem[{\citenamefont{Ticknor et~al.}(2019)\citenamefont{Ticknor, Ruban, and
  Kevrekidis}}]{PhysRevA.99.063604}
\bibinfo{author}{\bibfnamefont{C.}~\bibnamefont{Ticknor}},
  \bibinfo{author}{\bibfnamefont{V.~P.} \bibnamefont{Ruban}}, \bibnamefont{and}
  \bibinfo{author}{\bibfnamefont{P.~G.} \bibnamefont{Kevrekidis}},
  \emph{\bibinfo{title}{{Quasistable quantum vortex knots and links in
  anisotropic harmonically trapped Bose-Einstein condensates}}},
  \bibinfo{journal}{Phys. Rev. A} \textbf{\bibinfo{volume}{99}},
  \bibinfo{pages}{063604} (\bibinfo{year}{2019}).

\bibitem[{\citenamefont{Kevrekidis and
  Frantzeskakis}(2016)}]{KEVREKIDIS2016140}
\bibinfo{author}{\bibfnamefont{P.~G.} \bibnamefont{Kevrekidis}}
  \bibnamefont{and} \bibinfo{author}{\bibfnamefont{D.~J.}
  \bibnamefont{Frantzeskakis}}, \emph{\bibinfo{title}{{Solitons in coupled
  nonlinear Schr\"odinger models: A survey of recent developments}}},
  \bibinfo{journal}{Reviews in Physics} \textbf{\bibinfo{volume}{1}},
  \bibinfo{pages}{140 } (\bibinfo{year}{2016}).

\bibitem[{\citenamefont{Kawaguchi and Ueda}(2012)}]{kawaguchi2012spinor}
\bibinfo{author}{\bibfnamefont{Y.}~\bibnamefont{Kawaguchi}} \bibnamefont{and}
  \bibinfo{author}{\bibfnamefont{M.}~\bibnamefont{Ueda}},
  \emph{\bibinfo{title}{{Spinor Bose-Einstein condensates}}},
  \bibinfo{journal}{Physics Reports} \textbf{\bibinfo{volume}{520}},
  \bibinfo{pages}{253 } (\bibinfo{year}{2012}).

\bibitem[{\citenamefont{Stamper-Kurn and Ueda}(2013)}]{stamper2013spinor}
\bibinfo{author}{\bibfnamefont{D.~M.} \bibnamefont{Stamper-Kurn}}
  \bibnamefont{and} \bibinfo{author}{\bibfnamefont{M.}~\bibnamefont{Ueda}},
  \emph{\bibinfo{title}{{Spinor Bose gases: Symmetries, magnetism, and quantum
  dynamics}}}, \bibinfo{journal}{Rev. Mod. Phys.}
  \textbf{\bibinfo{volume}{85}}, \bibinfo{pages}{1191} (\bibinfo{year}{2013}).

\bibitem[{\citenamefont{Trippenbach et~al.}(2000)\citenamefont{Trippenbach,
  G{\'{o}}ral, Rzazewski, Malomed, and Band}}]{Trippenbach_2000}
\bibinfo{author}{\bibfnamefont{M.}~\bibnamefont{Trippenbach}},
  \bibinfo{author}{\bibfnamefont{K.}~\bibnamefont{G{\'{o}}ral}},
  \bibinfo{author}{\bibfnamefont{K.}~\bibnamefont{Rzazewski}},
  \bibinfo{author}{\bibfnamefont{B.}~\bibnamefont{Malomed}}, \bibnamefont{and}
  \bibinfo{author}{\bibfnamefont{Y.~B.} \bibnamefont{Band}},
  \emph{\bibinfo{title}{{Structure of binary Bose-Einstein condensates}}},
  \bibinfo{journal}{J. Phys. B: At. Mol. and Opt. Phys}
  \textbf{\bibinfo{volume}{33}}, \bibinfo{pages}{4017} (\bibinfo{year}{2000}).

\bibitem[{\citenamefont{Barankov}(2002)}]{Barankov2002}
\bibinfo{author}{\bibfnamefont{R.~A.} \bibnamefont{Barankov}},
  \emph{\bibinfo{title}{{Boundary of two mixed Bose-Einstein condensates}}},
  \bibinfo{journal}{Phys. Rev. A} \textbf{\bibinfo{volume}{66}},
  \bibinfo{pages}{013612} (\bibinfo{year}{2002}).

\bibitem[{\citenamefont{Lee et~al.}(2016)\citenamefont{Lee, J\o{}rgensen, Liu,
  Wacker, Arlt, and Proukakis}}]{Lee2016}
\bibinfo{author}{\bibfnamefont{K.~L.} \bibnamefont{Lee}},
  \bibinfo{author}{\bibfnamefont{N.~B.} \bibnamefont{J\o{}rgensen}},
  \bibinfo{author}{\bibfnamefont{I.-K.} \bibnamefont{Liu}},
  \bibinfo{author}{\bibfnamefont{L.}~\bibnamefont{Wacker}},
  \bibinfo{author}{\bibfnamefont{J.~J.} \bibnamefont{Arlt}}, \bibnamefont{and}
  \bibinfo{author}{\bibfnamefont{N.~P.} \bibnamefont{Proukakis}},
  \emph{\bibinfo{title}{{Phase separation and dynamics of two-component
  Bose-Einstein condensates}}}, \bibinfo{journal}{Phys. Rev. A}
  \textbf{\bibinfo{volume}{94}}, \bibinfo{pages}{013602}
  (\bibinfo{year}{2016}).

\bibitem[{\citenamefont{Indekeu et~al.}(2015)\citenamefont{Indekeu, Lin,
  Van~Thu, Van~Schaeybroeck, and Phat}}]{Indekeu2015}
\bibinfo{author}{\bibfnamefont{J.~O.} \bibnamefont{Indekeu}},
  \bibinfo{author}{\bibfnamefont{C.-Y.} \bibnamefont{Lin}},
  \bibinfo{author}{\bibfnamefont{N.}~\bibnamefont{Van~Thu}},
  \bibinfo{author}{\bibfnamefont{B.}~\bibnamefont{Van~Schaeybroeck}},
  \bibnamefont{and} \bibinfo{author}{\bibfnamefont{T.~H.} \bibnamefont{Phat}},
  \emph{\bibinfo{title}{{Static interfacial properties of
  Bose-Einstein-condensate mixtures}}}, \bibinfo{journal}{Phys. Rev. A}
  \textbf{\bibinfo{volume}{91}}, \bibinfo{pages}{033615}
  (\bibinfo{year}{2015}).

\bibitem[{\citenamefont{Sasaki et~al.}(2009)\citenamefont{Sasaki, Suzuki,
  Akamatsu, and Saito}}]{Sasaki}
\bibinfo{author}{\bibfnamefont{K.}~\bibnamefont{Sasaki}},
  \bibinfo{author}{\bibfnamefont{N.}~\bibnamefont{Suzuki}},
  \bibinfo{author}{\bibfnamefont{D.}~\bibnamefont{Akamatsu}}, \bibnamefont{and}
  \bibinfo{author}{\bibfnamefont{H.}~\bibnamefont{Saito}},
  \emph{\bibinfo{title}{{Rayleigh-Taylor instability and mushroom-pattern
  formation in a two-component Bose-Einstein condensate}}},
  \bibinfo{journal}{Phys. Rev. A} \textbf{\bibinfo{volume}{80}},
  \bibinfo{pages}{063611} (\bibinfo{year}{2009}).

\bibitem[{\citenamefont{Gautam and Angom}(2010)}]{Gautam}
\bibinfo{author}{\bibfnamefont{S.}~\bibnamefont{Gautam}} \bibnamefont{and}
  \bibinfo{author}{\bibfnamefont{D.}~\bibnamefont{Angom}},
  \emph{\bibinfo{title}{{Rayleigh-Taylor instability in binary condensates}}},
  \bibinfo{journal}{Phys. Rev. A} \textbf{\bibinfo{volume}{81}},
  \bibinfo{pages}{053616} (\bibinfo{year}{2010}).

\bibitem[{\citenamefont{Kadokura et~al.}(2012)\citenamefont{Kadokura, Aioi,
  Sasaki, Kishimoto, and Saito}}]{Kadokura}
\bibinfo{author}{\bibfnamefont{T.}~\bibnamefont{Kadokura}},
  \bibinfo{author}{\bibfnamefont{T.}~\bibnamefont{Aioi}},
  \bibinfo{author}{\bibfnamefont{K.}~\bibnamefont{Sasaki}},
  \bibinfo{author}{\bibfnamefont{T.}~\bibnamefont{Kishimoto}},
  \bibnamefont{and} \bibinfo{author}{\bibfnamefont{H.}~\bibnamefont{Saito}},
  \emph{\bibinfo{title}{{Rayleigh-Taylor instability in a two-component
  Bose-Einstein condensate with rotational symmetry}}}, \bibinfo{journal}{Phys.
  Rev. A} \textbf{\bibinfo{volume}{85}}, \bibinfo{pages}{013602}
  (\bibinfo{year}{2012}).

\bibitem[{\citenamefont{Takeuchi
  et~al.}(2010{\natexlab{a}})\citenamefont{Takeuchi, Suzuki, Kasamatsu, Saito,
  and Tsubota}}]{Takeuchi}
\bibinfo{author}{\bibfnamefont{H.}~\bibnamefont{Takeuchi}},
  \bibinfo{author}{\bibfnamefont{N.}~\bibnamefont{Suzuki}},
  \bibinfo{author}{\bibfnamefont{K.}~\bibnamefont{Kasamatsu}},
  \bibinfo{author}{\bibfnamefont{H.}~\bibnamefont{Saito}}, \bibnamefont{and}
  \bibinfo{author}{\bibfnamefont{M.}~\bibnamefont{Tsubota}},
  \emph{\bibinfo{title}{{Quantum Kelvin-Helmholtz instability in
  phase-separated two-component Bose-Einstein condensates}}},
  \bibinfo{journal}{Phys. Rev. B} \textbf{\bibinfo{volume}{81}},
  \bibinfo{pages}{094517} (\bibinfo{year}{2010}{\natexlab{a}}).

\bibitem[{\citenamefont{Suzuki et~al.}(2010)\citenamefont{Suzuki, Takeuchi,
  Kasamatsu, Tsubota, and Saito}}]{Suzuki}
\bibinfo{author}{\bibfnamefont{N.}~\bibnamefont{Suzuki}},
  \bibinfo{author}{\bibfnamefont{H.}~\bibnamefont{Takeuchi}},
  \bibinfo{author}{\bibfnamefont{K.}~\bibnamefont{Kasamatsu}},
  \bibinfo{author}{\bibfnamefont{M.}~\bibnamefont{Tsubota}}, \bibnamefont{and}
  \bibinfo{author}{\bibfnamefont{H.}~\bibnamefont{Saito}},
  \emph{\bibinfo{title}{{Crossover between Kelvin-Helmholtz and
  counter-superflow instabilities in two-component Bose-Einstein
  condensates}}}, \bibinfo{journal}{Phys. Rev. A}
  \textbf{\bibinfo{volume}{82}}, \bibinfo{pages}{063604}
  (\bibinfo{year}{2010}).

\bibitem[{\citenamefont{Baggaley and Parker}(2018)}]{Baggaley}
\bibinfo{author}{\bibfnamefont{A.~W.} \bibnamefont{Baggaley}} \bibnamefont{and}
  \bibinfo{author}{\bibfnamefont{N.~G.} \bibnamefont{Parker}},
  \emph{\bibinfo{title}{{Kelvin-Helmholtz instability in a single-component
  atomic superfluid}}}, \bibinfo{journal}{Phys. Rev. A}
  \textbf{\bibinfo{volume}{97}}, \bibinfo{pages}{053608}
  (\bibinfo{year}{2018}).

\bibitem[{\citenamefont{Sasaki et~al.}(2011)\citenamefont{Sasaki, Suzuki, and
  Saito}}]{Sasaki_K}
\bibinfo{author}{\bibfnamefont{K.}~\bibnamefont{Sasaki}},
  \bibinfo{author}{\bibfnamefont{N.}~\bibnamefont{Suzuki}}, \bibnamefont{and}
  \bibinfo{author}{\bibfnamefont{H.}~\bibnamefont{Saito}},
  \emph{\bibinfo{title}{{Capillary instability in a two-component Bose-Einstein
  condensate}}}, \bibinfo{journal}{Phys. Rev. A} \textbf{\bibinfo{volume}{83}},
  \bibinfo{pages}{053606} (\bibinfo{year}{2011}).

\bibitem[{\citenamefont{Indekeu et~al.}(2018)\citenamefont{Indekeu, Van~Thu,
  Lin, and Phat}}]{Indekeu2018}
\bibinfo{author}{\bibfnamefont{J.~O.} \bibnamefont{Indekeu}},
  \bibinfo{author}{\bibfnamefont{N.}~\bibnamefont{Van~Thu}},
  \bibinfo{author}{\bibfnamefont{C.-Y.} \bibnamefont{Lin}}, \bibnamefont{and}
  \bibinfo{author}{\bibfnamefont{T.~H.} \bibnamefont{Phat}},
  \emph{\bibinfo{title}{{Capillary-wave dynamics and interface structure
  modulation in binary Bose-Einstein condensate mixtures}}},
  \bibinfo{journal}{Phys. Rev. A} \textbf{\bibinfo{volume}{97}},
  \bibinfo{pages}{043605} (\bibinfo{year}{2018}).

\bibitem[{\citenamefont{Bezett et~al.}(2010)\citenamefont{Bezett, Bychkov,
  Lundh, Kobyakov, and Marklund}}]{Bezett}
\bibinfo{author}{\bibfnamefont{A.}~\bibnamefont{Bezett}},
  \bibinfo{author}{\bibfnamefont{V.}~\bibnamefont{Bychkov}},
  \bibinfo{author}{\bibfnamefont{E.}~\bibnamefont{Lundh}},
  \bibinfo{author}{\bibfnamefont{D.}~\bibnamefont{Kobyakov}}, \bibnamefont{and}
  \bibinfo{author}{\bibfnamefont{M.}~\bibnamefont{Marklund}},
  \emph{\bibinfo{title}{{Magnetic Richtmyer-Meshkov instability in a
  two-component Bose-Einstein condensate}}}, \bibinfo{journal}{Phys. Rev. A}
  \textbf{\bibinfo{volume}{82}}, \bibinfo{pages}{043608}
  (\bibinfo{year}{2010}).

\bibitem[{\citenamefont{Law et~al.}(2001)\citenamefont{Law, Chan, Leung, and
  Chu}}]{Law}
\bibinfo{author}{\bibfnamefont{C.~K.} \bibnamefont{Law}},
  \bibinfo{author}{\bibfnamefont{C.~M.} \bibnamefont{Chan}},
  \bibinfo{author}{\bibfnamefont{P.~T.} \bibnamefont{Leung}}, \bibnamefont{and}
  \bibinfo{author}{\bibfnamefont{M.-C.} \bibnamefont{Chu}},
  \emph{\bibinfo{title}{{Critical velocity in a binary mixture of moving Bose
  condensates}}}, \bibinfo{journal}{Phys. Rev. A}
  \textbf{\bibinfo{volume}{63}}, \bibinfo{pages}{063612}
  (\bibinfo{year}{2001}).

\bibitem[{\citenamefont{Yukalov and Yukalova}(2004)}]{Yukalov}
\bibinfo{author}{\bibfnamefont{V.~I.} \bibnamefont{Yukalov}} \bibnamefont{and}
  \bibinfo{author}{\bibfnamefont{E.~P.} \bibnamefont{Yukalova}},
  \emph{\bibinfo{title}{{Stratification of moving multicomponent Bose-Einstein
  condensates}}}, \bibinfo{journal}{Laser Physics Letters}
  \textbf{\bibinfo{volume}{1}}, \bibinfo{pages}{50} (\bibinfo{year}{2004}).

\bibitem[{\citenamefont{Takeuchi
  et~al.}(2010{\natexlab{b}})\citenamefont{Takeuchi, Ishino, and
  Tsubota}}]{Takeuchi_2}
\bibinfo{author}{\bibfnamefont{H.}~\bibnamefont{Takeuchi}},
  \bibinfo{author}{\bibfnamefont{S.}~\bibnamefont{Ishino}}, \bibnamefont{and}
  \bibinfo{author}{\bibfnamefont{M.}~\bibnamefont{Tsubota}},
  \emph{\bibinfo{title}{{Binary Quantum Turbulence Arising from
  Countersuperflow Instability in Two-Component Bose-Einstein Condensates}}},
  \bibinfo{journal}{Phys. Rev. Lett.} \textbf{\bibinfo{volume}{105}},
  \bibinfo{pages}{205301} (\bibinfo{year}{2010}{\natexlab{b}}).

\bibitem[{\citenamefont{Hamner et~al.}(2011)\citenamefont{Hamner, Chang,
  Engels, and Hoefer}}]{Hamner}
\bibinfo{author}{\bibfnamefont{C.}~\bibnamefont{Hamner}},
  \bibinfo{author}{\bibfnamefont{J.~J.} \bibnamefont{Chang}},
  \bibinfo{author}{\bibfnamefont{P.}~\bibnamefont{Engels}}, \bibnamefont{and}
  \bibinfo{author}{\bibfnamefont{M.~A.} \bibnamefont{Hoefer}},
  \emph{\bibinfo{title}{{Generation of Dark-Bright Soliton Trains in
  Superfluid-Superfluid Counterflow}}}, \bibinfo{journal}{Phys. Rev. Lett.}
  \textbf{\bibinfo{volume}{106}}, \bibinfo{pages}{065302}
  (\bibinfo{year}{2011}).

\bibitem[{\citenamefont{Saito et~al.}(2009)\citenamefont{Saito, Kawaguchi, and
  Ueda}}]{Saito}
\bibinfo{author}{\bibfnamefont{H.}~\bibnamefont{Saito}},
  \bibinfo{author}{\bibfnamefont{Y.}~\bibnamefont{Kawaguchi}},
  \bibnamefont{and} \bibinfo{author}{\bibfnamefont{M.}~\bibnamefont{Ueda}},
  \emph{\bibinfo{title}{{Ferrofluidity in a Two-Component Dipolar Bose-Einstein
  Condensate}}}, \bibinfo{journal}{Phys. Rev. Lett.}
  \textbf{\bibinfo{volume}{102}}, \bibinfo{pages}{230403}
  (\bibinfo{year}{2009}).

\bibitem[{\citenamefont{Bisset et~al.}(2015{\natexlab{a}})\citenamefont{Bisset,
  Wang, Ticknor, Carretero-Gonz\'alez, Frantzeskakis, Collins, and
  Kevrekidis}}]{Wang:hopfions}
\bibinfo{author}{\bibfnamefont{R.~N.} \bibnamefont{Bisset}},
  \bibinfo{author}{\bibfnamefont{W.}~\bibnamefont{Wang}},
  \bibinfo{author}{\bibfnamefont{C.}~\bibnamefont{Ticknor}},
  \bibinfo{author}{\bibfnamefont{R.}~\bibnamefont{Carretero-Gonz\'alez}},
  \bibinfo{author}{\bibfnamefont{D.~J.} \bibnamefont{Frantzeskakis}},
  \bibinfo{author}{\bibfnamefont{L.~A.} \bibnamefont{Collins}},
  \bibnamefont{and} \bibinfo{author}{\bibfnamefont{P.~G.}
  \bibnamefont{Kevrekidis}}, \emph{\bibinfo{title}{{Robust vortex lines, vortex
  rings, and hopfions in three-dimensional Bose-Einstein condensates}}},
  \bibinfo{journal}{Phys. Rev. A} \textbf{\bibinfo{volume}{92}},
  \bibinfo{pages}{063611} (\bibinfo{year}{2015}{\natexlab{a}}).

\bibitem[{\citenamefont{Wang et~al.}(2017)\citenamefont{Wang, Bisset, Ticknor,
  Carretero-Gonz\'alez, Frantzeskakis, Collins, and Kevrekidis}}]{Wang:VR}
\bibinfo{author}{\bibfnamefont{W.}~\bibnamefont{Wang}},
  \bibinfo{author}{\bibfnamefont{R.~N.} \bibnamefont{Bisset}},
  \bibinfo{author}{\bibfnamefont{C.}~\bibnamefont{Ticknor}},
  \bibinfo{author}{\bibfnamefont{R.}~\bibnamefont{Carretero-Gonz\'alez}},
  \bibinfo{author}{\bibfnamefont{D.~J.} \bibnamefont{Frantzeskakis}},
  \bibinfo{author}{\bibfnamefont{L.~A.} \bibnamefont{Collins}},
  \bibnamefont{and} \bibinfo{author}{\bibfnamefont{P.~G.}
  \bibnamefont{Kevrekidis}}, \emph{\bibinfo{title}{{Single and multiple vortex
  rings in three-dimensional Bose-Einstein condensates: Existence, stability,
  and dynamics}}}, \bibinfo{journal}{Phys. Rev. A}
  \textbf{\bibinfo{volume}{95}}, \bibinfo{pages}{043638}
  (\bibinfo{year}{2017}).

\bibitem[{\citenamefont{Bisset et~al.}(2015{\natexlab{b}})\citenamefont{Bisset,
  Wang, Ticknor, Carretero-Gonz\'alez, Frantzeskakis, Collins, and
  Kevrekidis}}]{Wang:DSVR}
\bibinfo{author}{\bibfnamefont{R.~N.} \bibnamefont{Bisset}},
  \bibinfo{author}{\bibfnamefont{W.}~\bibnamefont{Wang}},
  \bibinfo{author}{\bibfnamefont{C.}~\bibnamefont{Ticknor}},
  \bibinfo{author}{\bibfnamefont{R.}~\bibnamefont{Carretero-Gonz\'alez}},
  \bibinfo{author}{\bibfnamefont{D.~J.} \bibnamefont{Frantzeskakis}},
  \bibinfo{author}{\bibfnamefont{L.~A.} \bibnamefont{Collins}},
  \bibnamefont{and} \bibinfo{author}{\bibfnamefont{P.~G.}
  \bibnamefont{Kevrekidis}}, \emph{\bibinfo{title}{{Bifurcation and stability
  of single and multiple vortex rings in three-dimensional Bose-Einstein
  condensates}}}, \bibinfo{journal}{Phys. Rev. A}
  \textbf{\bibinfo{volume}{92}}, \bibinfo{pages}{043601}
  (\bibinfo{year}{2015}{\natexlab{b}}).

\bibitem[{\citenamefont{Ruban}(2017{\natexlab{a}})}]{Ruban2017}
\bibinfo{author}{\bibfnamefont{V.~P.} \bibnamefont{Ruban}},
  \emph{\bibinfo{title}{{Parametric instability of oscillations of a vortex
  ring in a z-periodic Bose condensate and return to the initial state}}},
  \bibinfo{journal}{J. Exp. Theor. Phys. Lett.} \textbf{\bibinfo{volume}{106}},
  \bibinfo{pages}{223} (\bibinfo{year}{2017}{\natexlab{a}}).

\bibitem[{\citenamefont{Law et~al.}(2010)\citenamefont{Law, Kevrekidis, and
  Tuckerman}}]{VB1}
\bibinfo{author}{\bibfnamefont{K.~J.~H.} \bibnamefont{Law}},
  \bibinfo{author}{\bibfnamefont{P.~G.} \bibnamefont{Kevrekidis}},
  \bibnamefont{and} \bibinfo{author}{\bibfnamefont{L.~S.}
  \bibnamefont{Tuckerman}}, \emph{\bibinfo{title}{{Stable
  Vortex--Bright-Soliton Structures in Two-Component Bose-Einstein
  Condensates}}}, \bibinfo{journal}{Phys. Rev. Lett.}
  \textbf{\bibinfo{volume}{105}}, \bibinfo{pages}{160405}
  (\bibinfo{year}{2010}).

\bibitem[{\citenamefont{Pola et~al.}(2012)\citenamefont{Pola, Stockhofe,
  Schmelcher, and Kevrekidis}}]{VB2}
\bibinfo{author}{\bibfnamefont{M.}~\bibnamefont{Pola}},
  \bibinfo{author}{\bibfnamefont{J.}~\bibnamefont{Stockhofe}},
  \bibinfo{author}{\bibfnamefont{P.}~\bibnamefont{Schmelcher}},
  \bibnamefont{and} \bibinfo{author}{\bibfnamefont{P.~G.}
  \bibnamefont{Kevrekidis}}, \emph{\bibinfo{title}{{Vortex--bright-soliton
  dipoles: Bifurcations, symmetry breaking, and soliton tunneling in a
  vortex-induced double well}}}, \bibinfo{journal}{Phys. Rev. A}
  \textbf{\bibinfo{volume}{86}}, \bibinfo{pages}{053601}
  (\bibinfo{year}{2012}).

\bibitem[{\citenamefont{Hayashi et~al.}(2013)\citenamefont{Hayashi, Tsubota,
  and Takeuchi}}]{VB3}
\bibinfo{author}{\bibfnamefont{S.}~\bibnamefont{Hayashi}},
  \bibinfo{author}{\bibfnamefont{M.}~\bibnamefont{Tsubota}}, \bibnamefont{and}
  \bibinfo{author}{\bibfnamefont{H.}~\bibnamefont{Takeuchi}},
  \emph{\bibinfo{title}{{Instability crossover of helical shear flow in
  segregated Bose-Einstein condensates}}}, \bibinfo{journal}{Phys. Rev. A}
  \textbf{\bibinfo{volume}{87}}, \bibinfo{pages}{063628}
  (\bibinfo{year}{2013}).

\bibitem[{\citenamefont{Ruban}(2021{\natexlab{a}})}]{viktor_rec1}
\bibinfo{author}{\bibfnamefont{V.~P.} \bibnamefont{Ruban}},
  \emph{\bibinfo{title}{{ Instabilities of a Filled Vortex in a Two-Component
  Bose-Einstein Condensate}}}, \bibinfo{journal}{JETP Lett.}
  \textbf{\bibinfo{volume}{113}}, \bibinfo{pages}{532}
  (\bibinfo{year}{2021}{\natexlab{a}}).

\bibitem[{\citenamefont{Ruban}(2021{\natexlab{b}})}]{viktor_rec2}
\bibinfo{author}{\bibfnamefont{V.~P.} \bibnamefont{Ruban}},
  \emph{\bibinfo{title}{{Bubbles with attached quantum vortices in trapped
  binary Bose-Einstein condensates }}} (\bibinfo{year}{2021}{\natexlab{b}}),
  \bibinfo{note}{(arXiv:2104.05296)}.

\bibitem[{\citenamefont{Richaud et~al.}(2020)\citenamefont{Richaud, Penna,
  Mayol, and Guilleumas}}]{richaud1}
\bibinfo{author}{\bibfnamefont{A.}~\bibnamefont{Richaud}},
  \bibinfo{author}{\bibfnamefont{V.}~\bibnamefont{Penna}},
  \bibinfo{author}{\bibfnamefont{R.}~\bibnamefont{Mayol}}, \bibnamefont{and}
  \bibinfo{author}{\bibfnamefont{M.}~\bibnamefont{Guilleumas}},
  \emph{\bibinfo{title}{{Vortices with massive cores in a binary mixture of
  Bose-Einstein condensates}}}, \bibinfo{journal}{Phys. Rev. A}
  \textbf{\bibinfo{volume}{101}}, \bibinfo{pages}{013630}
  (\bibinfo{year}{2020}).

\bibitem[{\citenamefont{Richaud et~al.}(2021)\citenamefont{Richaud, Penna, and
  Fetter}}]{richaud2}
\bibinfo{author}{\bibfnamefont{A.}~\bibnamefont{Richaud}},
  \bibinfo{author}{\bibfnamefont{V.}~\bibnamefont{Penna}}, \bibnamefont{and}
  \bibinfo{author}{\bibfnamefont{A.~L.} \bibnamefont{Fetter}},
  \emph{\bibinfo{title}{{Dynamics of massive point vortices in a binary mixture
  of Bose-Einstein condensates}}}, \bibinfo{journal}{Phys. Rev. A}
  \textbf{\bibinfo{volume}{103}}, \bibinfo{pages}{023311}
  (\bibinfo{year}{2021}).

\bibitem[{\citenamefont{Wang}(2021)}]{wenlong_rec3}
\bibinfo{author}{\bibfnamefont{W.}~\bibnamefont{Wang}},
  \emph{\bibinfo{title}{{Controlled engineering of a vortex-bright soliton
  dynamics using a constant driving force}}} (\bibinfo{year}{2021}),
  \bibinfo{note}{arXiv preprint arXiv:2107.08247}.

\bibitem[{\citenamefont{Ticknor et~al.}(2018)\citenamefont{Ticknor, Wang, and
  Kevrekidis}}]{Wang:AI3}
\bibinfo{author}{\bibfnamefont{C.}~\bibnamefont{Ticknor}},
  \bibinfo{author}{\bibfnamefont{W.}~\bibnamefont{Wang}}, \bibnamefont{and}
  \bibinfo{author}{\bibfnamefont{P.~G.} \bibnamefont{Kevrekidis}},
  \emph{\bibinfo{title}{{Spectral and dynamical analysis of a single vortex
  ring in anisotropic harmonically trapped three-dimensional Bose-Einstein
  condensates}}}, \bibinfo{journal}{Phys. Rev. A}
  \textbf{\bibinfo{volume}{98}}, \bibinfo{pages}{033609}
  (\bibinfo{year}{2018}).

\bibitem[{\citenamefont{Horng et~al.}(2006)\citenamefont{Horng, Gou, and
  Lin}}]{Horng:VR}
\bibinfo{author}{\bibfnamefont{T.-L.} \bibnamefont{Horng}},
  \bibinfo{author}{\bibfnamefont{S.-C.} \bibnamefont{Gou}}, \bibnamefont{and}
  \bibinfo{author}{\bibfnamefont{T.-C.} \bibnamefont{Lin}},
  \emph{\bibinfo{title}{{Bending-wave instability of a vortex ring in a trapped
  Bose-Einstein condensate}}}, \bibinfo{journal}{Phys. Rev. A}
  \textbf{\bibinfo{volume}{74}}, \bibinfo{pages}{041603}
  (\bibinfo{year}{2006}).

\bibitem[{\citenamefont{Pu and Bigelow}(1998)}]{Pu-Bigelow-1998}
\bibinfo{author}{\bibfnamefont{H.}~\bibnamefont{Pu}} \bibnamefont{and}
  \bibinfo{author}{\bibfnamefont{N.~P.} \bibnamefont{Bigelow}},
  \emph{\bibinfo{title}{{Properties of Two-Species Bose Condensates}}},
  \bibinfo{journal}{Phys. Rev. Lett.} \textbf{\bibinfo{volume}{80}},
  \bibinfo{pages}{1130} (\bibinfo{year}{1998}).

\bibitem[{\citenamefont{Egorov et~al.}(2013)\citenamefont{Egorov, Opanchuk,
  Drummond, Hall, Hannaford, and Sidorov}}]{drummond}
\bibinfo{author}{\bibfnamefont{M.}~\bibnamefont{Egorov}},
  \bibinfo{author}{\bibfnamefont{B.}~\bibnamefont{Opanchuk}},
  \bibinfo{author}{\bibfnamefont{P.}~\bibnamefont{Drummond}},
  \bibinfo{author}{\bibfnamefont{B.~V.} \bibnamefont{Hall}},
  \bibinfo{author}{\bibfnamefont{P.}~\bibnamefont{Hannaford}},
  \bibnamefont{and} \bibinfo{author}{\bibfnamefont{A.~I.}
  \bibnamefont{Sidorov}}, \emph{\bibinfo{title}{{Measurement of $s$-wave
  scattering lengths in a two-component Bose-Einstein condensate}}},
  \bibinfo{journal}{Phys. Rev. A} \textbf{\bibinfo{volume}{87}},
  \bibinfo{pages}{053614} (\bibinfo{year}{2013}).

\bibitem[{\citenamefont{Lannig et~al.}(2020)\citenamefont{Lannig, Schmied,
  Pr\"ufer, Kunkel, Strohmaier, Strobel, Gasenzer, Kevrekidis, and
  Oberthaler}}]{lannig}
\bibinfo{author}{\bibfnamefont{S.}~\bibnamefont{Lannig}},
  \bibinfo{author}{\bibfnamefont{C.-M.} \bibnamefont{Schmied}},
  \bibinfo{author}{\bibfnamefont{M.}~\bibnamefont{Pr\"ufer}},
  \bibinfo{author}{\bibfnamefont{P.}~\bibnamefont{Kunkel}},
  \bibinfo{author}{\bibfnamefont{R.}~\bibnamefont{Strohmaier}},
  \bibinfo{author}{\bibfnamefont{H.}~\bibnamefont{Strobel}},
  \bibinfo{author}{\bibfnamefont{T.}~\bibnamefont{Gasenzer}},
  \bibinfo{author}{\bibfnamefont{P.~G.} \bibnamefont{Kevrekidis}},
  \bibnamefont{and} \bibinfo{author}{\bibfnamefont{M.~K.}
  \bibnamefont{Oberthaler}}, \emph{\bibinfo{title}{{Collisions of
  Three-Component Vector Solitons in Bose-Einstein Condensates}}},
  \bibinfo{journal}{Phys. Rev. Lett.} \textbf{\bibinfo{volume}{125}},
  \bibinfo{pages}{170401} (\bibinfo{year}{2020}).

\bibitem[{\citenamefont{Timmermans}(1998)}]{tim}
\bibinfo{author}{\bibfnamefont{E.}~\bibnamefont{Timmermans}},
  \emph{\bibinfo{title}{{Phase Separation of Bose-Einstein Condensates}}},
  \bibinfo{journal}{Phys. Rev. Lett.} \textbf{\bibinfo{volume}{81}},
  \bibinfo{pages}{5718} (\bibinfo{year}{1998}).

\bibitem[{\citenamefont{Ao and Chui}(1998)}]{aochui}
\bibinfo{author}{\bibfnamefont{P.}~\bibnamefont{Ao}} \bibnamefont{and}
  \bibinfo{author}{\bibfnamefont{S.~T.} \bibnamefont{Chui}},
  \emph{\bibinfo{title}{{Binary Bose-Einstein condensate mixtures in weakly and
  strongly segregated phases}}}, \bibinfo{journal}{Phys. Rev. A}
  \textbf{\bibinfo{volume}{58}}, \bibinfo{pages}{4836} (\bibinfo{year}{1998}).

\bibitem[{\citenamefont{Papp et~al.}(2008)\citenamefont{Papp, Pino, and
  Wieman}}]{Papp}
\bibinfo{author}{\bibfnamefont{S.~B.} \bibnamefont{Papp}},
  \bibinfo{author}{\bibfnamefont{J.~M.} \bibnamefont{Pino}}, \bibnamefont{and}
  \bibinfo{author}{\bibfnamefont{C.~E.} \bibnamefont{Wieman}},
  \emph{\bibinfo{title}{{Tunable Miscibility in a Dual-Species Bose-Einstein
  Condensate}}}, \bibinfo{journal}{Phys. Rev. Lett.}
  \textbf{\bibinfo{volume}{101}}, \bibinfo{pages}{040402}
  (\bibinfo{year}{2008}).

\bibitem[{\citenamefont{Busch and Anglin}(2001)}]{DBS1}
\bibinfo{author}{\bibfnamefont{T.}~\bibnamefont{Busch}} \bibnamefont{and}
  \bibinfo{author}{\bibfnamefont{J.~R.} \bibnamefont{Anglin}},
  \emph{\bibinfo{title}{{Dark-Bright Solitons in Inhomogeneous Bose-Einstein
  Condensates}}}, \bibinfo{journal}{Phys. Rev. Lett.}
  \textbf{\bibinfo{volume}{87}}, \bibinfo{pages}{010401}
  (\bibinfo{year}{2001}).

\bibitem[{\citenamefont{Ruban}(2001)}]{Ruban2001}
\bibinfo{author}{\bibfnamefont{V.~P.} \bibnamefont{Ruban}},
  \emph{\bibinfo{title}{{Slow inviscid flows of a compressible fluid in
  spatially inhomogeneous systems}}}, \bibinfo{journal}{Phys. Rev. E}
  \textbf{\bibinfo{volume}{64}}, \bibinfo{pages}{036305}
  (\bibinfo{year}{2001}).

\bibitem[{\citenamefont{Ruban}(2018{\natexlab{b}})}]{Ruban2018}
\bibinfo{author}{\bibfnamefont{V.~P.} \bibnamefont{Ruban}},
  \emph{\bibinfo{title}{{Stable and Unstable Vortex Knots in a Trapped Bose
  Condensate}}}, \bibinfo{journal}{J. Exp. Theor. Phys.}
  \textbf{\bibinfo{volume}{126}}, \bibinfo{pages}{397}
  (\bibinfo{year}{2018}{\natexlab{b}}).

\bibitem[{\citenamefont{Ruban}(2017{\natexlab{b}})}]{Ruban2017-2}
\bibinfo{author}{\bibfnamefont{V.~P.} \bibnamefont{Ruban}},
  \emph{\bibinfo{title}{{Dynamics of straight vortex filaments in a
  Bose-Einstein condensate with the Gaussian density profile}}},
  \bibinfo{journal}{J. Exp. Theor. Phys.} \textbf{\bibinfo{volume}{124}},
  \bibinfo{pages}{932} (\bibinfo{year}{2017}{\natexlab{b}}).

\bibitem[{\citenamefont{Ponstein}(1959)}]{Ponstein}
\bibinfo{author}{\bibfnamefont{J.}~\bibnamefont{Ponstein}},
  \emph{\bibinfo{title}{{Instability of rotating cylindrical jets}}},
  \bibinfo{journal}{Appl. Sci. Res.} \textbf{\bibinfo{volume}{8}},
  \bibinfo{pages}{425} (\bibinfo{year}{1959}).

\bibitem[{\citenamefont{Van~Schaeybroeck}(2008)}]{tension}
\bibinfo{author}{\bibfnamefont{B.}~\bibnamefont{Van~Schaeybroeck}},
  \emph{\bibinfo{title}{{Interface tension of Bose-Einstein condensates}}},
  \bibinfo{journal}{Phys. Rev. A} \textbf{\bibinfo{volume}{78}},
  \bibinfo{pages}{023624} (\bibinfo{year}{2008}).

\bibitem[{\citenamefont{Pitaevskii and Stringari}(2003)}]{becbook1}
\bibinfo{author}{\bibfnamefont{L.}~\bibnamefont{Pitaevskii}} \bibnamefont{and}
  \bibinfo{author}{\bibfnamefont{S.}~\bibnamefont{Stringari}},
  \emph{\bibinfo{title}{{B}ose--{E}instein Condensation}}
  (\bibinfo{publisher}{Oxford University Press}, \bibinfo{address}{Oxford, UK},
  \bibinfo{year}{2003}).

\bibitem[{\citenamefont{Wang et~al.}(2016)\citenamefont{Wang, Kevrekidis,
  Carretero-Gonz\'alez, and Frantzeskakis}}]{Wang:DSS}
\bibinfo{author}{\bibfnamefont{W.}~\bibnamefont{Wang}},
  \bibinfo{author}{\bibfnamefont{P.~G.} \bibnamefont{Kevrekidis}},
  \bibinfo{author}{\bibfnamefont{R.}~\bibnamefont{Carretero-Gonz\'alez}},
  \bibnamefont{and} \bibinfo{author}{\bibfnamefont{D.~J.}
  \bibnamefont{Frantzeskakis}}, \emph{\bibinfo{title}{{Dark spherical shell
  solitons in three-dimensional Bose-Einstein condensates: Existence,
  stability, and dynamics}}}, \bibinfo{journal}{Phys. Rev. A}
  \textbf{\bibinfo{volume}{93}}, \bibinfo{pages}{023630}
  (\bibinfo{year}{2016}).

\bibitem[{\citenamefont{Wang and Kevrekidis}(2017)}]{Wang:DBS}
\bibinfo{author}{\bibfnamefont{W.}~\bibnamefont{Wang}} \bibnamefont{and}
  \bibinfo{author}{\bibfnamefont{P.~G.} \bibnamefont{Kevrekidis}},
  \emph{\bibinfo{title}{{Two-component dark-bright solitons in
  three-dimensional atomic Bose-Einstein condensates}}},
  \bibinfo{journal}{Phys. Rev. E} \textbf{\bibinfo{volume}{95}},
  \bibinfo{pages}{032201} (\bibinfo{year}{2017}).

\bibitem[{\citenamefont{Kollár and Pego}(2011)}]{10.1093/amrx/abr007}
\bibinfo{author}{\bibfnamefont{R.}~\bibnamefont{Kollár}} \bibnamefont{and}
  \bibinfo{author}{\bibfnamefont{R.~L.} \bibnamefont{Pego}},
  \emph{\bibinfo{title}{{Spectral Stability of Vortices in Two-Dimensional
  Bose-Einstein Condensates via the Evans Function and Krein Signature}}},
  \bibinfo{journal}{Applied Mathematics Research eXpress}
  \textbf{\bibinfo{volume}{2012}}, \bibinfo{pages}{1} (\bibinfo{year}{2011}),
  ISSN \bibinfo{issn}{1687-1200}.

\bibitem[{VRB({\natexlab{a}})}]{VRBmv1}
\bibinfo{note}{Please see the relevant movie:
  \url{https://www.youtube.com/watch?v=2MOc4o_Zn14}}.

\bibitem[{VRB({\natexlab{b}})}]{VRBmv2}
\bibinfo{note}{Please see the relevant movie:
  \url{https://www.youtube.com/watch?v=oLbMDMiPUQY}}.

\bibitem[{\citenamefont{Boull\'e et~al.}(2020)\citenamefont{Boull\'e,
  Charalampidis, Farrell, and Kevrekidis}}]{deflation}
\bibinfo{author}{\bibfnamefont{N.}~\bibnamefont{Boull\'e}},
  \bibinfo{author}{\bibfnamefont{E.~G.} \bibnamefont{Charalampidis}},
  \bibinfo{author}{\bibfnamefont{P.~E.} \bibnamefont{Farrell}},
  \bibnamefont{and} \bibinfo{author}{\bibfnamefont{P.~G.}
  \bibnamefont{Kevrekidis}}, \emph{\bibinfo{title}{{Deflation-based
  identification of nonlinear excitations of the three-dimensional
  Gross-Pitaevskii equation}}}, \bibinfo{journal}{Phys. Rev. A}
  \textbf{\bibinfo{volume}{102}}, \bibinfo{pages}{053307}
  (\bibinfo{year}{2020}).

\end{thebibliography}

\end{document}